\documentclass[12pt]{report}

\usepackage{pdfx}
\usepackage{pdfpages}

\newdimen \jot \jot=5mm
\usepackage[utf8]{inputenc} 
\DeclareUnicodeCharacter{00B0}{$\degree$}
\DeclareUnicodeCharacter{00B1}{$\pm$}
\DeclareUnicodeCharacter{00D7}{$\times$}
\usepackage[T1]{fontenc}
\usepackage{lmodern} 
\usepackage{tocbibind} 
\usepackage{setspace}
\usepackage[hypcap=true]{caption}
\hypersetup{linktocpage}
\usepackage{amsmath,amssymb,array}
\usepackage{booktabs}
\usepackage{subfig}

\usepackage{graphicx}
\usepackage{float}
\usepackage{tikz}
\usepackage{graphics} 
\usetikzlibrary{positioning}
\usetikzlibrary{shapes,arrows}
\usepackage{dcolumn}

\usepackage{color}
\makeatletter
\def\maxwidth{ %
  \ifdim\Gin@nat@width>\linewidth
    \linewidth
  \else
    \Gin@nat@width
  \fi
}
\makeatother

\definecolor{fgcolor}{rgb}{0.345, 0.345, 0.345}
\usepackage{framed}
\makeatletter
 {\par\unskip\endMakeFramed%
 \at@end@of@kframe}
\makeatother

\definecolor{shadecolor}{rgb}{.97, .97, .97}
\definecolor{messagecolor}{rgb}{0, 0, 0}
\definecolor{warningcolor}{rgb}{1, 0, 1}
\definecolor{errorcolor}{rgb}{1, 0, 0}
\makeatletter
\newcommand\gobblepars{%
    \@ifnextchar\par%
        {\expandafter\gobblepars\@gobble}%
        {}}
\makeatother

\usepackage[round]{natbib}
\usepackage{multirow}
\usepackage{gensymb}

\usepackage{enumerate}

\usepackage[raggedright]{titlesec}

\usepackage{ragged2e}
\usepackage{pdflscape}

\graphicspath{{rnw_chapter/figure/}{rnw_chapter/}} 

\begin{document}

\newcommand{\bm}[1]{ \mbox{\boldmath $ #1 $} }
\newcommand{\bin}[2]{\left(\begin{array}{@{}c@{}} #1 \\ #2
             \end{array}\right) }
\renewcommand{\contentsname}{Table of Contents}
\baselineskip=24pt
 
\pagenumbering{roman}
\thispagestyle{empty}
\begin{center}
\vspace*{.25in}
{\bf\large{KINETIC ENERGY FLUCTUATION-DRIVEN LOCOMOTOR TRANSITIONS 
ON POTENTIAL ENERGY LANDSCAPES OF BEAM OBSTACLE TRAVERSAL AND SELF-RIGHTING 
}}\\ 
\vspace*{.75in}
{\bf by} \\*[18pt]
\vspace*{.2in}
{\bf Ratan Sadanand Othayoth Mullankandy}\\ 
\vspace*{1in}
{\bf A dissertation submitted to Johns Hopkins University\\
in conformity with the requirements for the degree of\\
Doctor of Philosophy }\\
\vspace*{.75in}
{\bf Baltimore, Maryland} \\
{\bf October, 2021} \\     
\vspace*{.5in}
\begin{small}
{\bf \copyright{ }2021 by Ratan Sadanand Othayoth Mullankandy} \\ 
{\bf All rights reserved}
\end{small}
\end{center}
\newpage 

\pagestyle{plain}
\pagenumbering{roman}
\setcounter{page}{2}
\chapter*{Abstract}

When moving in nature, animal contend with constraints imposed by their environment, morphology, and physiology. Despite this seeming difficulty, animals move amazingly well by making effective physical interaction with the environment to use and transition between different modes of locomotion such as walking, running, climbing, and self-righting. By contrast, robots struggle to do so in real world. Understanding the principles of how locomotor transitions emerge from constrained physical interaction is necessary for robots to move robustly in nature by using and transitioning between locomotor modes. 

Recent studies of physical interaction with environment discovered that discoid cockroaches use and transition between diverse locomotor modes to traverse beams and self-right on ground. For both systems, animals probabilistically transitioned between modes via multiple pathways, while its self-propulsion created seemingly wasteful kinetic energy fluctuation. In this dissertation, we seek mechanistic explanations for these observations by adopting a physics-based approach that integrates biological and robotic studies. 

We discovered that animal and robot locomotor transitions during beam obstacle traversal and ground self-righting are barrier-crossing transitions on potential energy landscapes. Whereas animals and robot traversed stiff beams by rolling their body between beam, they pushed across flimsy beams, suggesting a concept of terradynamic favorability where modes with easier physical interaction are more likely to occur. Robotic beam traversal revealed that, system state either remains in a favorable mode or probabilistically transitions to one when kinetic energy fluctuation is comparable to the transition barrier. Robotic self-righting transitions occurred similarly and additionally revealed that changing system parameters (wing opening) lowers landscape barriers over which comparable kinetic energy fluctuation can induce probabilistic transitions. Animals’ transitions in both systems mostly occurred similarly, but sensory feedback may facilitate its beam traversal. Finally, we developed a method to measure animal movement across large spatiotemporal scales in an existing terrain treadmill. 

\chapter*{Dissertation Readers}

\section*{}

\begin{singlespace}

\indent Chen Li (Advisor)\\
\indent \indent Assistant Professor \\
\indent \indent Department of Mechanical Engineering\\
\indent \indent Johns Hopkins Whiting School of Engineering \\

\smallskip 

\noindent Noah J. Cowan \\
\indent \indent Professor\\
\indent \indent Department of Mechanical Engineering\\
\indent \indent Johns Hopkins Whiting School of Engineering \\

\smallskip 

\noindent Louis L. Whitcomb\\
\indent \indent Professor\\
\indent \indent Department of Mechanical Engineering\\
\indent \indent Johns Hopkins Whiting School of Engineering \\

\end{singlespace}


\pagestyle{plain}
\baselineskip=24pt
\tableofcontents

\listoftables
\listoffigures

\cleardoublepage 
\pagenumbering{arabic}

\chapter{Introduction}
\label{chap:intro}

\section{Motivation and Significance}
\begin{figure}
    \centering
    \includegraphics[width=1.0\linewidth]{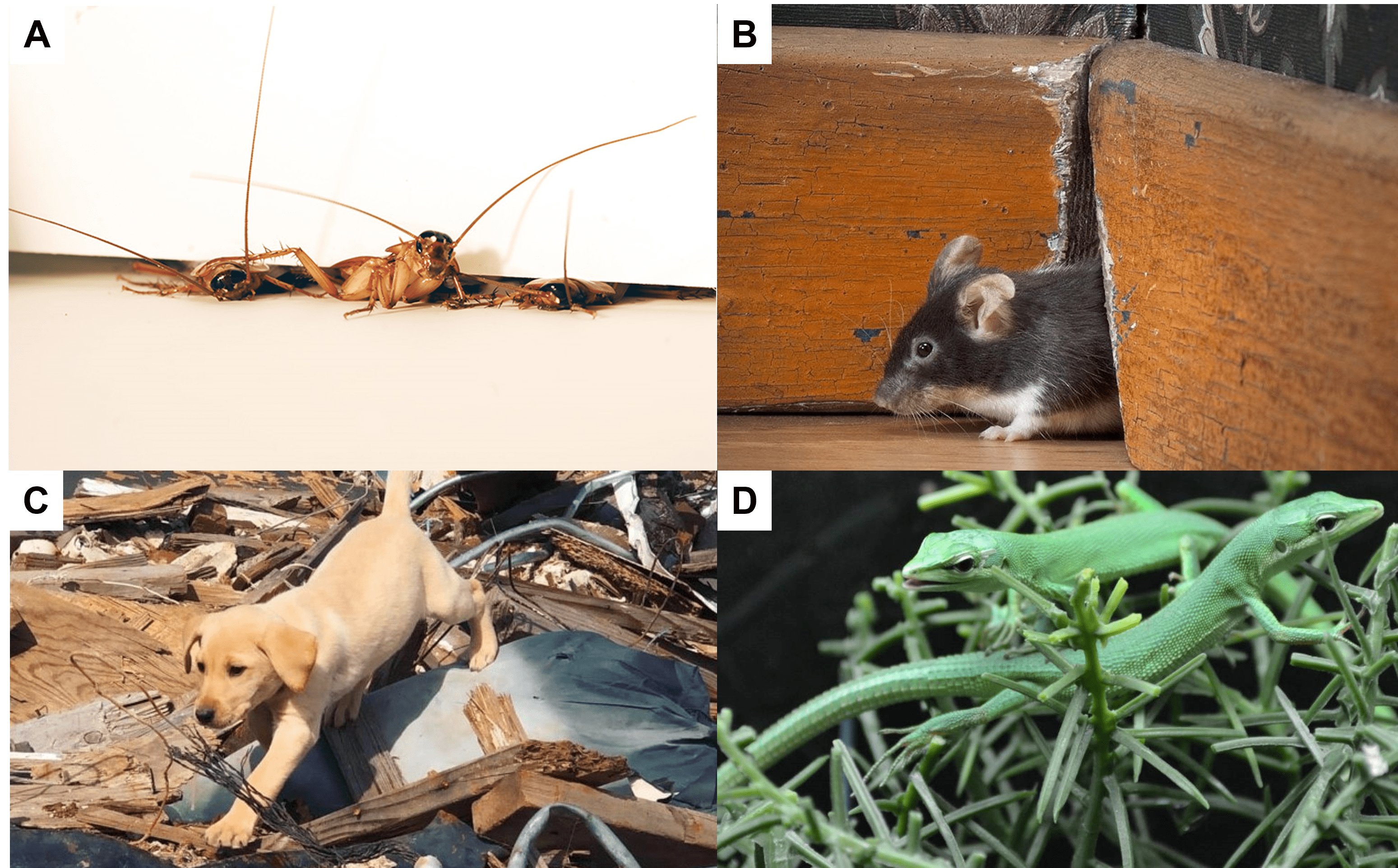}
    \caption[Physical interaction during terrestrial locomotion is constrained by environmental, morphological, and physiological constraints] %
    {Physical interaction during terrestrial locomotion is constrained by environmental, morphological, and physiological constraints. Examples of physical interaction during locomotion. (A) American cockroach exiting a crevice. (B) Mouse passing trough a vertical gap. (C) Search and Rescue dog moving in rubble. (D) Lizards moving through flexible shrubs. Images courtesy of (A) PolyPEDAL Lab, (B) Getty Images, (C) Ground Zero Emergency Canine Training, (D) LLL Reptile and Supply Co Inc.}
    \label{fig:i1_animal_example}
\end{figure}

Movement is one of the most ubiquitous and conspicuous features of animals \citep{alexander2006a,biewener2003a,tinbergen1955a} and occurs across diverse environments ranging from rainforests to deserts to flat plains, occurring over large spatiotemporal scale. For effective terrestrial locomotion in such natural environments, animals must generate necessary forces \citep{taylor1982a} to both support and propel themselves, which requires significant physical interaction of limbs and often the body with the terrain \citep{alexander2006a,biewener2003a,dickinson2000a}. However, when moving in real world, physical interaction between animal and environment during is often complex \citep{dickinson2000a,holmes2006a} even in simple, homogeneous environments. Because the environment is spatially and temporally variable, so is the physical interaction with it. In addition, to the heterogeneity and variation in physical properties such as shape, size, and stiffness of terrain, as well as continual animal-terrain contact animals must operate under the constraints imposed by their morphology, physiology, and environment (Figure \ref{fig:i1_animal_example}. These may limit the propulsive that the animal can generate \citep{dickinson2000a,taylor1982a} to move in environment.

\begin{figure}
    \centering
    \includegraphics[width=1.0\linewidth]{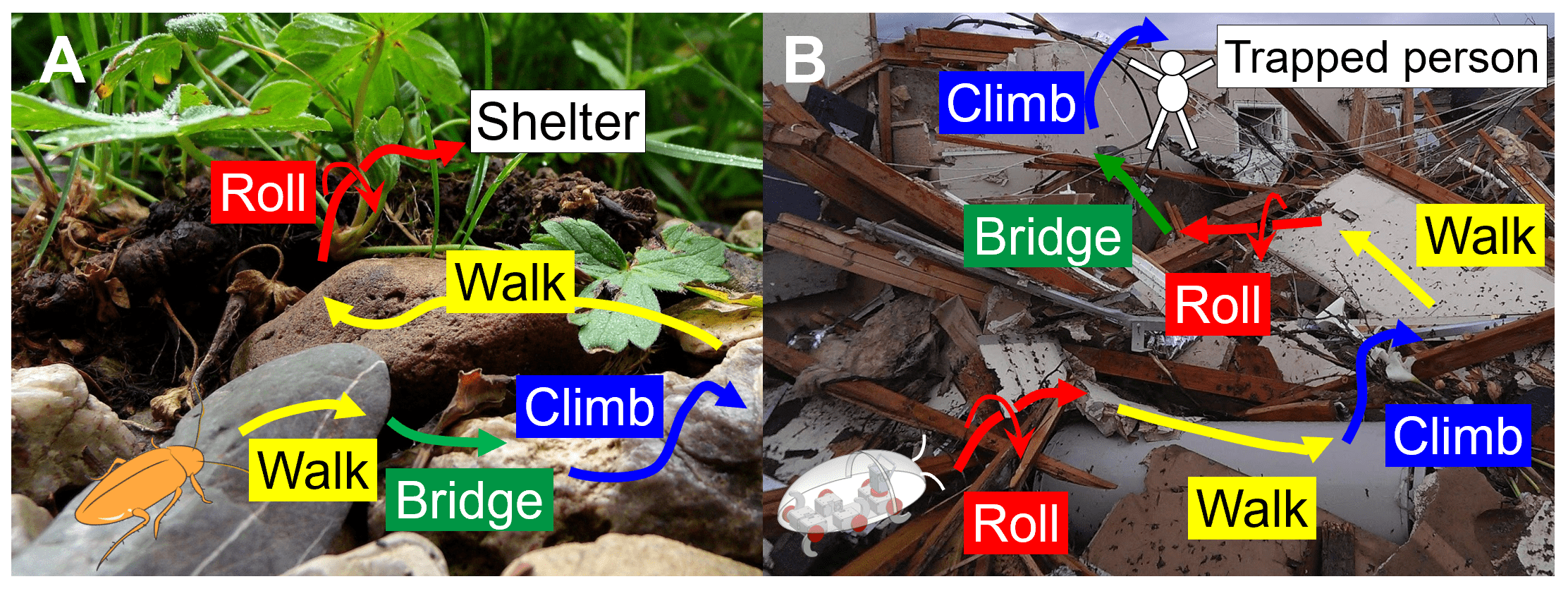}
    \caption[Illustrative examples of animal and robot locomotor transitions.] %
    {Illustrative examples of animal and robot locomotor transitions. Reproduced from \citep{othayoth2020a}}
    \label{fig:i2_transition_schematic}
\end{figure}

Despite these seeming difficulties during physical interaction, terrestrial locomotion in biological organisms is robust and agile. Even with perturbations from continual contact and instabilities in environment \citep{sponberg2008a,biewener2007a} and the biomechanical constraints \citep{biewener2003a,holmes2006a}, animals adjust their physical interaction to maintain dynamic running \citep{jindrich2002a,full2006a} and walking. Furthermore, in the extreme case of losing foothold and flipping over on their back, animals self-right to get back on their feet and continue moving \citep{ashe1970a,full1995a,li2019a}. Often, running or walking alone cannot accomplish effective locomotion in varying environments (e.g., cluttered forest floor, sparse branches in canopy, etc.) and animals must transition \citep{lock2013a,low2015a} between different modes of locomotion such as climbing \citep{goldman2006a}, rolling \citep{domokos2008a}, burrowing \citep{winter2014a}, or even self-righting \citep{li2019a} (Figure \ref{fig:i2_transition_schematic}).

By contrast, although advancements in robotics over the past decades \citep{altendorfer2001a,raibert1986a,raibert2008a} have enabled robots to walk and run stably even across surfaces that are rigid \citep{raibert1986a,raibert2008a}, rugged \citep{altendorfer2001a}, and yielding \citep{li2009a,li2013a,aguilar2016a}, they either lack the ability or struggle to robustly transition between modes beyond stable walking and running \citep{guizzo2015a,yang2018a}. It is crucial to be able to robustly transition between different modes of locomotion for robots to move effectively in natural, artificial or extraterrestrial environments (which often have physical constraints) to assist or autonomously perform tasks such as search and rescue \citep{murphy2008a}, environmental monitoring \citep{dunbabin2012a}, extraterrestrial exploration \citep{titus2021a,Li2022}, and home service \citep{forlizzi2006a}.

A major hurdle towards this grand vision in robotics is that beyond stable running and walking, we do not yet know how to effectively generate and use forces to move in environment when physical interaction is constrained by the robot’s environment and morphology. For example, how should the robot move its body and legs to exert forces against its environment to enter a cave opening \citep{titus2021a} narrower than its body, or regain foothold when fallen over? Without such capabilities, robots struggle or even fail to move in the real world. To enable these capabilities, we need general physics models that can inform us how to predict and generate the necessary forces \citep{koditschek2021a} to interact with environment for transitioning to, avoiding, or escaping from locomotor modes beyond walking and running. Improving robotic mobility in extreme environments is a grand challenge in robotics \citep{yang2018a}, and understanding the physics of locomotor transitions is an essential ingredient for addressing this challenge.

\begin{figure}[h]
    \centering
    \includegraphics[width=1.0\linewidth]{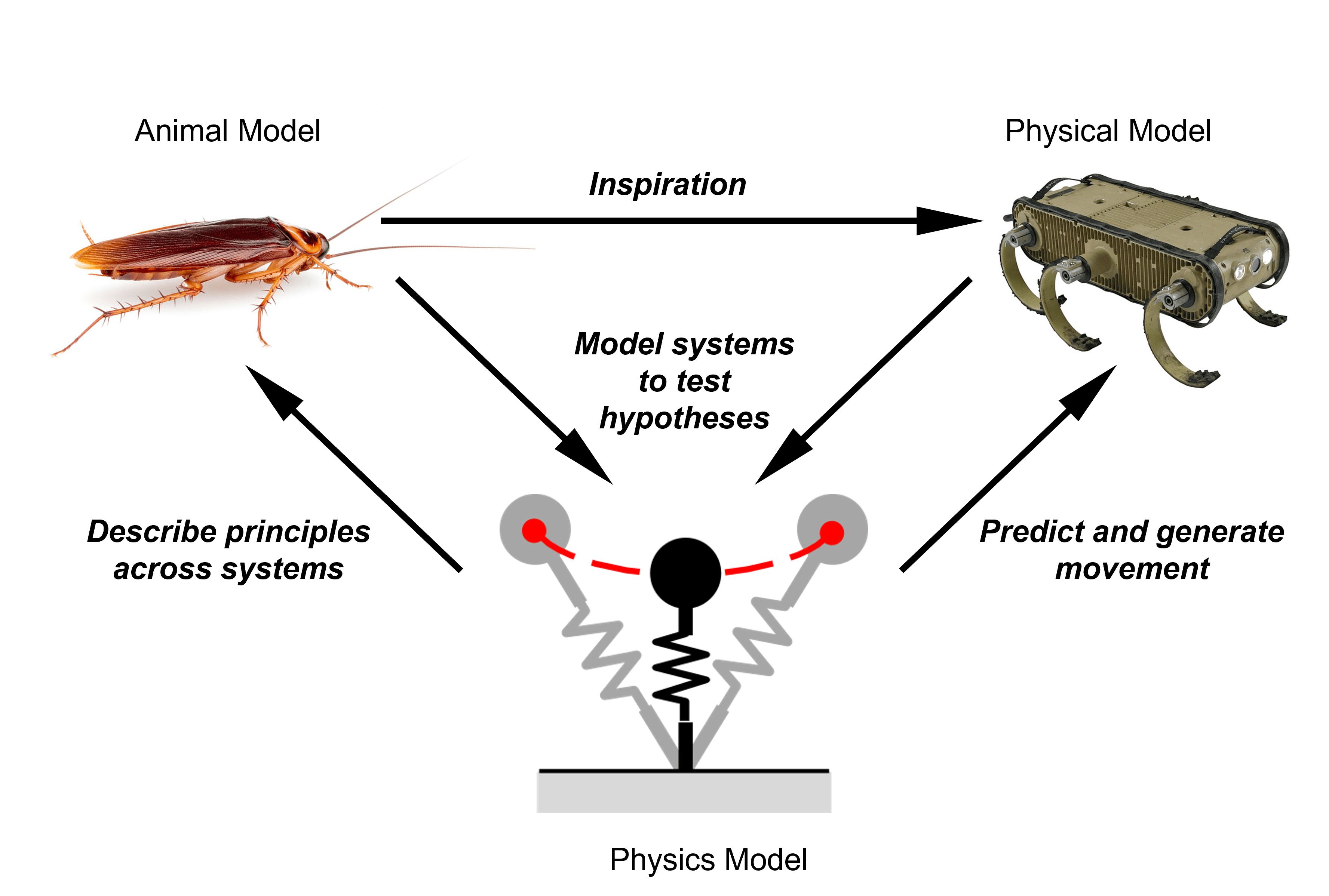}
    \caption[Bioinspired approach in legged locomotion. ] %
    {Bioinspired approach in legged locomotion. }
    \label{fig:i3_bioinspired}
\end{figure}

While the challenge is formidable, a bioinspired approach \citep{sharbafi2017a,lynch2012a,pfeifer2007a} proved tractable and successful. In this approach (Figure \ref{fig:i3_bioinspired}), the principles learnt from biological systems were translated to create engineering design rules and general physics models that predicted forces and informed robot design to improve their mobility. Such simplified physics models \citep{blickhan1993a,dickinson1999a,hu2009a,li2013a}, either derived analytically or synthesized from systematic experimental studies of animals and their robotic physical models \citep{aguilar2016b,aydin2019a}, have been successful in revealing principles of generating and maintaining steady state locomotion in modes such as walking \citep{kuo2007a}, running \citep{blickhan1993a}, vertical climbing \citep{goldman2006a}. For example, the simplest model of running on ground, spring-loaded inverted pendulum model (SLIP \citep{blickhan1993a}) was inspired from studying legged animals and have advanced the capability of robots to run stably and autonomously in moderately rugged environments \citep{altendorfer2001a}. It also described the fundamental dynamics of running and hopping in two-, four-, six- and eight-legged animals \citep{blickhan1993a}. In addition to improving robotic mobility, these physics models help understand general principles spanning different biological systems.

\clearpage
\section{Background}
\label{sec:background}
\begin{figure}
    \centering
    \includegraphics[width=1.0\linewidth]{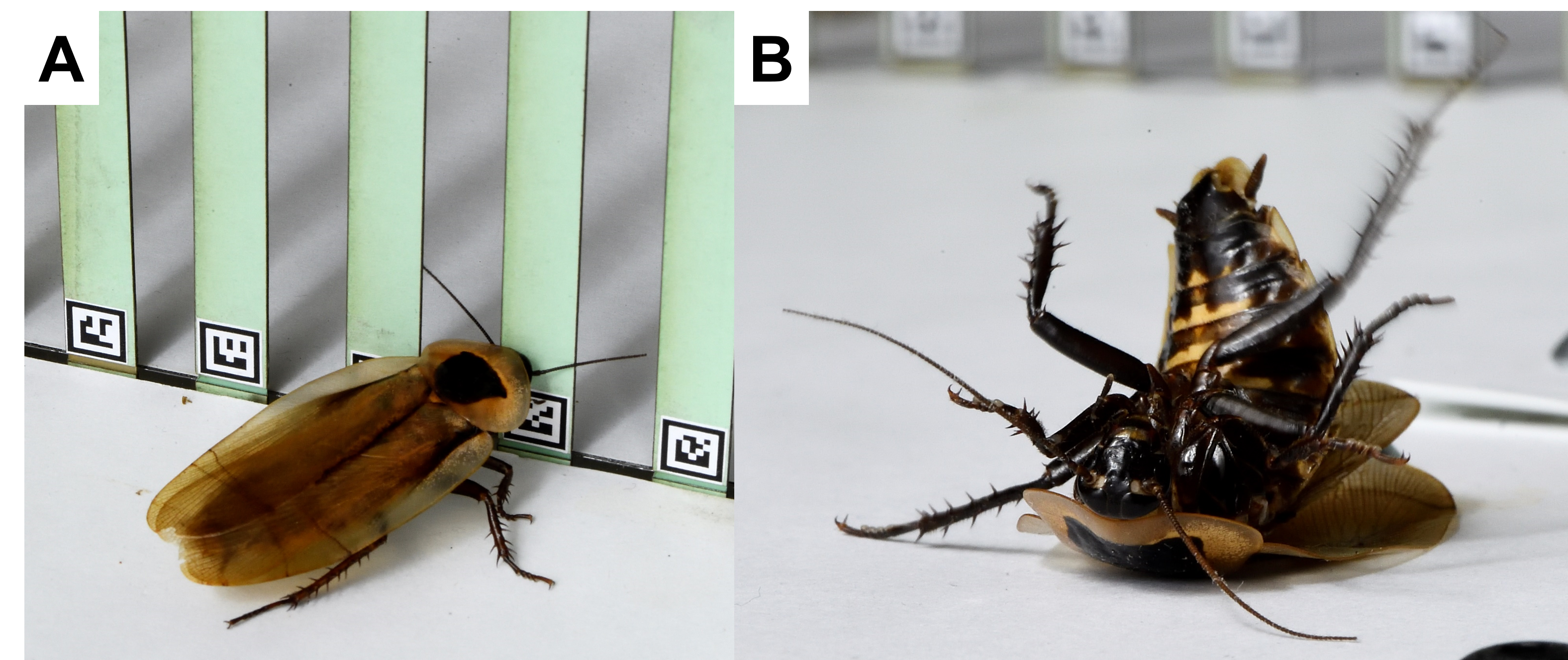}
    \caption[Model organism and model systems] %
    {Model organism and model systems. Discoid cockroach (\textit{Blaberus discoidalis}) (A) traversing flexible beams and (B) self-righting on flat ground. Images courtesy of Will Kirk.}
    \label{fig:i4_blab}
\end{figure}

Previous studies predominantly focused on generating \citep{blickhan1993a,goldman2006a,kuo2007a,li2012a}, stabilizing \citep{biewener2007a,couzin-fuchs2015a,revzen2013a} or transitioning between \citep{bramble2004a,diedrich1995a,hoyt1981a,li2000a} steady-state between walking and running. But insights from these studies do not translate to scenarios where the animal or robot must make locomotor transitions using physical interaction while operating under environmental or biomechanical constraints, which is often the case when moving in real world. The studies in this dissertation are motivated by the recent observations of physical interaction during traversal of flexible beam-like obstacles \citep{li2015a} and self-righting on flat ground \citep{li2019a} in discoid cockroaches (\textit{Blaberus discoidalis}, Figure \ref{fig:i4_blab}). 

In both model systems, the animal displayed diverse, probabilistic locomotor transitions that emerged via constrained physical interaction with its environment. To traverse flexible beam obstacles or self-right, animals must physically interact with the environment, which is often constrained or strenuous. For example, traversing layers of adjacent beams with a gap narrower than their body width is difficult for animals—so is pushing against stiff beams which may not deflect easily due to large restoring forces. Similarly, to self-right, animals must overcome potential energy barriers that are seven times greater than the mechanical energy required per stride for steady-state, medium speed running \citep{kram1997a} or, exert ground reaction forces eight times greater than that during steady-state medium speed running \citep{full1995a}. In the next sections, we briefly summarize the results from the previous studies of locomotor transitions in beam obstacle traversal \citep{li2015a} and ground self-righting of cockroaches \citep{li2019a}.

\subsection{Model system I: Beam obstacle traversal}
\begin{figure}
    \centering
    \includegraphics[width=1.0\linewidth]{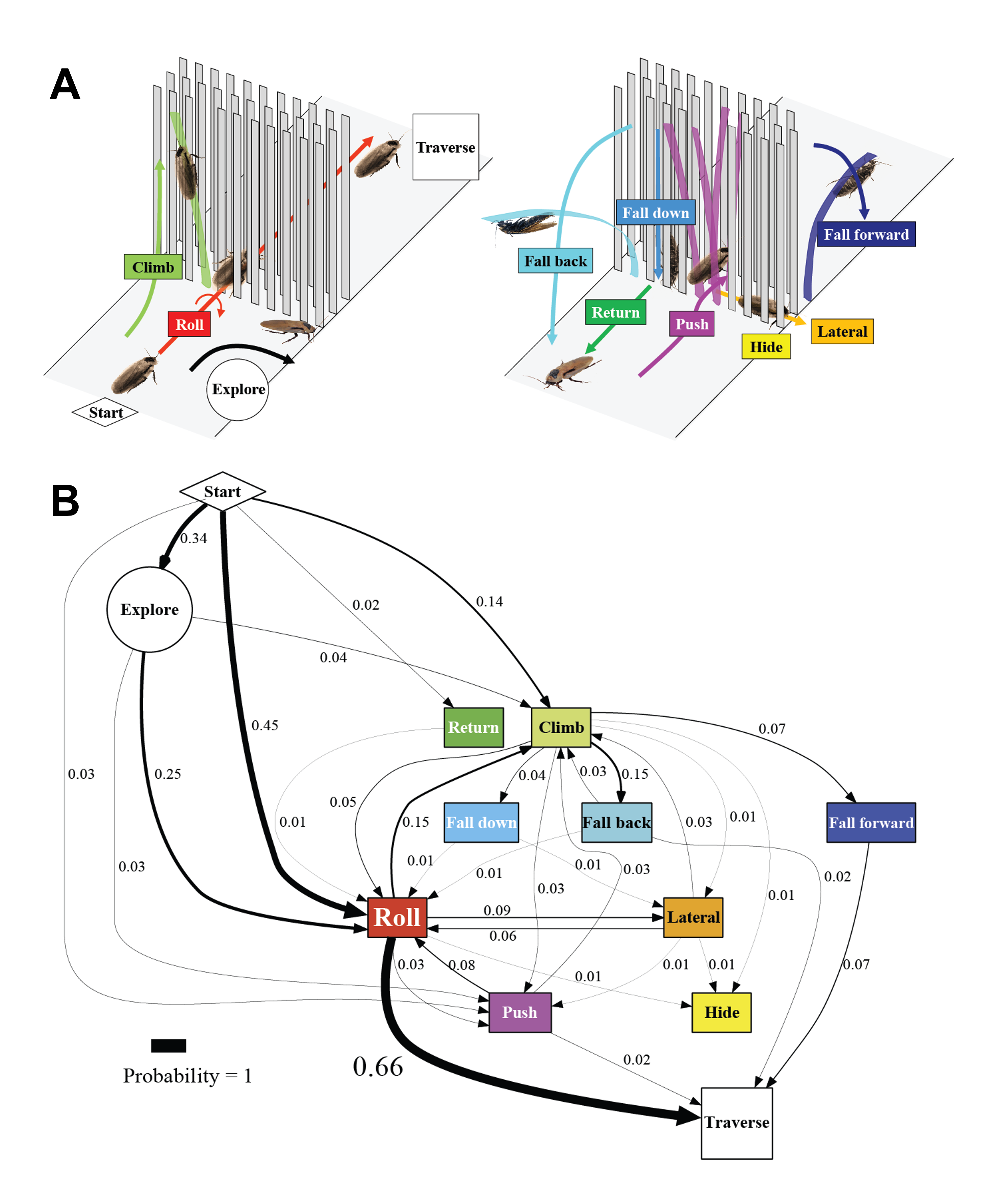}
    \caption[Model system I : Beam obstacle traversal] %
    {Model system I : Beam obstacle traversal. (A) Schematic of diverse locomotor modes observed during beam traversal. (B) Multi-pathway locomotor transitions and their probabilities during beam traversal. Reproduced from \citep{li2015a}}
    \label{fig:i5_chen_beams}
\end{figure}

To begin to advance terradynamics, (the study of locomotor-terrain interactions \citep{li2013a}) into three-dimensions, beyond relatively uniform granular media, \cite{li2015a} challenged cockroaches to traverse flexible beams with gaps smaller than their body width. The study discovered that during the physical interaction to traverse the beam obstacles, cockroaches used different modes such as climbing up the beams, rolling in between the beam gaps, pushing against the beams, falling forward, or moving laterally (Figure \ref{fig:i5_chen_beams}A). Animals probabilistically transitioned between different modes and did so via multiple pathways (Figure \ref{fig:i5_chen_beams}B) during traversal attempts. Some modes and transitions were more probable than other, with traversal most likely to occur via body rolling. This fact was attributed to the streamlined ellipsoidal shape of the cockroach, which induced body rolling from passive mechanical interactions. The animal also experienced constant body vibrations due to intermittent ground contact. A minimal potential energy landscape was used to speculate that during traversal, the animal must overcome a potential energy barrier (which varied with modes) (see Section \ref{sec:i_pel} for details).

\subsection{Model system II: Winged self-righting on flat ground}
\begin{figure}
    \centering
    \includegraphics[width=1.0\linewidth]{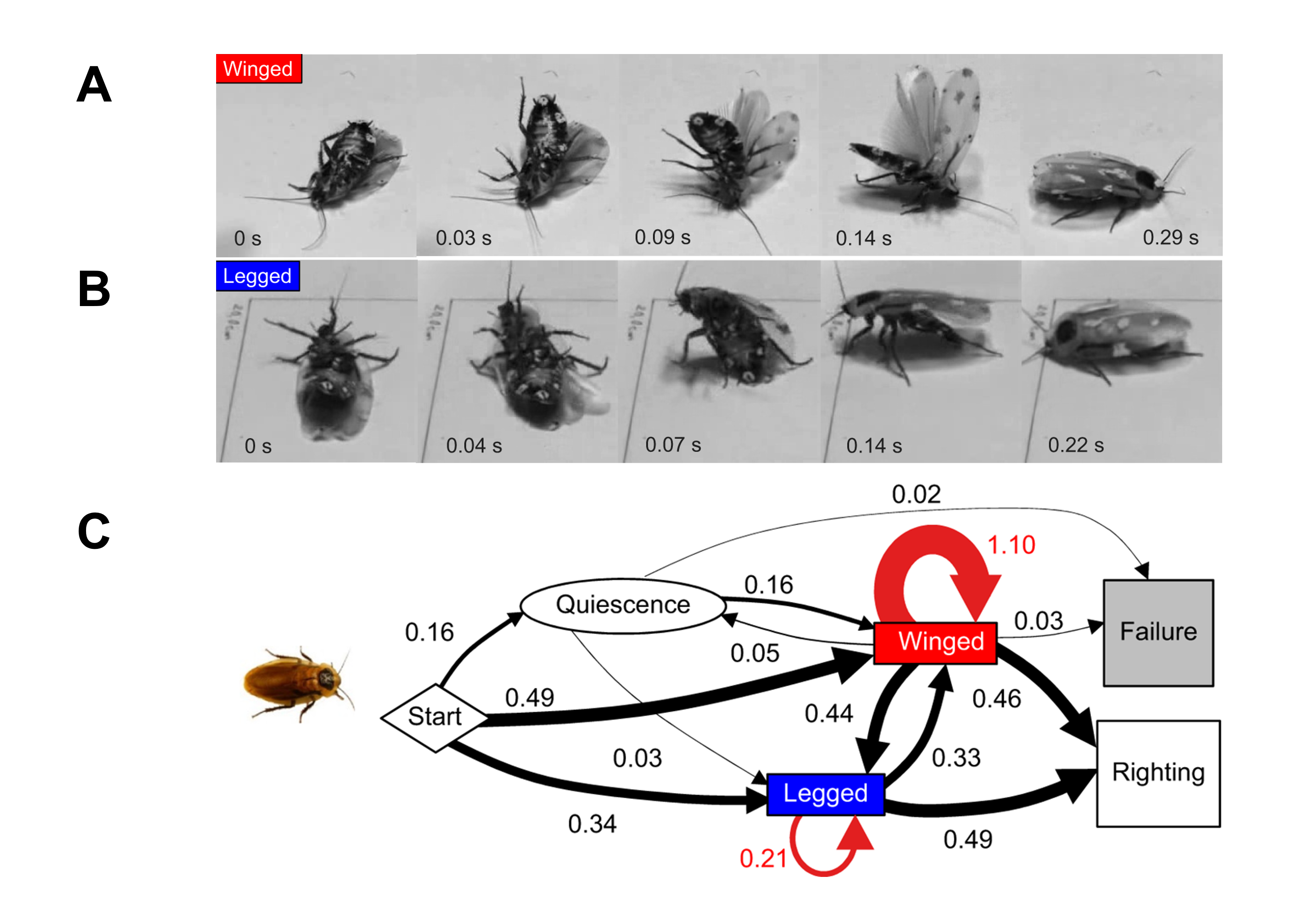}
    \caption[Model system II : Ground self-righting] %
    {Model system II : Ground self-righting (A) Winged and legged modes observed during self-righting. (B) Multi-pathway locomotor transitions and their probabilities during beam traversal. Reproduced from \citep{li2019a}}
    \label{fig:i6_chen_right}
\end{figure}

An interesting observation \citep{li2015a} in beam obstacle traversal led to a more detailed study on self-righting \citep{li2019a}. As the cockroach traversed and exited the beam obstacle field, it occasionally became unstable, lost its foothold, and fell on its back but almost always self-righted to continue moving. \cite{li2019a} discovered that cockroaches self-right on flat ground via diverse strategies such as by opening their wings against ground (winged self-righting) or by pushing their legs against the ground (legged self-righting) (Figure \ref{fig:i6_chen_right}A). It was also observed that physical interaction with flat ground resulted in probabilistic transitions between self-righting modes via multiple pathways Figure \ref{fig:i6_chen_right}. Certain modes and transitions were more likely to occur. In addition, the animal frequently and desperately flailed its legs during its self-righting attempts; it was presumed that these created small kinetic energy fluctuation. Finally, a static potential energy landscape model (that considered body rotation but not wing opening; see Section \ref{sec:i_pel} for details) observed that different modes overcame varying potential energy barriers.

These common observations suggest that despite their differences, studying both model systems together may help understand how the locomotor transitions emerge from physical interactions in beam traversal and self-righting.

\section{Knowledge gap and challenges}
\label{sec:kgap_chall}
The common themes across both model systems such as probabilistic, multi-pathway transitions, likelihood of some modes over others, could not be satisfactorily explained by the potential energy landscape modelling. More broadly, there exists a knowledge gap in our understanding of how animals make direct physical interaction with 3-D terrain to transition between locomotor modes, and how robots should do so too. For example, although the existing models such as spring-loaded inverted pendulum (SLIP) are effective and generalize over a broad range of locomotor-terrain parameters for dynamic running or walking, they do not inform or extrapolate to locomotor transitions. Bridging this knowledge gap, in addition to advancing to our understanding of how biological organisms move \citep{dickinson2000a,padilla2014a}, will also enable robots to move robustly in nature for relevant societal applications \citep{yang2018a}. 

Given the common observations across both model systems (see Section \ref{sec:background}) it is possible, as previous studies of beam traversal and self-righting have \textit{posited}, that the potential energy landscape could be a conceptual framework for thinking about how to generate and control locomotor transitions. However, it has not matured sufficiently to quantitatively reason about our observations of locomotor transitions or provide answers to above questions. This dissertation is a step towards advancing the potential energy landscape approach to be able to provide explanations for the above questions. 

However, such an effort has its challenges \citep{holmes2006a}—physical interaction during locomotor transitions often involves intermittent contact of the animal or robot with the environment, and their combined degrees of freedom are large. In addition, continual collisions and frictional contact introduce nonlinearities. Although the physical interaction obeys Newton’s laws of motion, solving (or even deriving) the equation for such complex systems is often intractable. Given these complexities, and the fact that transitions are probabilistic in both model systems, a statistical physics-like approach may prove useful. A statistical physics treatment has advanced understanding of complex, stochastic, macroscopic phase transitions in self-propelled living systems, such as animal foraging \citep{viswanathan2011a}, traffic \citep{helbing2001a}, and active matter \citep{fodor2018a,ramaswamy2010a}. 

Beginning to answer these questions requires an interdisciplinary approach integrating biology, robotics, and physics.

\clearpage
\section{Integrative approach for bridging the knowledge gap}
\subsection{Rationale and challenges in studying physical interaction in animals}
“The essential function of a robot is to perform work on its environment specified by its user \citep{koditschek2021a}”. In the same vein, at its most fundamental level, animals are mechanical systems that physically interact with the environment to propel themselves \citep{dickinson2000a}.  Investigating the mechanics of physical interaction with environment is a seemingly obvious first step towards understanding how organism, and how robot can, elicit (or avoid) a desired locomotor transition. 

\begin{figure}
    \centering
    \includegraphics[width=1.0\linewidth]{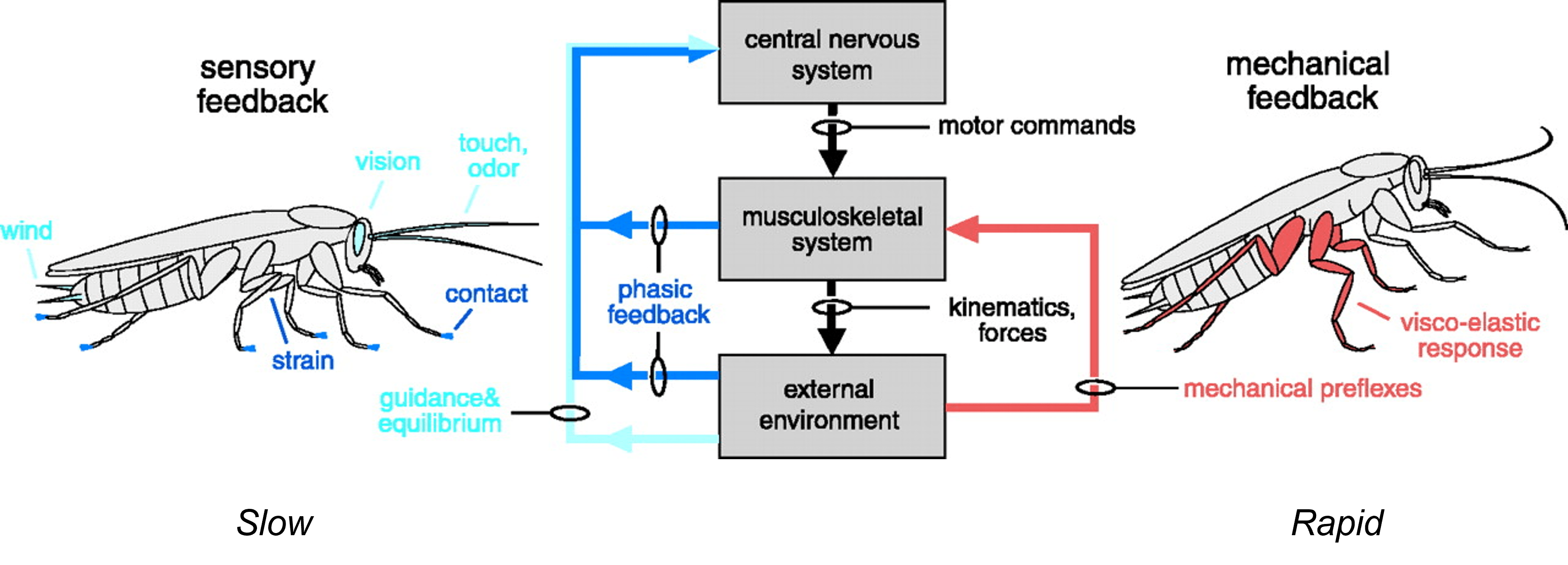}
    \caption[Neuromechanical feedback during locomotion] %
    {Neuromechanical feedback during locomotion. Neural sensory feedback (blue lines) is minimal  during rapid bandwidth-limited locomotion. Image modified from \cite{dickinson2000a}.}
    \label{fig:i7_neuro}
\end{figure}

However, locomotion in biological systems emerges \citep{dickinson2000a,anderson1972a} not just from mechanical interactions; animals can also sense and adjust its mechanical interaction in response to the sensed information via feedback, both neural and mechanical \citep{dickinson2000a} (Figure \ref{fig:i7_neuro}). For example, a when fetching balls, a dog adjusts its running speed and direction based on the sensory information from its eyes. As a result, even if our focus is to tease apart the role of passive physical interaction alone, we must consider the possibility that effects of sensory feedback may not be entirely avoidable.

Here, we first focus on understanding passive mechanical interaction, which provides a foundation for understanding sensory feedback control. This approach is inspired from early studies of aerodynamics of passive airfoils \citep{cayley1876a}. Although airfoils were extremely simplified models of bird wings, these studies a provided physics insights about flight control, which were lacking in detailed anatomical studies and observation of birds at the time. We follow a similar approach here by studying not just the animal, but also its simplified robotic model and compare them to gain physical insights. To make this comparison meaningful, we will study locomotor transitions of the animal during its rapid, bandwidth-limited locomotion during which sensory feedback is minimal due to delays in neuronal transmission speeds (Figure \ref{fig:i7_neuro}).

\subsection{Integrative approach to studying physical interaction during locomotion}
\begin{figure}[p]
    \centering
    \includegraphics[width=1.0\linewidth]{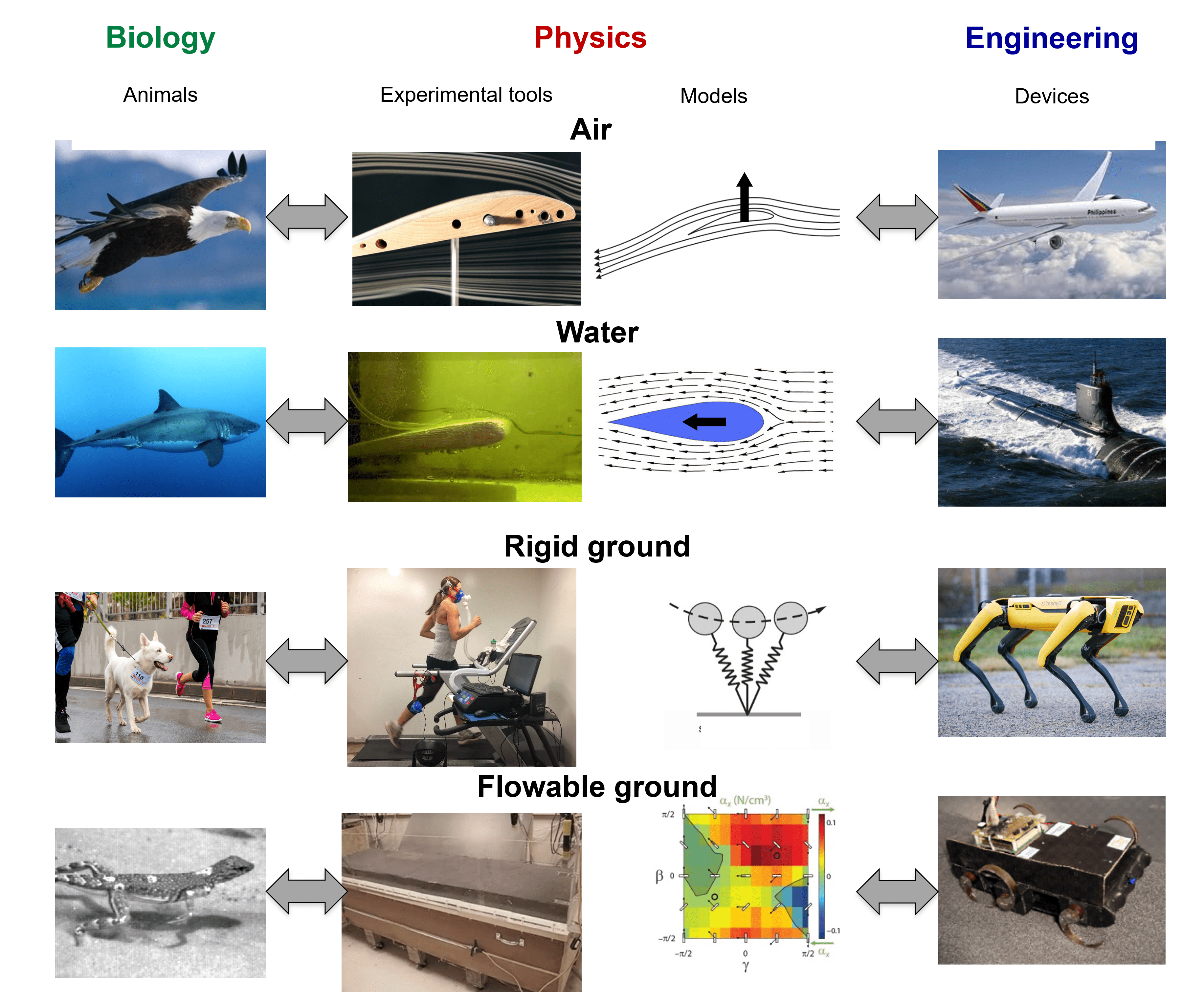}
    \caption[Integrative approach to understand physical interaction in different media] %
    {Integrative approach to understand physical interaction in different media.}
    \label{fig:i7_approach}
\end{figure}

Considering the challenges discussed in Section \ref{sec:kgap_chall} and the coupled neuromechanical interactions, studying physical interaction during locomotor transitions seems daunting at first, but a physics-based approach that has begun to advance terradynamics \citep{li2013a,li2015a} (the study of locomotor-terrain interactions) may prove beneficial. Such an approach, rooted in physics and integrating biology and robots has previously helped understand fluid–structure physical interaction in aerial and aquatic locomotion of animals \citep{dickinson1999a,lauder2002a} and robots \citep{teoh2013a,zhu2019a} , we understand fairly well their thanks to well-established experimental, theoretical and computational tools, such as wind tunnel and water channel, airfoil and hydrofoil, aero- and hydrodynamic theories, and computational fluid dynamics techniques \citep{vogel1996a}(Figure \ref{fig:i7_approach}). 

By creating controlled granular media testbeds, robotic physical models, and theoretical and computational models, recent studies elucidated how animals (and how robots should) use physical interaction with granular media to move effectively both on and within sandy terrain. (see \citep{goldman2014a} for a review). The general physical principles \citep{goldman2014a} and predictive physics models \citep{goldman2014a,li2013a} not only advanced understanding of functional morphology \citep{li2012a,maladen2011a,sharpe2015a}, muscular control \citep{ding2013a,sharpe2013a}, and evolution \citep{mcinroe2016a} of animals, but also led to new design and control strategies \citep{aguilar2016b,goldman2014a,li2009a,li2010a,marvi2014a,shrivastava2020a} that enabled a diversity of robots to traverse granular environments.

\clearpage
Such an integrated approach offers several advantages. 
\begin{itemize}
    \item Observations of model organisms inspire robot design and action. For example, insights from studying walking and running of cockroaches inspired the design of the RHex robot \citep{altendorfer2001a}.
    \item Simplified robots serve as physical models for testing biological hypotheses or generating new ones and allow control and variation of parameters to discover general principles. For example, the RHex-like robot was used to demonstrate that physical interaction of robots and animals with vertical pillar obstacles depends sensitively on robot body shape but not the pillar shape or geometry \citep{han2021a}.
    \item 	Physical principles and predictive models from this empirical approach provide mechanistic explanations for animal locomotion and design tools and action strategies for robots. For example, a robotic testbed (Robofly) provided mechanistic insights into how insects fly by flapping their wings \citep{dickinson1999a}, which later informed design of miniature flying robot – Robobee \citep{teoh2013a}.
\end{itemize}

Inspired by these successes, we will adopt a similar physics-based approach by integrating biological and robotic studies of beam traversal and self-righting with potential energy landscape modelling of the physical interaction.

\clearpage
\section{Potential energy landscapes to model physical interaction}
\label{sec:i_pel}
\begin{figure}[h]
    \centering
    \includegraphics[width=1.0\linewidth]{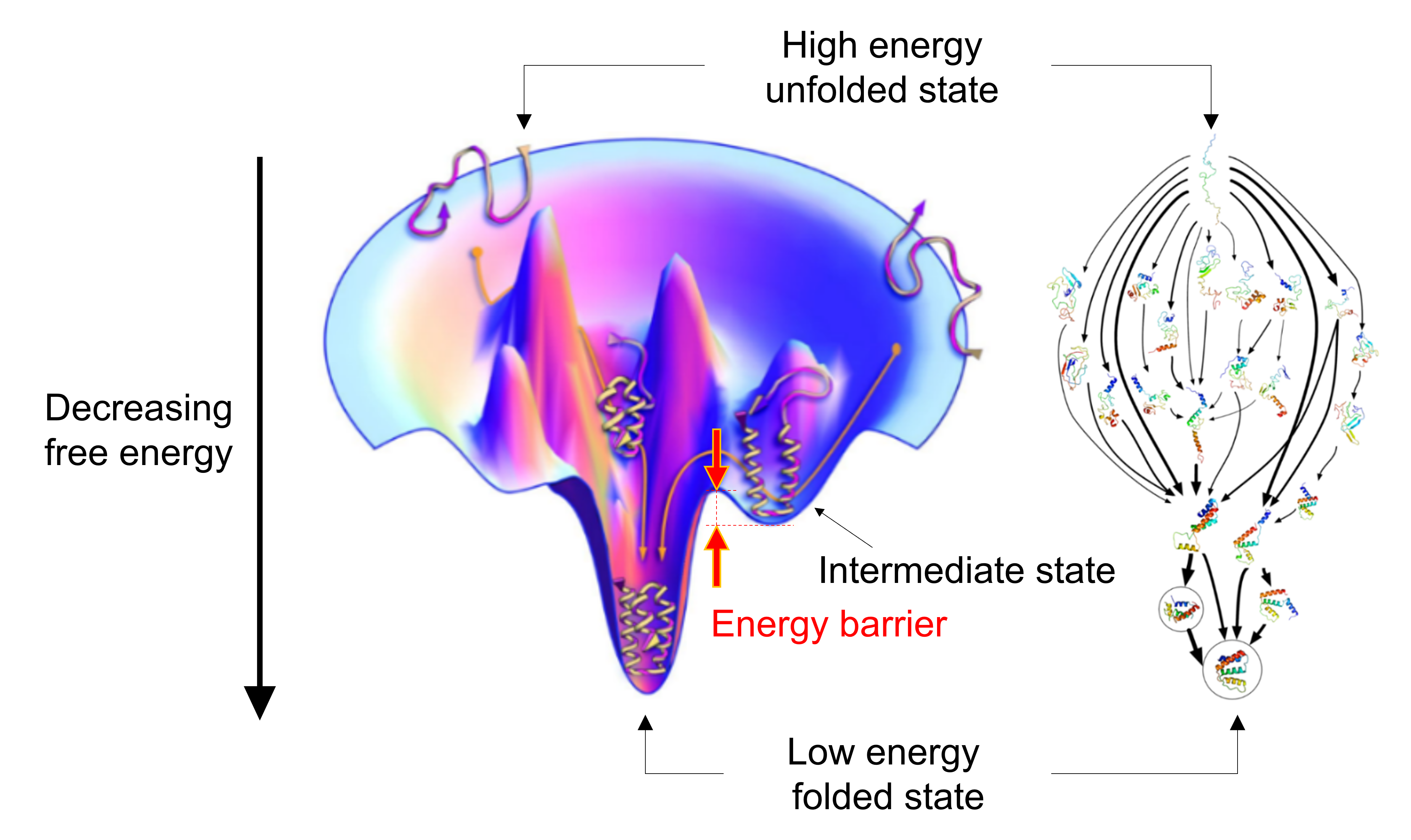}
    \caption[Free energy landscapes of protein folding transitions.] %
    {Free energy landscapes of protein folding transitions. Left: Funnel shaped free energy landscapes of protein folding. As proteins fold, their state moves transitions from a high energy state to a low energy state by overcoming free energy barriers. Modified from \citep{dill2012a}. Right: Schematic of stochastic, multi-pathway protein folding transitions. Thicker arrows show transitions that are more probable. Modified from \citep{voelz2012a}}
    \label{fig:i9_protein}
\end{figure}

\subsection{Inspiration: Free energy landscapes of protein folding transitions}
The potential energy landscapes for modelling physical interaction during locomotor transition are directly inspired by free energy landscapes of multi-pathway protein folding transitions \citep{dill1997a,dill2008a,onuchic2004a,wales2003a}. Free energy landscape is a function of all possible confirmations (molecular structure) of a protein molecule (Figure \ref{fig:i9_protein}). Microscopic, near-equilibrium proteins have a three-dimensional chain-like conformation which is initially unfolded and has a high free energy. To achieve its biological function, the unfolded protein chain must fold into a specific confirmation called the native state, which has the lowest possible free energy. However, the transition from unfolded to native state does not occur in a single step; instead, unfolded states progressively transition to various intermediate states that are reaching the native state. Lower free energy states are more stable and hence thermodynamically favorable. When the protein folding problem is seen through the lens of free energy landscapes (Figure \ref{fig:i9_protein}), following observations emerge:
\begin{enumerate}

    \item 	Free energy landscapes have peaks (local free energy maxima) and basins (local free energy minima). 
    \item 	When proteins fold, they from higher energy states/basins to lower energy states/basins. In other words, proteins transition towards thermodynamically favorable states. 
    \item The transitions from high energy state/basin to a low energy state/basin requires overcoming a substantial free energy barrier separating the basins, which is enabled by the random thermal energy fluctuations. Hence, transitions are probabilistic. 
    \item 	Proteins can probabilistically transition via multiple pathways. However, some transitions are more likely than others depending on the thermodynamic favorability, barrier height, and available thermal energy fluctuation.
    \item 	In addition to thermal energy fluctuation, transitions can also be enabled modifying the landscape to lower the transition barrier.

\end{enumerate}

Although our model systems \citep{li2015a,li2019a} of beam traversal and self-righting are macroscopic, self-propelled, and far-from-equilibrium, their locomotor transitions share several similarities such as diverse locomotor modes, multi-pathway transitions between locomotor modes that occur probabilistically, and preference of some modes over others Inspired by the seeming similarities of our system to them, we contend that the potential energy landscape approach helps understand how self-propelled, far-from-equilibrium macroscopic animals’ and robots’ probabilistic locomotor transitions during traversal of flexible beam obstacles and self-righting on flat ground emerge from physical interaction, whose equations of motion are unknown or intractable \citep{aguilar2016b,han2021a}. 

\subsection{Hypotheses}
Having reasoned about the validity of using potential energy landscape approach, we present the hypotheses that we seek to resolve in this dissertation:
\begin{enumerate}

    \item 	Are locomotor transitions of beam traversal and self-righting system barrier-crossing transitions on evolving potential energy landscapes? 
    \item 	When it is comparable to the potential energy barriers between basins, can the kinetic energy fluctuation observed during beam traversal and self-righting help escape from a basin to make locomotor transitions for traversal and self-righting? 
    \item 	When kinetic energy fluctuation is not sufficient to escape barriers along certain direction, is it possible to alter the landscape to lower the barriers to be comparable to available kinetic energy fluctuation and induce transitions?
    \item 	Analogous to thermodynamic favorability of protein states, do locomotor modes have terradynamic favorability? If so, is the locomotor-terrain system more likely to transition to a terradynamically favorable modes.
    \item 	How can we begin to quantify physical interaction and transitions at larger spatiotemporal scales?

\end{enumerate}

These hypotheses have only been speculated and not tested in previous studies of beam traversal and self-righting.

\subsection{Drawbacks of previous potential energy landscapes}
\label{subsec:pel_drawbacks}
\begin{figure}
    \centering
    \includegraphics[width=1.0\linewidth]{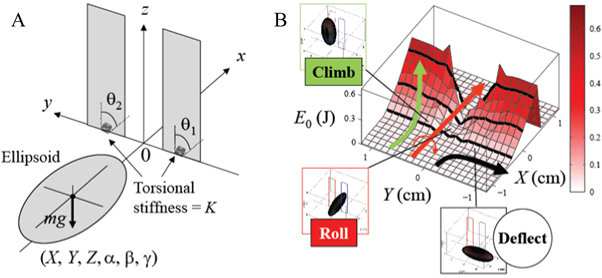}
    \caption[Minimal potential energy landscape of flexible beam traversal] %
    {Minimal potential energy landscape of flexible beam traversal.(A) Simplified physics model to calculate system potential energy. (B) Potential energy landscape in the space of forward ($X$) and lateral ($Y$) body center of mass position. Colored arrows indicate hypothesized systems state trajectories on landscape. Roll mode (red arrow) overcomes lower barrier on landscape during traversal compared to climb mode (green arrow). Deflect mode (black arrow) does not result in traversal.  Reproduced from \citep{li2015a}.}
    \label{fig:i10_chen_pel_beam}
\end{figure}

\begin{figure}
    \centering
    \includegraphics[width=1.0\linewidth]{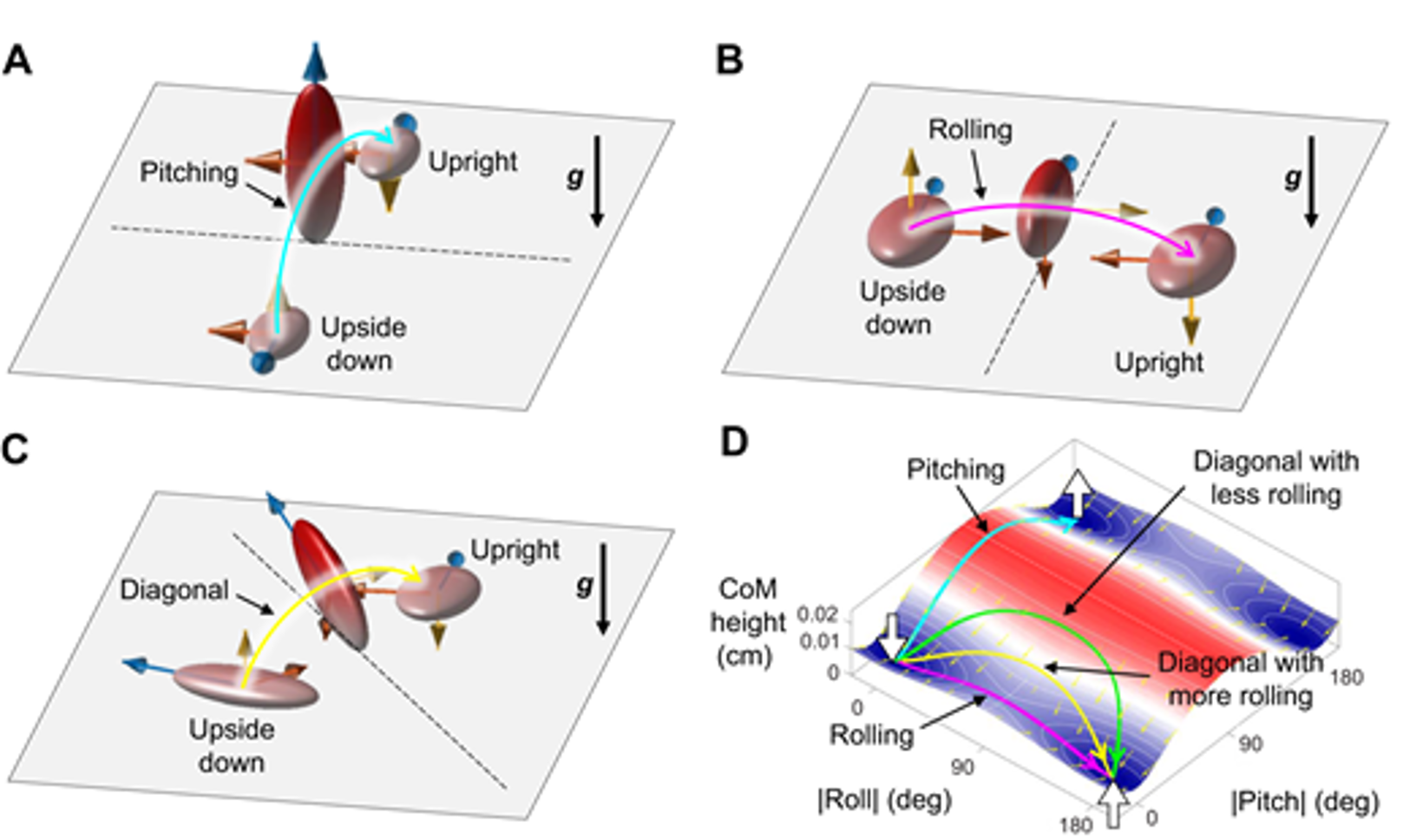}
    \caption[Static potential energy landscape of ground self-righting.] %
    {Static potential energy landscape of ground self-righting. (A-C) An ellipsoid approximating the animal body in contact with the ground, pitching (A), rolling (B) or rotating diagonally (simultaneous pitching and rolling; (D), Potential energy landscape, shown as the center of mass height in the space of body pitch and roll. Colored arrows indicate hypothesized system state trajectories for body rotations of modes shown in (A)-(C)  Reproduced from \citep{li2019a}}
    \label{fig:i11_chen_pel_right}
\end{figure}

Previous studies of beam obstacle traversal \citep{li2015a} and self-righting \citep{li2019a} used a simple physics model to obtain the system potential energy and visualize the potential energy landscape. To calculate system potential energy, following approximations were used to obtain simple physics models. In both model systems, the animal was approximated as a rigid ellipsoid with uniform density while ignoring legs and wings. The lowest point on the rigid ellipsoid was always assumed to be in contact with ground. In addition, the beams were approximated as rigid plates attached to the ground through torsional spring joints (Figure \ref{fig:i10_chen_pel_beam}). For both systems, friction and other non-conservative effects were not considered. System potential energy for self-righting system was the gravitational potential energy of the body (Figure \ref{fig:i11_chen_pel_right}), whereas that for beam traversal was the sum of body gravitational potential energy and beam deflection elastic energy. While these landscapes provided initial qualitative explanations, they had a few shortcomings.

In both these simplified potential energy landscape models (Figures \ref{fig:i10_chen_pel_beam}, \ref{fig:i11_chen_pel_right}), the systems potential energy depends on more than two parameters. The previous beam traversal landscape was a minimal potential energy landscape; \cite{li2015a} simply calculated the minimal potential energy over all rotational degrees of freedom for positions in the vicinity of the beam obstacles. Similarly in self-righting landscape, the animal’s wing opening, which substantially changes potential energy, are not considered. Finally, resolving these hypotheses will provide quantitative observations of how system state behaves on potential energy landscapes during physical interaction, which have been only hypothesized in previous studies. 

\clearpage
\section{Organization of Chapters}

\begin{itemize}
\item Chapter 2 details the biological, robotic, and physics studies of physical interaction during beam traversal.
\item Chapter 3 details the biological, robotic, and physics studies of physical interaction during self-righting on flat ground.
\item Chapter 4 discusses methods to track and analyze animal kinematics data collected during movement over large spatiotemporal scales using a previously built terrain treadmill.
\item Chapter 5 summarize the discoveries from Chapters 2-4 and their implications.
\end{itemize}

\cleardoublepage

\chapter{Kinetic energy fluctuation from oscillatory self-propulsion facilitates barrier-crossing locomotor transitions during beam traversal}
\label{chap:pnas}

\let\thefootnote\relax\footnotetext{This chapter is a published paper by Ratan Othayoth, George Thoms, and Chen Li in \textit{The Proceedings of the National Academy of Sciences} (2020) (\cite{othayoth2020a})}

\section{Summary}
Effective locomotion in nature happens by transitioning across multiple modes (e.g., walk, run, climb). Despite this, far more mechanistic understanding of terrestrial locomotion has been on how to generate and stabilize around near-steady-state movement in a single mode. We still know little about how locomotor transitions emerge from physical interaction with complex terrain. Consequently, robots largely rely on geometric maps to avoid obstacles, not traverse them. Recent studies revealed that locomotor transitions in complex 3-D terrain occur probabilistically via multiple pathways. Here, we show that an energy landscape approach elucidates the underlying physical principles. We discovered that locomotor transitions of animals and robots self-propelled through complex 3-D terrain correspond to barrier-crossing transitions on a potential energy landscape. Locomotor modes are attracted to landscape basins separated by potential energy barriers. Kinetic energy fluctuation from oscillatory self-propulsion helps the system stochastically escape from one basin and reach another to make transitions. Escape is more likely towards lower barrier direction. These principles are surprisingly similar to those of near-equilibrium, microscopic systems. Analogous to free energy landscapes for multi-pathway protein folding transitions, our energy landscape approach from first principles is the beginning of a statistical physics theory of multi-pathway locomotor transitions in complex terrain. This will not only help understand how the organization of animal behavior emerges from multi-scale interactions between their neural and mechanical systems and the physical environment, but also guide robot design, control, and planning over the large, intractable locomotor-terrain parameter space to generate robust locomotor transitions through the real world.

\section{Author contributions}
Ratan Othayoth designed study, developed robotic physical model, performed animal and robot experiments, analyzed data, developed energy landscape model, drafted and revised the paper; George Thoms developed robotic physical model and performed preliminary robot experiments; Chen Li designed and oversaw study, defined analyses, and wrote and revised the paper.
\clearpage
\section{Introduction}
To move about in the environment, animals can use many modes  of locomotion (e.g., walk, run, crawl, climb, fly, swim, jump, burrow) \citep{alexander2006a,biewener2003a,dickinson2000a} and must often transition across them \citep{lock2013a,low2015a} (e.g., Figure \ref{fig:i2_transition_schematic}A). Despite this, far more of our mechanistic understanding of terrestrial locomotion has been on how animals generate \citep{blickhan1993a,goldman2006a,hu2009a,kuo2007a,li2012a} and stabilize \citep{biewener2007a,couzin-fuchs2015a,revzen2013a} steady-state, limit-cycle-like locomotion using a single mode.

Recent studies begin to reveal how terrestrial animals transition across locomotor modes in complex environments. Locomotor transitions, like other animal behavior, emerge from multi-scale interactions of the animal and external environment across the neural, postural, navigational, and ecological levels \citep{berman2018a,brown2018a,nathan2008a}. At the neural level, terrestrial animals can use central pattern generators \citep{ijspeert2008a} and sensory information \citep{blaesing2004a,kohlsdorf2006a,ritzmann2012a} to switch locomotor modes to traverse different media or overcome obstacles. At the ecological level, terrestrial animals foraging across natural landscapes switch locomotor modes to minimize metabolic cost \citep{shepard2013a}. At the intermediate level, terrestrial animals also transition between walking and running to save energy \citep{bramble2004a}. However, there remains a knowledge gap in how locomotor transitions in complex terrain emerge from direct physical interaction (i.e., terradynamics \citep{li2013a}) of an animal’s body and appendages with the environment. In particular, we lack theoretical concepts for thinking about how to generate and control locomotor transitions in complex terrain that are on the same level of limit cycles for single-mode locomotion (25). For example, locomotion in irregular terrain with repeated perturbations is rarely near steady state and requires an animal to continually modify its behavior, which cannot be well described by limit cycles \citep{spagna2007a,sponberg2008a}.

Understanding of how to make use of physical interaction with complex terrain (environmental affordance \citep{gibson2014a,roberts2020a}) to generate and control locomotor transitions is also critical to advancing mobile robotics. Similar to personal computers decades ago, mobile robots are on the verge of becoming a part of society. Some robots (e.g., robot vacuums, self-driving cars) already excel at navigating flat surfaces, by transitioning across driving modes (e.g., forward drive, U-turn, stop, park \citep{thrun2010a}) to avoid sparse obstacles using a geometric map of the environment \citep{latombe2012a}. However, many critical applications, such as search and rescue in rubble, inspection and monitoring in buildings, extraterrestrial exploration through rocks, and even drug delivery inside a human body, require robots to transition across diverse locomotor modes to traverse unavoidable obstacles in complex terrain \citep{hu2018a,lock2013a,low2015a} (Figure \ref{fig:i2_transition_schematic}B)). Yet, terrestrial robots still struggle to do so robustly \citep{guizzo2015a}, because we do not understand well how locomotor transitions (or lack thereof) emerge from physical interaction with complex terrain.

\begin{figure}
    \centering
    \includegraphics[width=1.0\linewidth]{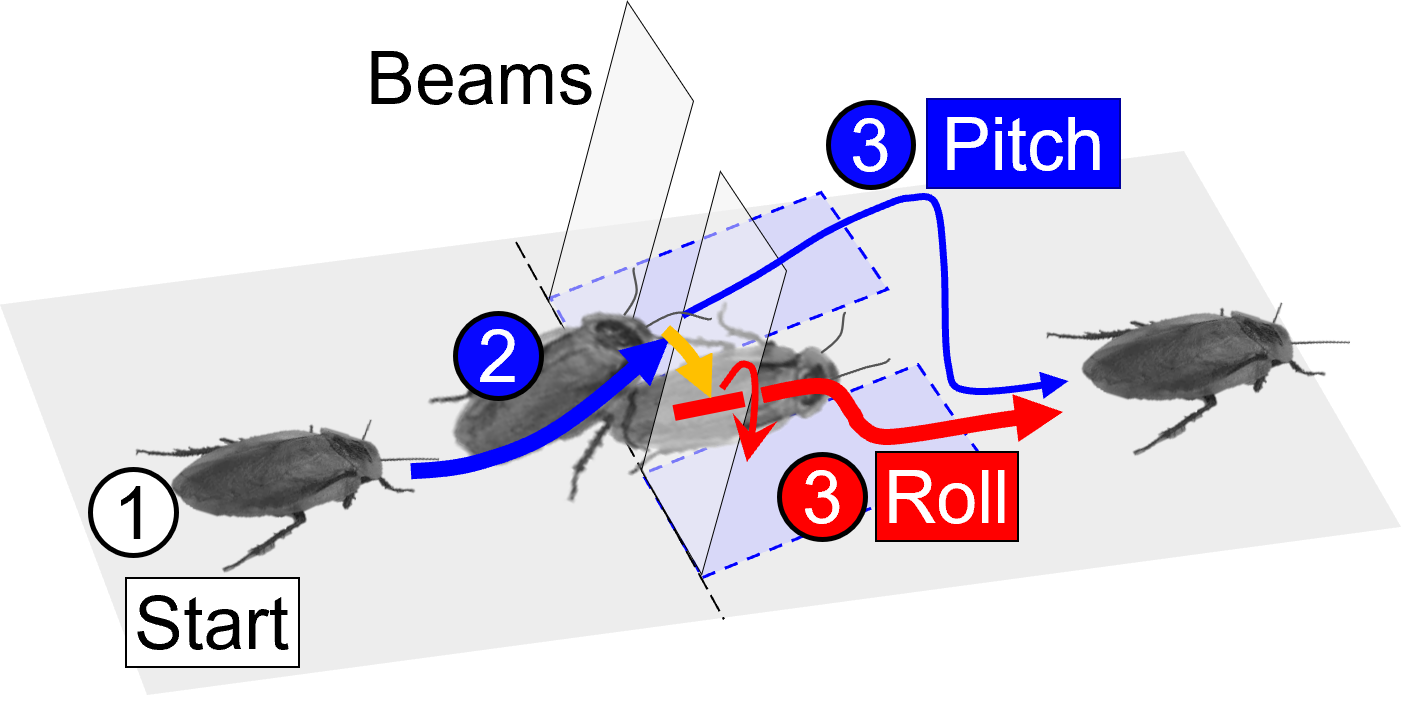}
    \caption[A cockroach transitioning (orange arrow)
from pitch to roll mode to traverse grass-like beam obstacles.]{A cockroach transitioning (orange arrow) from pitch to roll mode to traverse grass-like beam obstacles.}
    \label{fig:p1_1C}
\end{figure}

Our study is motivated by recent observations in a model system of insects traversing complex 3-D terrain. The discoid cockroach, native to rainforest floor, can traverse flexible, grass-like beam obstacles using many locomotor modes, stochastically transitioning across them via multiple pathways \citep{li2015a}. For simplicity, hereafter we focus on the transition between two modes. The animal often first pushes against the beams, and beam elastic restoring forces lead the animal body to pitch up (Figure \ref{fig:p1_1C}), blue). After this, though, the animal rarely pushes across (3\% probability) but often rolls (Figure \ref{fig:p1_1C}), red) to maneuver through beam gaps (45\% probability). We define these as “pitch” and “roll” modes. Note that we use “locomotor mode” here in the general sense, not confined to limit-cycle locomotor behavior. The pitch mode is more challenging than the roll mode because the animal has to lift its weight and deflect the beams more (this is only true when beams are stiff, though; see Results). Thus, the animal appears to statistically transition from less to more favorable modes. In addition, the animal’s body oscillates as its legs continually pushed against the ground when trying to traverse. Besides in obstacle traversal, similar multi-pathway locomotor transitions, preference of some modes over others, and seemingly wasteful body oscillation were observed in self-righting of insects \citep{li2019a}.

In the field of protein folding, adopting a statistical physics view and using an energy landscape approach led researchers to recognize that proteins fold via multiple pathways and understand the physical principles \citep{dill2008a,onuchic2004a,wales2003a}. These near-equilibrium, microscopic systems statistically transition from higher to lower energy states (local minima) on a free energy landscape (increasing thermodynamic favorability). Thermal fluctuation helps the system stochastically cross energy barriers at transition states (saddle points between local minimum basins). These physical principles operating on a rugged landscape leads to the multi-pathway protein folding transitions. Inspired by the seeming similarities of our system to them, we contend that an energy landscape approach helps understand how self-propelled, far-from-equilibrium macroscopic animals’ and robots’ probabilistic locomotor transitions in complex 3-D terrain emerge from physical interaction, whose equations of motion are unknown or intractable \citep{aguilar2016a,han2021a}. Specifically, we hypothesize that:
\begin{enumerate}
    \item The self-propelled system’s state is attracted to a local minimum basin on a potential energy landscape; locomotor transition from one mode to another can be viewed as the system state escaping from one basin and settling into another. (What governs transition?)
    \item When it is comparable to the potential barrier, kinetic energy fluctuation from oscillatory self-propulsion helps the system escape from a landscape basin to make locomotor transitions. (When does transition happen?)
    \item Escape from a basin is more likely towards a direction along which the escape barrier is lower. (How does transition happen?)
\end{enumerate}

To begin to establish an energy landscape approach of locomotor transitions across modes in complex 3-D terrain, we tested these hypotheses for the two representative modes (pitch and roll) of the model body-beam interaction system defined above. Although the previous study introduced an early energy landscape model to qualitatively explain why locomotor shape affected physical interaction and thus locomotion \citep{li2015a}, none of these hypotheses were proposed or tested. We emphasize that our potential energy landscape directly arises from locomotor-terrain interaction physics using first principles. This is unlike artificially defined potential functions to explain walk-to-run transition \citep{diedrich1995a} and other non-equilibrium biological phase transitions \citep{kelso2012a}, or metabolic energy landscapes inferred from oxygen consumption measurements to explain behavioral switching of locomotor modes \citep{shepard2013a}.

\begin{figure}
    \centering
    \includegraphics[width=0.7\linewidth]{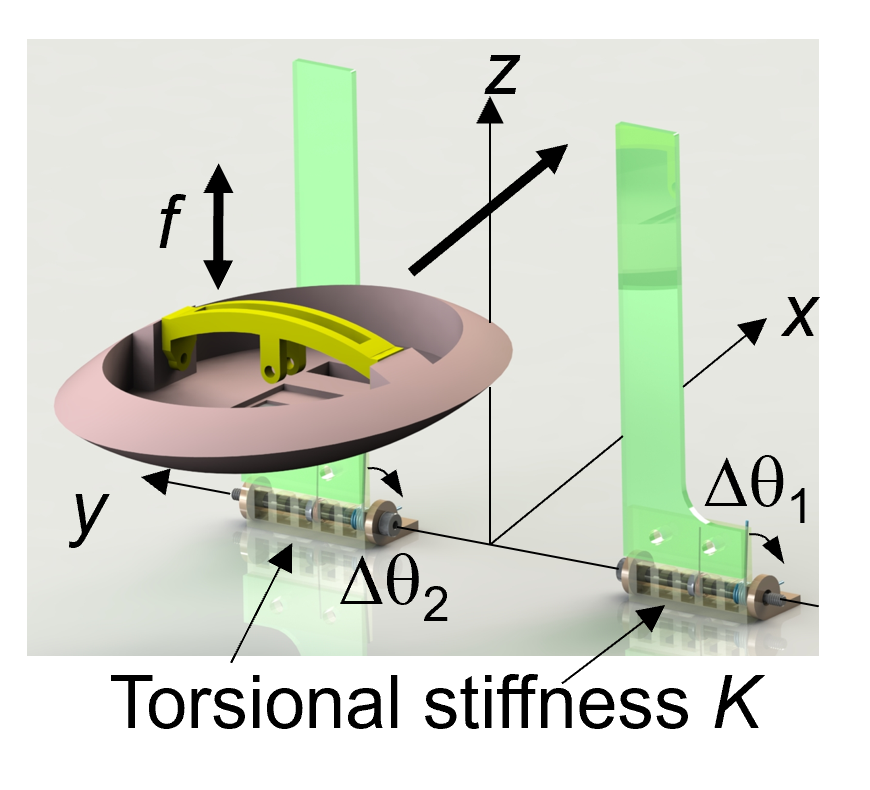}
    \caption[ Robotic physical model.]{Robotic physical model.}
    \label{fig:p2_1D}
\end{figure}

Because animal locomotion emerges from complex interactions of neural and physical mechanisms \citep{dickinson2000a}, to observe the outcome of pure physical interaction, we developed and tested a minimalistic robotic physical model (Figure \ref{fig:p2_1D})) with feedforward control. The robot had an ellipsoid-like body that was propelled forward at a constant speed and was free to pitch and roll (achieved through a gyroscope mechanism) in response to interaction with two beams. The body was constrained not to yaw or move laterally to simplify energy landscape modeling. We also performed experiments with the discoid cockroach traversing beams during escape response to study how physical interaction affects the animal’s locomotor transitions when neural control is bandwidth limited \citep{dickinson2000a}. Comparison of robot and animal observations can reveal aspects of the transitions that likely involve neural mechanisms.

To test the first hypothesis, in both robot and animal experiments, we used rigid “beams” with torsional joints at the base (Figures \ref{fig:p4_s1abc},\ref{fig:p4_s1de},\ref{fig:p5_s2ab},\ref{fig:p6_s2cd}) as one-degree-of-freedom 3-D terrain components to generate a simple potential energy landscape. We then reconstructed the potential energy landscape and 3-D motion of the robot or animal body and beams in high accuracy (as opposed to visual examination in the previous study \citep{li2015a}) (Figures \ref{fig:p14_3}, \ref{fig:p15_4}) for the entire traversal. This allowed us to quantify how the system state behaved on the landscape during each observed locomotor mode and transition between modes. To test the second hypothesis, for the robot, we applied controlled oscillation with variable frequency f to vary kinetic energy fluctuation (Figure \ref{fig:p10_s5}). Because we could not vary the animal’s naturally occurring body oscillation, in animal experiments we changed the barrier relative to kinetic energy oscillation by varying beam torsional joint stiffness K by over an order of magnitude in the range of natural flexible terrain elements (Table S2). \textit{K} was also varied by over an order of magnitude for robot experiments and, together with animal experiments, helped elucidate how transition depended on terrain properties. Because the potential energy landscape consists of not only beam elastic energy but also body and beam gravitational energy, variation of K also changed how escape barrier compared in different directions, allowing the third hypothesis to be tested. See Methods and Supplementary Methods for technical detail and Table \ref{table:p_t1} for sample sizes.

\clearpage
\section{Methods}

\subsection{Robotic physical model}
\begin{figure}
    \centering
    \includegraphics[width=1.0\linewidth]{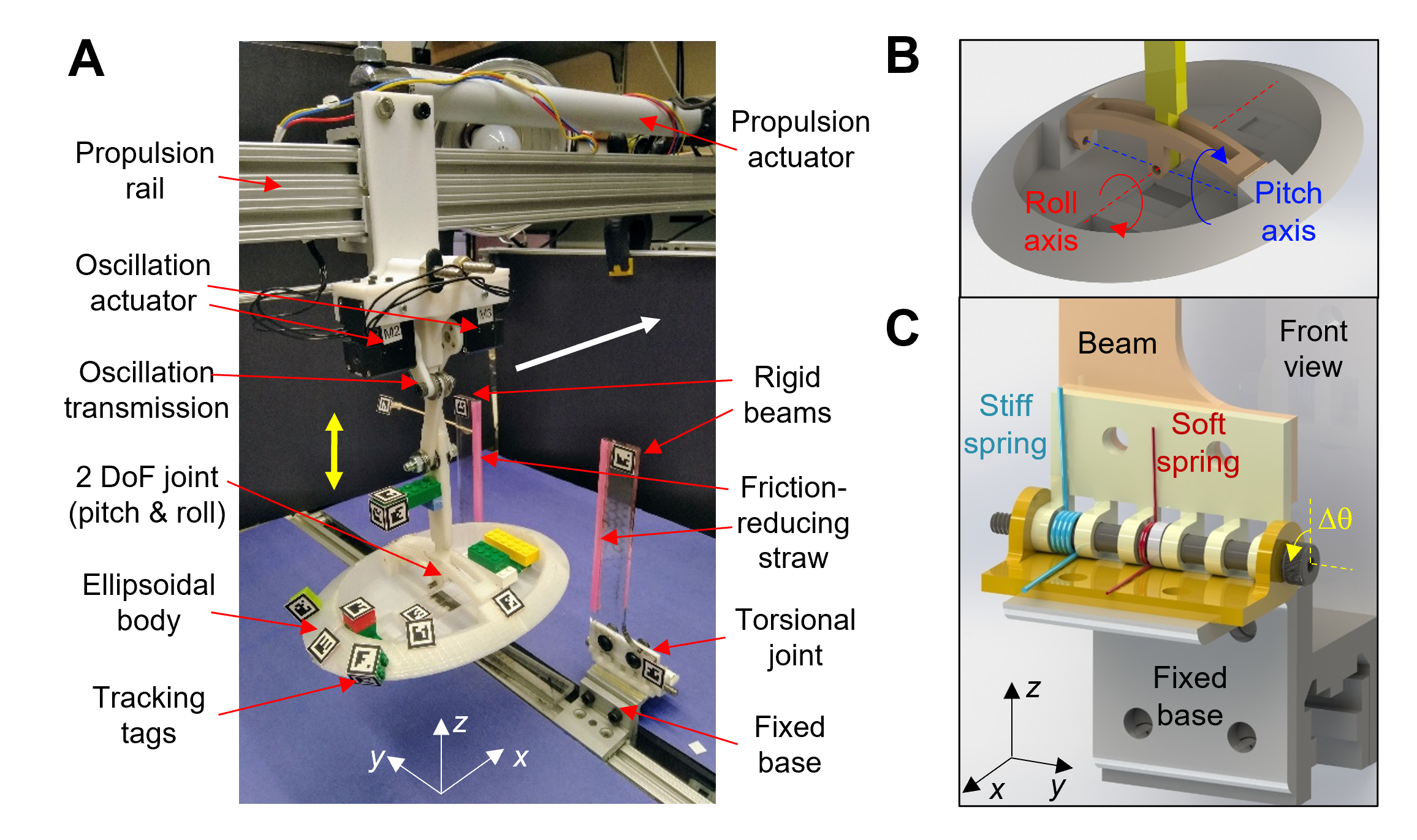}
    \caption[Design of robotic physical model and rigid beams with torsional springs at base.]{Design of robotic physical model and rigid beams with torsional springs at base. (A) Photo of robot body and beams. Body is propelled forward at a constant speed (white arrow) and can be oscillated vertically (yellow arrows). Body can freely pitch and roll in response to interaction with beams. (B) CAD model of body, showing design of pitch and roll joints and axes. Body center of mass is below geometric center due to added weight. Pitch (blue) and roll (red) axes cross geometric center. (C) CAD model of beam base, showing design of torsional joint. Rigid beams rotate about an axis parallel to y-axis (yellow arrow). \textit{K} is varied by using different combinations of soft (red) and stiff (cyan) springs. }
    \label{fig:p4_s1abc}
\end{figure}

To approximate the body shape of the discoid cockroach \citep{li2015a}, we 3-D printed an ellipsoid-like body, PLA plastic using UPBOX+, Tiertime, CA, USA), whose top and bottom halves were slices of an ellipsoid. The body was suspended (center of mass at 10 cm above the ground) via a custom gyroscope mechanism that allowed free body pitching and rolling (Figure \ref{fig:p4_s1abc}A). We added mass to the body so that it is bottom heavy, with body center of mass at 1.1 cm below the pitch axis and 1.6 cm below the roll axis. Body pitch and roll at static equilibrium for a freely suspended body without beam contact were near zero (pitch = 3.3$\degree$ $\pm$ 0.4$\degree$, roll = 1.7$\degree$ $\pm$ 0.8$\degree$; note that positive pitch is pitching downward). See Table \ref{table:p_t1} for geometric dimensions and physical properties of the body.

We used a linear actuator (Firgelli FA-HF-100-12-12, Firgelli Automation, WA, USA) to propel the body forward towards the obstacles. To introduce body kinetic energy fluctuation, we oscillated the body vertically using two DC servo motors (XM430-W350T, Dynamixel, CA, USA) via a five-bar linkage mechanism 3-D printed from PLA plastic (UPBOX+, Tiertime, CA, USA). We varied kinetic energy fluctuation by varying oscillation frequency. Our preliminary experiments showed that body oscillation along different directions did not qualitatively affect the outcome. Thus, we chose vertical oscillation to better observe response in body pitch and roll.

The body oscillated vertically along the following triangular wave trajectory (fitted from the measured $z$ position):
\begin{align}
z &= z_0  + Aft + N(\mu,\sigma), &&0\leq t\leq \frac{T}{2} \\        
z &= z_0  + A(1-ft) + N(\mu,\sigma), && \frac{T}{2}<t\leq T
\end{align}
where $z$ is the vertical position of the body geometric center, $f$ is vertical oscillation frequency, $T = 1/f$ is vertical oscillation period, $A$ = 23.4 mm is the vertical oscillation amplitude, and $z_0$ = 102.4 mm is the average vertical position when there is no oscillation. To prevent the body from being stuck against beams due to friction, we added a small noise, $N$, which is normally distributed with a mean of $\mu$ = 0.7 mm and a standard deviation of $\sigma$ = 1.2 mm. Kinetic energy fluctuation from this noise was small compared to that from the vertical oscillation. The vertical oscillation induced small lateral oscillation (12\% of vertical oscillation amplitude). The motor angles were commanded using a microcontroller (Open CM 0.94, Robotis, CA, USA). We note that the animal’s body oscillation is much more complex, variable, and less periodic than the robot’s. It was difficult to use a wave oscillation with well-defined amplitude and frequency to approximate it. 

\clearpage
\begin{landscape}
\begin{table}[t!]
\centering
\caption{Geometric dimensions, physical properties, and sample sizes for animal and robot experiments.}
\begin{tabular}{|c|c|c|c|c|c|c|c|c|c|c|c|c|}
\hline
\multirow{2}{*}{}            &                                 & \multicolumn{6}{c|}{\textbf{Animal}}                          & \multicolumn{5}{c|}{\textbf{Robot}}                   \\ \cline{2-13} 
                             & Number of Individuals           & \multicolumn{6}{c|}{6}                               & \multicolumn{5}{c|}{N/A}                     \\ \hline
\multirow{4}{*}{\textbf{Body}}        & Mass $m_{\text{body}}$ (g)                   & \multicolumn{6}{c|}{2.6 ± 0.3}                             & \multicolumn{5}{c|}{233}                     \\ \cline{2-13} 
                             & Length (cm)                     & \multicolumn{6}{c|}{5.3 ± 0.1}                             & \multicolumn{5}{c|}{22.1}                    \\ \cline{2-13} 
                             & Width (cm)                      & \multicolumn{6}{c|}{2.4 ± 0.1}                             & \multicolumn{5}{c|}{15.8}                    \\ \cline{2-13} 
                             & Thickness (cm)                  & \multicolumn{6}{c|}{0.8 ± 0.1}                             & \multicolumn{5}{c|}{5.8}                     \\ \hline
\multirow{7}{*}{\textbf{Beam}}        & Lateral spacing(cm)                    & \multicolumn{6}{c|}{1.0}                               & \multicolumn{5}{c|}{12.7}                    \\ \cline{2-13} 
                             & Width (cm)                      & \multicolumn{6}{c|}{1.0}                               & \multicolumn{5}{c|}{2.8}                     \\ \cline{2-13} 
                             & Mass $m_{\text{beam}}$ (g)                  & \multirow{5}{*}{} & 0.33 & 0.42 & 0.63 & 0.70  & 1.03 & \multirow{5}{*}{} & \multicolumn{4}{c|}{38}  \\ \cline{2-2} \cline{4-8} \cline{10-13} 
                             & Inner layer thickness (mm)      &                   & 0.04 & 0.05 & 0.07 & 0.10  & 0.25 &                   & \multicolumn{4}{c|}{N/A} \\ \cline{2-2} \cline{4-8} \cline{10-13} 
                             & Total thickness (mm)            &                   & 0.54 & 0.55 & 0.72 & 0.75 & 0.85 &                   & \multicolumn{4}{c|}{6}   \\ \cline{2-2} \cline{4-8} \cline{10-13} 
                             & Length $L$ (cm)                  &                   & 5.7  & 8.8  & 8.7  & 8.6  & 9.3  &                   & \multicolumn{4}{c|}{18}  \\ \cline{2-2} \cline{4-8} \cline{10-13} 
                             & Torsional stiffness \textit{K} (mN$\cdot$m/rad) &                   & 0.1  & 0.2  & 0.7  & 1.7  & 11.4 &                   & 28   & 55  & 255  & 344  \\ \hline
\multirow{8}{*}{\textbf{Sample size}} & \multirow{7}{*}{No. of trials}  & Ind. 1            & 11   & 10   & 9    & 11   & 10   & 0 Hz              & 10   & 10  & 10   & 10   \\ \cline{3-13} 
                             &                                 & Ind. 2            & 10   & 10   & 10   & 7    & 10   & 1 Hz              & 10   & 10  & 10   & 10   \\ \cline{3-13} 
                             &                                 & Ind. 3            & 11   & 10   & 10   & 11   & 11   & 2 Hz              & 10   & 10  & 10   & 10   \\ \cline{3-13} 
                             &                                 & Ind. 4            & 13   & 10   & 11   & 11   & 13   & 3 Hz              & 10   & 10  & 10   & 10   \\ \cline{3-13} 
                             &                                 & Ind. 5            & 10   & 10   & 10   & 12   & 10   & 4 Hz              & 10   & 10  & 10   & 10   \\ \cline{3-13} 
                             &                                 & Ind. 6            & 9    & 10   & 10   & 10   & 10   & 5 Hz              & 10   & 10  & 10   & 10   \\ \cline{3-13} 
                             &                                 & Total             & 64   & 60   & 60   & 62   & 64   & 6 Hz              & 10   & 10  & 10   & 10   \\ \cline{2-13} 
                             & Total no. of trials             & \multicolumn{6}{c|}{310}                             & \multicolumn{5}{c|}{280}                     \\ \hline
\end{tabular}
\label{table:p_t1}\\
\justifying
\end{table}
\noindent All data averages are mean ± s.d. $m_\text{beam}$ is the mass of one beam.
\end{landscape}

\subsection{Robot beam obstacles}
For robot experiments, we mounted two rigid beams to a fixed base (Figure \ref{fig:p4_s1abc}A) vertically using 3-D printed torsional spring joints (Figure \ref{fig:p4_s1abc}). We varied \textit{K} by using different combinations of soft and stiff torsional springs (McMaster Carr, NJ) (Figure \ref{fig:p4_s1abc}C, red and cyan) in parallel. The rigid beams were laser cut from acrylic plates (VLS60, Universal Laser \& McMaster-Carr, NJ, USA). We covered the beam edges using smooth plastic straw (6 mm diameter) to reduce friction between them and the body during interaction.

\begin{figure}[t]
    \centering
    \includegraphics[width=1.0\linewidth]{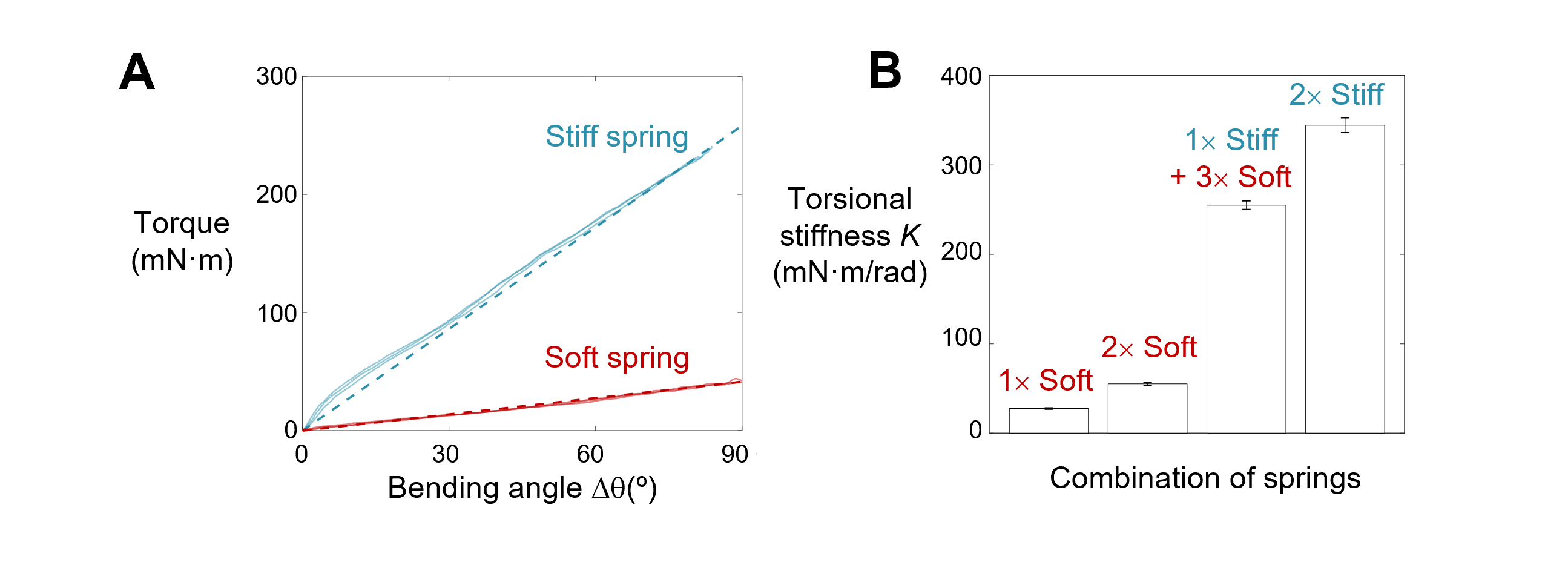}
    \caption[Robot beam stiffness characterization]{Robot beam stiffness characterization. Beam restoring torque as a function of bending angle $\Delta\theta$(defined in Figure \ref{fig:p4_s1abc}C). Red and cyan curves are data for soft and stiff spring. Dashed lines are linear fits (through the origin) of data of each \textit{K}, whose slope give \textit{K}. (B) \textit{K} for different combinations of springs used (mean $\pm$ s.d., \textit{n} = 3 springs, 3 loading cycles each).}%
    {}
    \label{fig:p4_s1de}
\end{figure}

We characterized torsional stiffness of the stiff and soft torsional springs by measuring the restoring torque about the torsional joint as a function of joint deflection angle (Figure \ref{fig:p4_s1de}A) using a 3-axis force sensor (Optoforce OMD-20-FG, OnRobot, Denmark). Torsional stiffness was calculated from the slope of the linear fit (across the origin) of torque as a function of deflection angle (Figure \ref{fig:p4_s1de}B). By combining the stiff and soft torsional springs, we varied \textit{K} by over an order of magnitude ([28, 55, 255, 344] mN$\cdot$m/rad). See Table \ref{table:p_t1} for geometric dimensions and physical properties of the beams.

\subsection{Robot experiment imaging}
Robot experiments were recorded using three synchronized high-speed cameras (IL5, Fastec Imaging, San Diego, CA) at 200 frames s$^{-1}$ and a resolution of 1920 $\times$ 1080 pixels. To automatically track the body and beams over the entire range of rotation, we attached BEEtags \citep{crall2015a} (18 mm $\times$ 18 mm) on the body (9 markers), vertical oscillation transmission (3 markers), right beam (2 markers), and left beam (5 markers). We used FasMotion software (Fastec Imaging, San Diego, CA) to save the videos to storage drives after recording for tracking and processing.

\subsection{Robot experiment protocol}
Before each trial, the body was positioned at a distance of 11 cm from the beams, and the beams were set to be vertical. We started video recording and body oscillation (for $f$ > 0), waited for 1 s, and then propelled the body forward at a constant speed of 0.7 cm$\cdot$s$^{-1}$ by a distance of 30 cm (maximum possible by the linear actuator). Body oscillation was applied (for $f$ > 0) until the end of forward translation. After forward translation completed, we stopped body oscillation and video recording and moved the body to its initial position for the next trial.

At each \textit{K}, we varied kinetic energy fluctuation by varying $f$ from 0 Hz to 6 Hz with an increment of 1 Hz. At each \textit{K} and each $f$, we performed 10 trials. This resulted in a total of 280 trials, with 70 trials at each \textit{K} across all $f$. See Table \ref{table:p_t1} for detailed sample size.

\subsection{Animals}
We chose to study the discoid cockroach, \textit{Blaberus discoidalis}, because it dwells on the floor of tropical rainforests with dense vegetation and litter and excels at traversing complex terrain \citep{li2015a}. We used adult male discoid cockroaches (Pinellas County Reptiles, St Petersburg, FL, USA), as females are often gravid and under different load bearing conditions. Prior to experiments, we kept the cockroaches in individual plastic containers at room temperature (24 $\degree$C) on a 12h:12h light:dark cycle. See Table \ref{table:p_t1} for dimensions and mass of the animals tested.

\subsection{Animal beam obstacles}
\begin{figure}[t]
    \centering
    \includegraphics[width=1.0\linewidth]{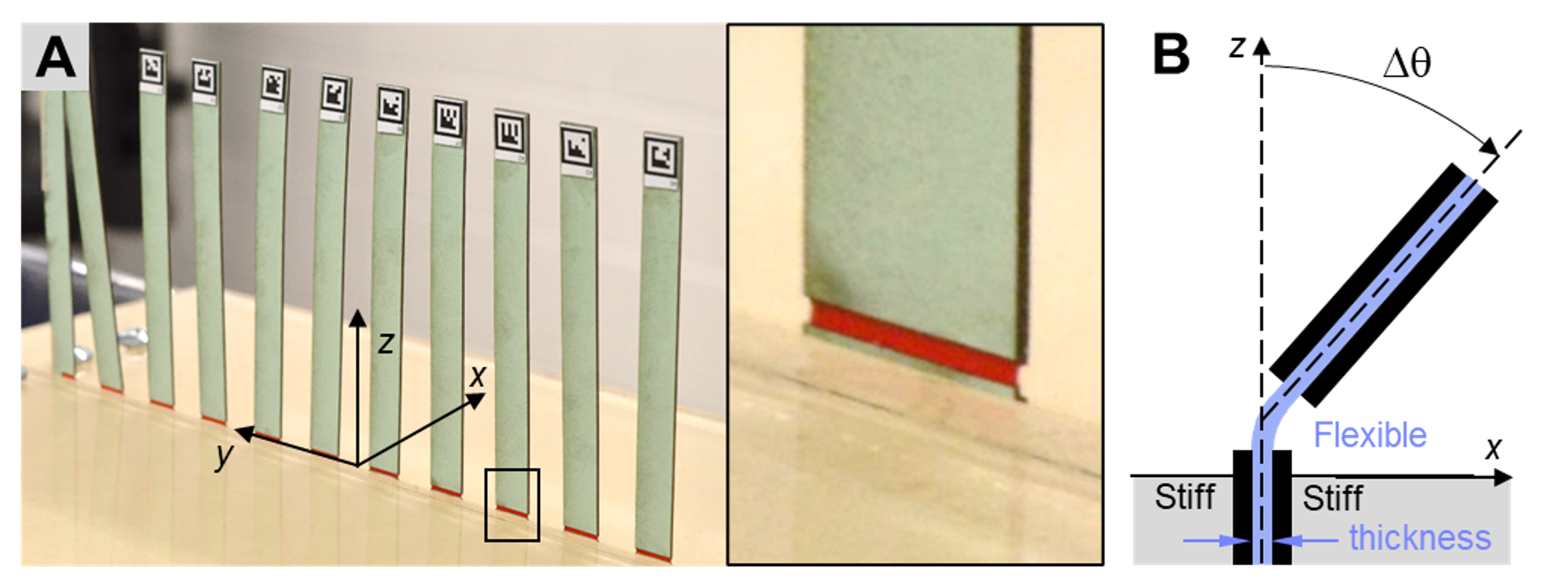}
    \caption[Design of rigid beams with torsional springs at base for animal experiments.] %
    {Design of rigid beams with torsional springs at base for animal experiments.(A) Photo of a layer of animal beams. Inset shows a closer view of torsional joint. (B) Side view schematic of beam design following \citep{haldane2015a}. Stiff outer layers (black) provide rigidity, and a small exposed section of flexible inner layer (blue) acts as a torsional spring joint. Dimensions not true to scale. }
    \label{fig:p5_s2ab}
\end{figure}

We custom made rigid “beams” with torsional springs at the base (Figure \ref{fig:p5_s2ab}A). For each beam, we sandwiched a flexible layer between two stiff layers and exposed a small portion of the flexible layer (Figure \ref{fig:p5_s2ab}B), which acted as torsional spring joint about which the beams deflect in the $x-z$ plane. We varied the thickness of the flexible layer ([0.04, 0.05, 0.07, 0.10, 0.25] mm) to vary the torsional stiffness \textit{K} of the torsional joint by over two orders of magnitude ([0.1, 0.2, 0.7, 1.7, 11.4] mN$\cdot$m/rad) in a similar range as natural obstacles like leaves, stalks, and grass. Polyethylene terephthalate plastic (McMaster Carr, NJ, USA) and cardstock (0.2 mm thickness, Neenah Inc., GA, USA) were used for the flexible and stiff layers and bonded using thermally bonding glue (Therm-O-Web, IL, USA) and a laminating machine (AmazonBasics, Amazon). The layer of 10 beams was laser cut (VLS60, Universal Laser Systems, AZ, USA) to have identical geometry and spacing. 

\begin{figure}[t]
    \centering
    \includegraphics[width=1.0\linewidth]{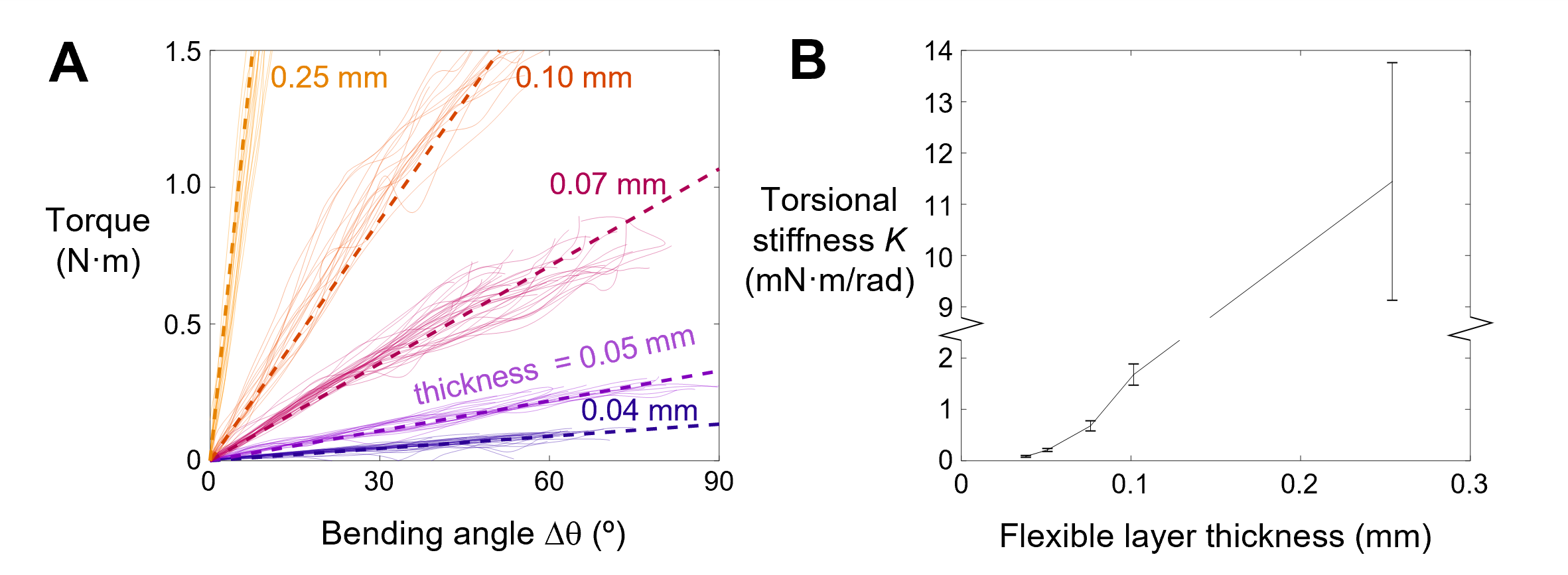}
    \caption[Characterization of rigid beams with torsional springs at base for animal experiments]{Characterization of rigid beams with torsional springs at base for animal experiments. (A) Beam restoring torque as a function of bending angle for different flexible layer thickness ([0.04, 0.05 0.07, 0.10, 0.25] mm). Dashed lines are linear fits (through the origin) of data, whose slopes give \textit{K}. (B) \textit{K} as a function of flexible layer thickness (mean $\pm$ s.d., \textit{n} = 62, 37, 76, 38, 32 loading cycles).} %
    {}
    \label{fig:p6_s2cd}
\end{figure}

We characterized \textit{K} by measuring the restoring torque about the torsional joint as a function of beam deflection angle (Figure \ref{fig:p6_s2cd}A) using a 6-axis force and torque sensor (Nano 43, ATI Industrial Automation, NC, USA). \textit{K} was calculated from the slope of the linear fit (across the origin) of torque as a function of deflection angle (Figure \ref{fig:p6_s2cd}B). See Table \ref{table:p_t1} for geometric dimensions and physical properties of the beams.

\subsection{Animal multi-camera imaging arena}
\begin{figure}
    \centering
    \includegraphics[width=0.6\linewidth]{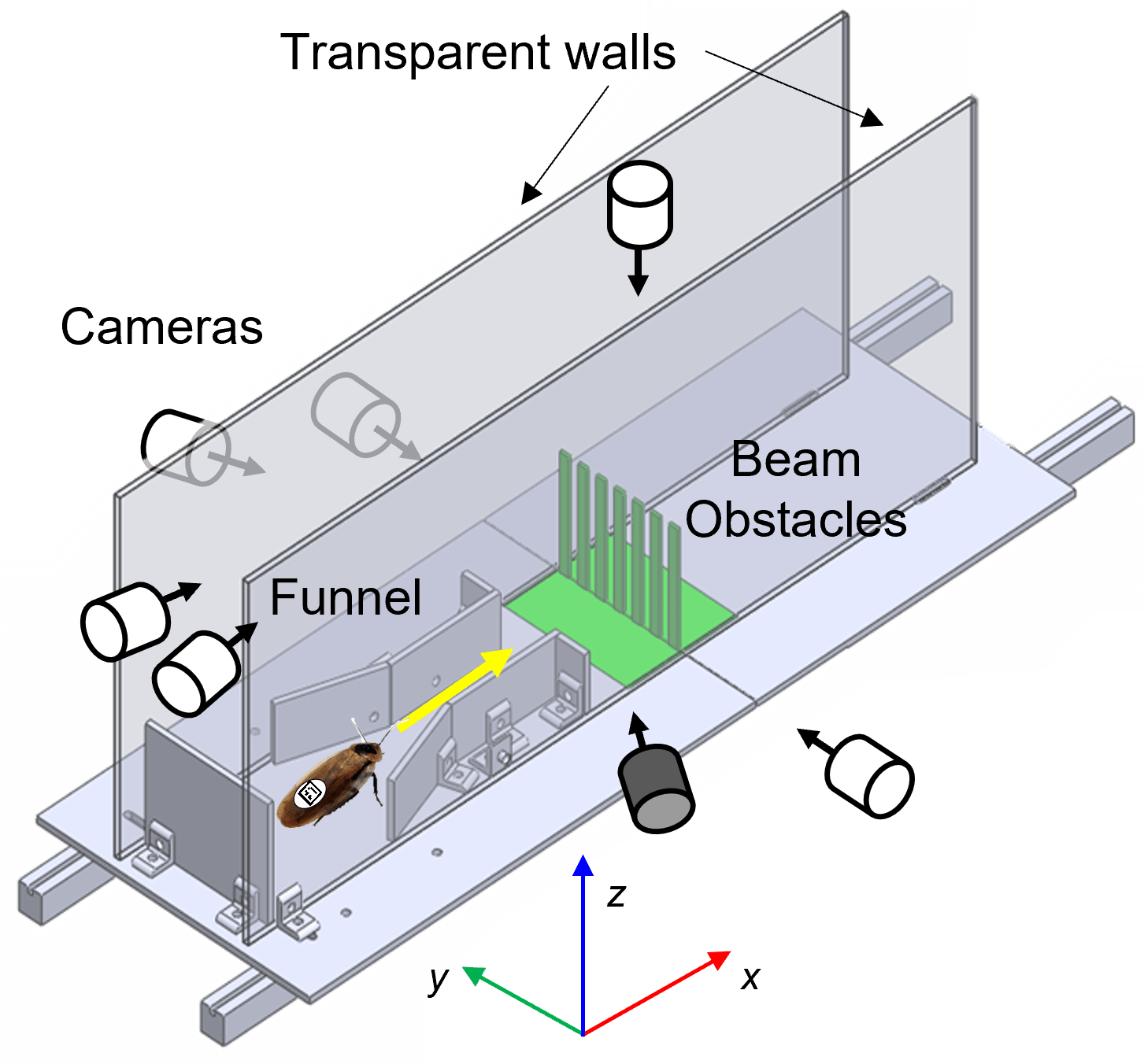}
    \caption[Animal locomotion arena with a layer of beam obstacles (green), with seven high-speed cameras.]{Animal locomotion arena with a layer of beam obstacles (green), with seven high-speed cameras. $x$, $y$,$z$ axes show lab frame. } %
    \label{fig:p7_s3a}
\end{figure}

We constructed an arena for animal experiments to measure locomotor transitions (Figure \ref{fig:p7_s3a}). Previous studies showed that animals often laterally explored beam obstacles before traversing \citep{li2015a}. To increase experimental yield, we used 10 identical beams in an obstacle layer, which presented nine gaps of 1 cm (narrower than animal body width of 2.4 cm, but larger than body thickness of 0.8 cm) for the animal to traverse. All the beams were vertical without external force from the animal. The beam obstacle layer was inserted into a slit cut in the flat ground between two transparent sidewalls made of acrylic sheets. A runway funneled the animal towards the middle of beam obstacle layer to minimize the interaction with the sidewall. To facilitate traversal with minimal body yaw (on average), we arranged the beam obstacle layer to be perpendicular to the direction of animal movement. The reduced body yaw allowed us to more accurately visualize how trials evolved on the potential energy landscape (see section below), which was calculated using the average body yaw from all trials. Paper cardstock covered the ground surface. We placed a dark shelter with food and water on the exit side of the obstacle layer for the animal to rest after each trial.

Animal experiments were recorded using seven synchronized high-speed cameras (N5A-100, Adimec, Netherlands) at 100 frames s$^{-1}$ and a resolution of 2592 $\times$ 2048 pixels. When interacting with the obstacles, animal body orientation varied substantially. We carefully positioned the cameras around the entire arena to cover the entire rotation range of motion, with two from back views, two side views, two isometric views, and one top view (Figure \ref{fig:p7_s3a}A). We used the StreamPix software (Norpix Inc., Montreal, Canada) to automatically save the videos to storage drives as they were being recorded, after which they were converted to AVI format for tracking and processing.

\begin{figure}[h]
    \centering
    \includegraphics[width=0.8\linewidth]{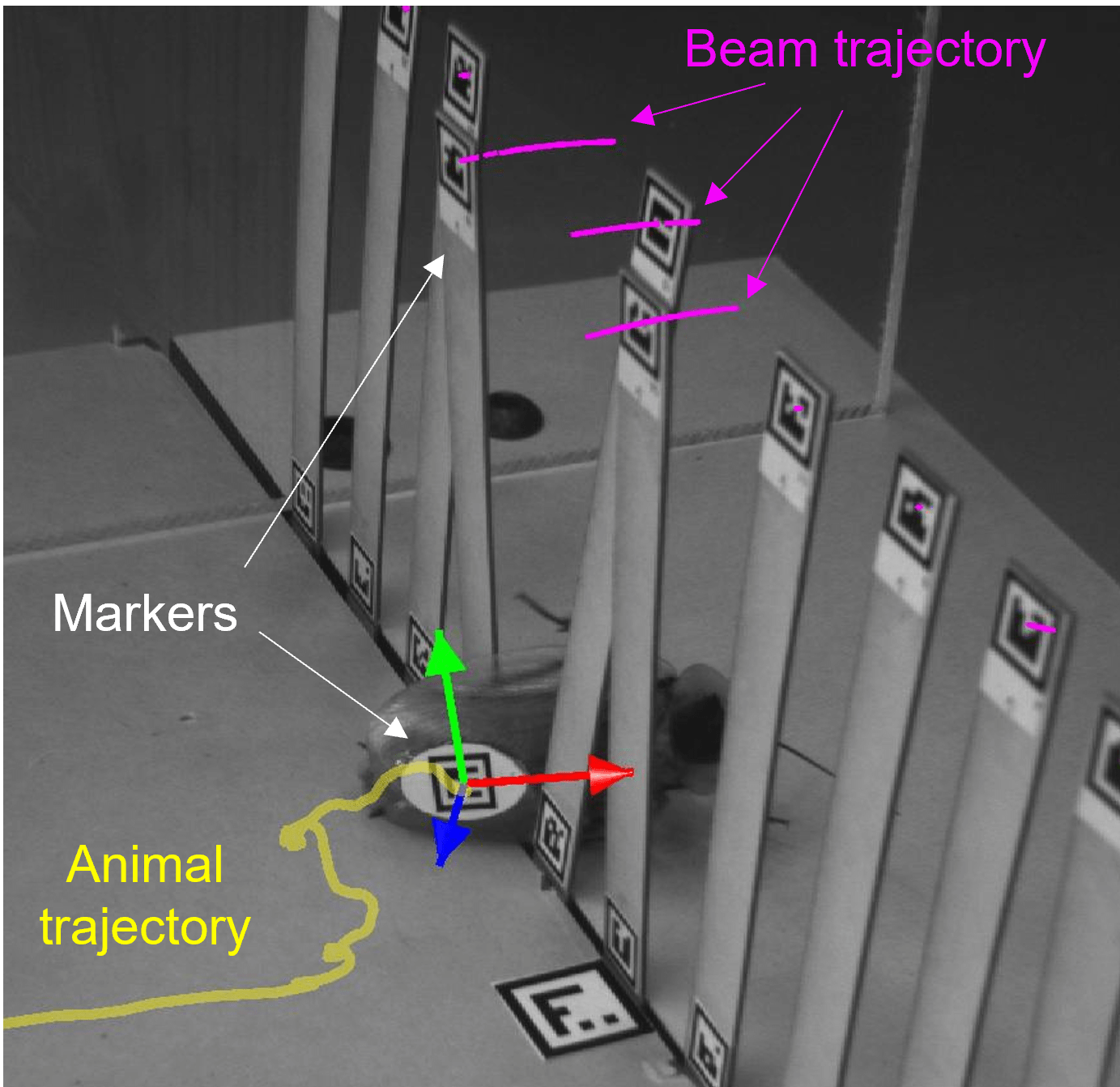}
    \caption[Snapshot of animal traversing beam obstacles] %
    {Snapshot of animal traversing beam obstacles (view from the shaded camera in Figure \ref{fig:p7_s3a}). Markers are attached to the animal body and beams to track their 3-D motion (yellow and magenta trajectories). Red, green, blue axes show body frame attached to markers.}
    \label{fig:p8_s3b}
\end{figure}

To automatically track the animal and beams, we attached a 7 mm $\times$ 7 mm BEEtag \citep{crall2015a} to the animal body and 9 mm $\times$ 9 mm BEEtags to the top and bottom ends of both sides of each beam (Figure \ref{fig:p8_s3b}). The animal BEEtag was much lighter (< 0.15 g) than the animal itself (2.6 g). It was printed onto a rounded oval cardboard to minimize interference with the obstacle traversal and attached to the dorsal surface of the abdomen using ultraviolet curing glue (Bondic, Aurora, Canada).

\subsection{Animal experiment protocol}
Before the experiment, the arena was illuminated and heated to about 43$\degree$C with six work lamps (Coleman Cable, Waukegan, IL, USA). Before each trial, the animal was placed in the starting end of the arena and allowed to settle down. We then started video recording and probed the animal with a stick with a soft tip (made from paper tapes) to induce it to run towards the obstacles. The animal did not always immediately traverse after running into beam obstacles. Instead, it often made multiple failed attempts to traverse and sometimes explored the obstacle layer laterally to attempt traversing at different beam gaps, before eventually traversing. Once the animal traversed and reached the shelter, we stopped video recording and allowed the animal to rest for $\sim$10 minutes before the next trial.

We tested six animal individuals and beams of five different torsional stiffness \textit{K} and collected a total of 337 trials. The same six individuals were tested across all \textit{K}. We discarded trials in which any of the following were observed: (1) the animal did not move within 10 s after it was probed; (2) the animal moved back to the starting area or did not attempt to traverse; (3) the animal used the sidewall to traverse; or (4) the animal climbed up the beams and its body and all six legs lost contact with the ground. This resulted in a total of 310 accepted trials, with approximately 10 trials for each animal at each \textit{K}. See Table \ref{table:p_t1} for detailed sample size.

\subsection{High accuracy 3-D motion reconstruction}
To calibrate the cameras over the working space for 3-D motion reconstruction, for both robot and animal experiments, we built a calibration object with multiple markers (47 for robot and 17 for animal) using Lego bricks (The Lego Group, Denmark). We then used the direct linear transformation software DLTcal5 \citep{hedrick2008a} to obtain intrinsic and extrinsic camera parameters. We used a custom MATLAB script to automatically track 2-D coordinates of the markers in each camera view using the BEEtag code \citep{crall2015a}.

Using the tracked 2-D marker coordinates from multiple camera views and camera calibration parameters, we obtained the 3-D position of the four corners of each BEEtag markers using the direct linear transformation software DLTdv5 \citep{hedrick2008a} , which was then used to obtain the marker frame (Figure \ref{fig:p8_s3b}). For the animal, we translated and rotated the marker frame by the measured translational ($\Delta x$ = 10 mm, $\Delta y$ = $-$0.2 mm, $\Delta z$ = $-$3 mm) and rotational (roll = 0$\degree$, pitch  = 10$\degree$, yaw = 1$\degree$) offsets to obtain 3-D position and orientation of the body frame at the body geometric center, which nearly overlapped with body center of mass \citep{kram1997a}. For the robot, we used a CAD model of the body to determine the location of center of mass relative to the markers fixed to the body. Depending on which body markers were reconstructed in each video frame, we translated and rotated the reconstructed marker frame by its measured translational and rotational offsets to obtain 3-D position and orientation of the body frame at the center of mass. For both the robot and animal, we used Euler angles (yaw $\alpha$, pitch $\beta$, and roll $\gamma$, \textit{Z}$-$\textit{Y}’$-$\textit{X}” Tait-Bryan convention) to define 3-D rotation. Note that with this convention, when the body pitches upward, pitch angle is negative.

To quantify the accuracy of 3-D reconstruction using BEEtag tracking combined with Direct Linear Transformation, we 3-D printed a high-precision calibration object. The calibration object had nine BEEtag markers mounted on a horizontal plate in a 3 $\times$ 3 grid with a 7 cm grid distance, each oriented at a pitch and yaw angle of 0$\degree$, 30$\degree$, and 60$\degree$. We measured the 3-D position and orientation of each marker from 3-D reconstruction (described above) and compared them to the designed values. This demonstrated that our imaging setup achieved high accuracy in 3-D position and orientation reconstruction (s.d. of position error = 0.6 mm; s.d. of orientation error = 1.1$\degree$). We also verified that lens distortion was minimal (< 1\%) using the checkboard distortion measurement method.

For each trial, we calculated body translational ($v_x$, $v_y$, $v_z$) and rotational ($\omega_\alpha$, $\omega_\beta$, $\omega_\gamma$) velocities and beam deflection angles from vertical ($\Delta\theta_i$) as a function of time. Beam angle was averaged from visible tags on each beam. Considering lateral symmetry, to simplify analysis of the roll mode, we flipped all trials in which the body rolled left to rolling right. For the animal, we offset the measured lateral positions (\textit{y}) of each trial so that \textit{y} = 0 in the middle of the gap that the animal traversed during the final, successful attempt.

\subsection{Definition of pitch and roll modes and pitch-to-roll transition}
\begin{figure}[p]
    \centering
    \includegraphics[width=1.0\linewidth]{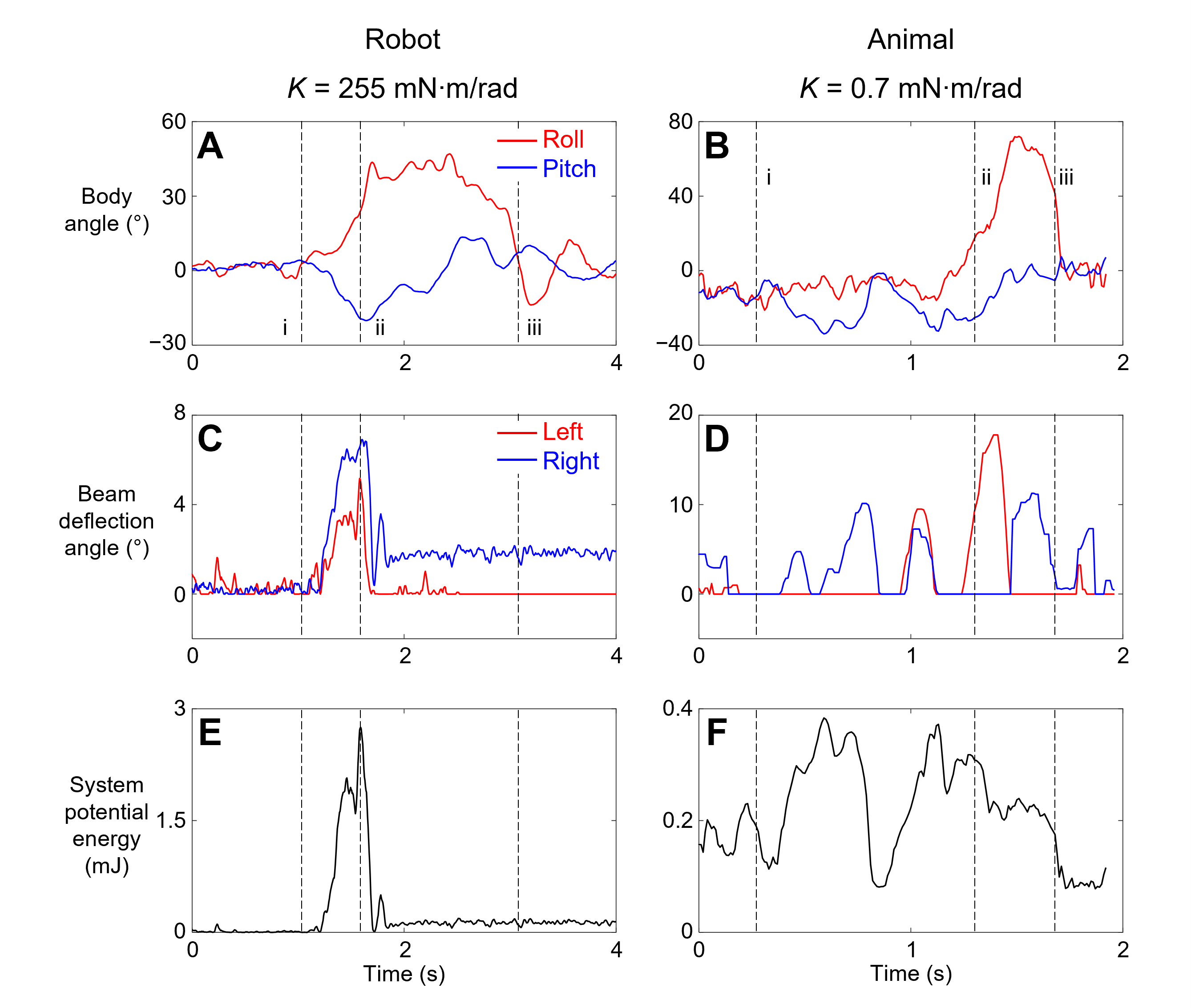}
    \caption[Representative motion of body and beams and system potential energy during interaction and definition of traversal and pitch-to-roll transition.] %
    {Representative motion of body and beams and system potential energy during interaction and definition of traversal and pitch-to-roll transition. (A, B) Body roll (red) and pitch (blue) as a function of time. (C, D) Left (red) and right (blue) beam deflection angle as a function of time. (E, F) System potential energy as a function of time. Data shown for a representative pitch-to-roll transition at \textit{K} = 255 mN$\cdot$m/rad for robot and \textit{K} = 0.7 mN$\cdot$m/rad for animal. For both the robot and animal, pitch-to-roll transition resulted in a reduction in system potential energy. Note that negative pitch is the body pitching head-up. Dashed lines (i) and (ii) are when body first contacts beams and when pitch-to-roll transition occurs. Dashed line (iii) is when the robot’s forward translation ends and when animal’s distal end crosses the beam (\textit{x} = 0).}
    \label{fig:p9_s4}
\end{figure}

We defined the robot to be in the roll mode if both beams lost contact with the body and bounced back to vertical before the distal end of the body crossed the beams ($x$ = 0), and we defined it to be in the pitch mode otherwise. For the robot, body motion was highly repeatable from trial to trial, and pitch-to-roll transition always resulted in a sharp decrease in system potential energy. Thus, we defined transition to occur when system potential energy reached a peak value (Figure \ref{fig:p9_s4}E, vertical dashed line ($ii$)), after which it immediately reduced.

We defined the animal to be in the roll mode if its body roll (absolute value) exceeded 62$\degree$, because from system geometry this was the minimal roll for the body to move through the gap between two adjacent beams without deflecting them. The animal was defined to be in pitch mode otherwise. For trials in which the animal transitioned from the pitch to roll mode, we defined transitions to occur when body roll (absolute value) exceeded 20$\degree$ (Figure \ref{fig:p9_s4}B, vertical dashed line ($ii$)). We verified that system potential energy (Figure \ref{fig:p9_s4}F) decreased at this moment.

\subsection{Data averaging}
Because the robot was propelled forward at a constant speed, its 3-D kinematics, potential energy, and kinetic energy were a function of body forward position $x$. To obtain average 3-D kinematics and potential energy as a function of $x$, we interpolated the measured position, orientation, and potential energy over $x$ and then averaged them across all trials at a given \textit{K}. For the robot, we averaged lateral position $y$ and body yaw $\alpha$ for all the trials at each \textit{K} for each $x$ and used this average trajectory of measured $x$, $y$, and $\alpha$ to calculate an average potential energy landscape. For the animal, because of the high variability in $y$ and  $\alpha$, for simplicity we set both to zero when calculating the average potential energy landscape at each $x$.

Because we focused on the pitch-to-roll transition (see definition in the next section), we considered only the animal’s final, successful attempt in which such a transition may occur. For the final, successful attempt, we analyzed the portion of the trial starting from five frames (0.05 s) before the animal’s head contacted the beams (Figure \ref{fig:p9_s4}, dashed vertical line ($i$)) to ten frames (0.1 s) after the entire body crossed the obstacle layer (at $x$ = 0, Figure \ref{fig:p9_s4}, dashed vertical line ($iii$)). Because the robot body was translated with a constant forward speed and always crossed the beams, for it we analyzed the portion of the trial starting from when the body first contacted the beams (Figure \ref{fig:p9_s4}, dashed vertical line ($i$)) until the end of forward translation (Figure \ref{fig:p9_s4}, dashed vertical line ($iii$)). 

\subsection{Kinetic energy fluctuation}
For both the robot and animal, we defined body kinetic energy fluctuation as the sum of kinetic energy due to translational and rotational velocity components other than forward motion of the body ($v_y$, $v_z$, $\omega_\alpha$, $\omega_\beta$, $\omega_\gamma$). To calculate moment of inertia, we approximated the animal body as an ellipsoid with uniform mass distribution, considering that legs only consist less than 15\% of total mass \citep{kram1997a}. For the robot, we calculated moment of inertia from a CAD model of the body with accurate geometry and mass distribution.

\begin{figure}
    \centering
    \includegraphics[width=1.0\linewidth]{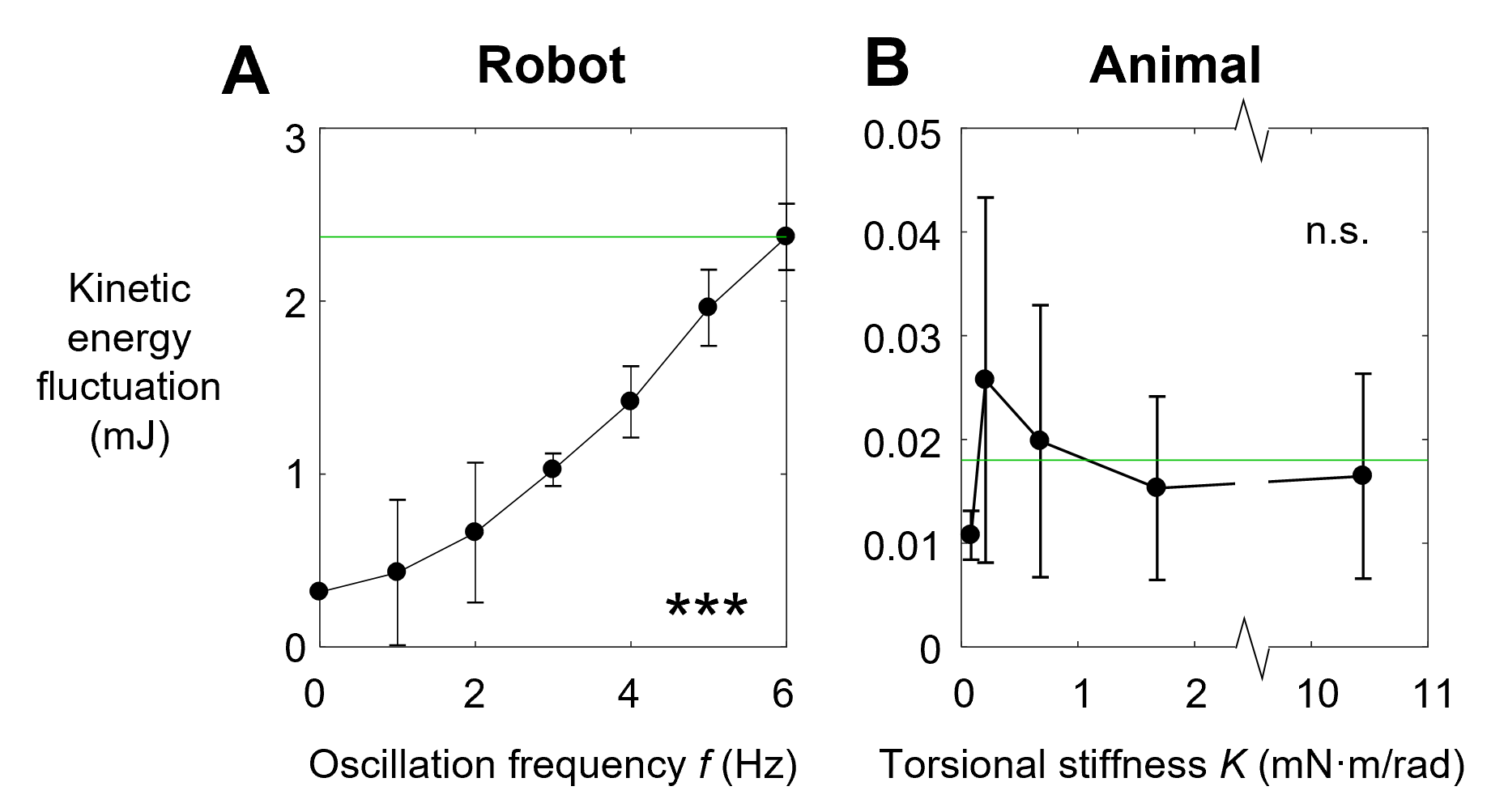}
    \caption[Kinetic energy fluctuation] %
    {Kinetic energy fluctuation. (A) Kinetic energy fluctuation of robot as function of \textit{f}. *** indicates a significant dependence (ANOVA, \textit{P} < 0.0001, \textit{F} = 520.99). (B) Kinetic energy fluctuation of animal as function of \textit{K}. n.s. indicate no significant difference (ANOVA, \textit{P} = 0.3835, \textit{F} = 0.9047). See Table \ref{table:p_t1} for sample size.}
    \label{fig:p10_s5}
\end{figure}

For both the robot and animal, we calculated average kinetic energy fluctuation from first beam contact (of the final successful attempt of each trial for the animal) to when transition occurred using the trials in which the body transitioned to the roll mode. This was because for the trials in which the body was trapped in the pitch mode, it was difficult to define the onset of pitching as can be readily done for the onset of rolling. Including these trials would add the substantial kinetic energy of continuous body pitching that resulted from the interaction, which was not part of the fluctuation that induced the transition. We verified that kinetic energy fluctuation differed little between before contact with the beams and from first contact to when transition occurred. We then averaged kinetic energy fluctuation over time for each trial, from when the body first contacted a beam (Figure \ref{fig:p9_s4}, dashed line ($i$)) (in the final successful attempt of each trial for the animal), to when it transitioned to roll mode (Figure \ref{fig:p9_s4}, dashed line ($ii$)). For the robot, we then averaged these trial averages across all trials at each $f$ in which the robot transitioned to the roll mode to obtain average kinetic energy fluctuation at each $f$ (Figure \ref{fig:p10_s5}A). For the animal, we averaged these trial averages across all trials at each \textit{K} to obtain average kinetic energy fluctuation at each \textit{K} (Figure \ref{fig:p10_s5}B). 

\subsection{Statistics}
All probabilities were calculated relative to the total number of accepted trials of each treatment. All average data are reported as mean $\pm$ s.d.  For the robot, we used a chi-square test to test whether pitch-to-roll transition probability depended on \textit{K}, with \textit{K} and $f$ as fixed factors. For the animal, we used a chi-square test to test whether pitch-to-roll transition probability depended on \textit{K}, with \textit{K} and individual as fixed factors and including their crossed effect. For the robot, we used an ANOVA to test whether kinetic energy fluctuation increased with $f$. To test whether the animal’s kinetic energy fluctuation depended on \textit{K}, we pooled data from all the trials in which pitch-to-roll transition occurred (see section above for explanation) for each \textit{K} and performed a mixed-effect ANOVA with  \textit{K} as a fixed factor and  individual as a random factor. We used a Student’s \textit{t}-test to test whether the robot's system state was attracted to the basin corresponding to the measured mode in all trials. All statistical tests were performed using JMP Pro 13 (SAS Institute Inc., NC, USA).

\clearpage
\subsection{Potential energy landscape}
\begin{figure}
    \centering
    \includegraphics[width=1.0\linewidth]{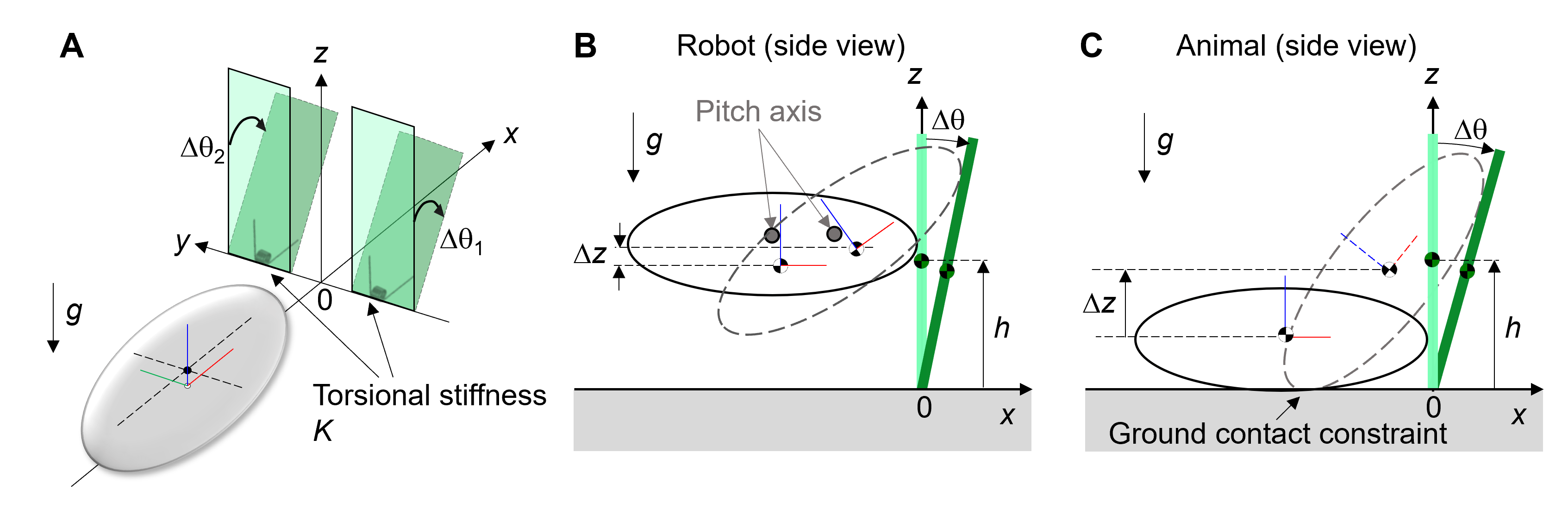}
    \caption[Potential energy landscape model, with definition of variables and parameters.] %
    {Potential energy landscape model, with definition of variables and parameters. (A) Oblique view schematic of body (a rigid ellipsoid) and beams (rigid rectangular plates with torsional joints at base) of torsional stiffness K. Without body contact, both beams are vertical (light green). With body contact, beams are deflected forward (dark green) by angles $\Delta\theta_{1,2}$. (B, C) Side view of model for robot (B) and animal (C) to show center of mass height changes with body pitching and beam deflection. Solid and dashed ellipses show body in static equilibrium and pitched-up, respectively. Center of mass of body and beams are shown.}
    \label{fig:p11_s6}
\end{figure}

In energy landscape modeling, we approximated the animal body as a rigid ellipsoid and obtained the robot body shape from a CAD model used for 3-D printing the body. The beams were modeled as rigid rectangular plates on torsional joints (Figure \ref{fig:p11_s6}A). Because the beams had a finite mass, forward deflection lowered beam center of mass and thus beam gravitational potential energy. Because the measured beam restoring torque was nearly proportional to deflection angle for both the robot (Figure \ref{fig:p4_s1de}B) and animal (Figure \ref{fig:p6_s2cd}A), we approximated the torsional joint at the base of each beam as a perfect Hookean torsional spring and assumed that there was no damping. Because the body only pushed forward against the beams, in the model we only allowed forward beam deflection ($\Delta\theta_{1,2} \geq 0$).

For robot modeling, we set center of mass to be below the pitch and roll axes as measured (Figure \ref{fig:p11_s6}B). For animal modeling, we constrained the lowest point of the body to always touch the ground (ground constraint, Figure \ref{fig:p11_s6}C), because the animal maintained ground contact during traversal (we rejected trials in which the animal climbed onto the beams) and we neglected the animal’s legs. Thus, for both the robot and animal, body pitching and rolling in response to interaction with the beams increased center of mass height and thus body gravitational potential energy. In addition, because the robot was suspended from and driven forward by a linear actuator, its center of mass height was constrained to move within a measured range of $z$ = [9.9 cm, 11.8 cm]. Because the robot’s controlled vertical oscillation was modeled as part of kinetic energy fluctuation, we used the average body center of mass vertical position before contacting the beams ($z$ = 10.8 cm, vertical height constraint) to calculate its initial body potential energy. We verified that at any given $x$, landscape shape remained similar within the $z$ range in which the robot was oscillated. For both the robot and animal, we offset system potential energy to zero when the body was not in contact with beams and in its static equilibrium (at zero pitch and zero roll) so system potential energy shown on the landscapes were relative to this initial equilibrium (\ref{fig:p11_s6}B, C).

The full potential energy landscape depended on body orientation (pitch, roll, yaw) and forward and lateral positions ($x$, $y$), given the vertical height and ground constraints on the robot and animal, respectively. Because we focused on body pitch and roll motions, for a given body position ($x$, $y$) and yaw, we varied body pitch and roll over [$-$180$\degree$, 180$\degree$] to calculate system potential energy landscape over pitch-roll space. In Figures \ref{fig:p13_2}B, \ref{fig:p14_3}, \ref{fig:p15_4}A, and \ref{fig:p18_6}A, we only show the landscape over a part of the entire pitch-roll space to better focus on the pitch and roll basins. We then calculated beam deflection due to body contact (only allowing $\Delta\theta _{1,2} \geq 0 $) and center of mass height increase ($\Delta z$) to obtain system potential energy as below.
\begin{equation}
  E = m_{body}g\Delta z + \frac{1}{2}m_{beam}gL(\cos\Delta\theta_1+\cos\Delta\theta_2-2) + \frac{1}{2}K(\Delta\theta^{2}_1+\Delta\theta^{2}_2) 
\end{equation}
where where $m_{body}$ is body mass, $g$ is gravitational acceleration, $\Delta z$ is body center of mass height increase from its equilibrium configuration (at near zero pitch and zero roll), $m_{beam}$ is beam mass, $L$ is beam length, $K$ is beam torsional stiffness, and $\Delta\theta_1$ and $\Delta\theta_2$ are beam deflection angles from vertical.

We note that our landscape did not model body-beam interaction after the beams bounced back. 

\subsection{Local minima and system state trajectories on potential energy landscape}
For each forward position $x$ of the body relative to the beams, we examined the landscape to determine the pitch and roll local minima and measured their potential energies. Note that for the robot their potential energies did not include height change due to controlled vertical oscillation (see section above). To visualize how the measured state of the system behaved on the landscape, we projected the measured body pitch and roll onto the landscape for each $x$ (Figure \ref{fig:p14_3}A, \ref{fig:p15_4}, \ref{fig:p18_6}A, blue and red dots for trials in which the system was trapped in the pitch mode and transitioned to the roll mode), which formed a system state trajectory over time as traversal progressed. Note that only the end points of the trajectory, which represent the current state, showed the actual potential energy of the system at the corresponding $x$. The rest of the visualized trajectory showed how body pitch and roll evolved but, for visualization purpose, was simply projected on the landscape surface. Because roll local minimum does not exist at \textit{K} = 28 mN$\cdot$m/rad for the robot, for comparison with other \textit{K}, we defined it to be at (pitch, roll) = (0$\degree$, $\pm$42$\degree$) based on the minimal body roll required to traverse without beam deflection. 

\subsection{Average potential energy landscape at each beam stiffness}
To facilitate observation of statistical trends, we calculated the average potential energy landscape at each \textit{K} and visualized all trials on it. Average landscape calculation used the average measured lateral position $y$ and body yaw for each $x$. For the robot, this average potential energy landscape was a good approximation of the actual landscape for each trial, because the robot was constrained by design to have minimal lateral motion or yawing. Despite this, when projected onto the average potential energy landscape, in some trials at high \textit{K}, a portion of the system state trajectory appeared to momentarily go out of the pitch basin and then re-entered it (Figure \ref{fig:p15_4}). This was an artifact from landscape averaging. In those trials, the robot body experienced larger yawing due to a slight lateral bending of the plastic pole that suspended the robot resulting from high beam restoring forces. Because such trials are rare in the robot experiment, the average landscape basin was close to that without body yawing. Examination of the actual landscape for each robot trial (see next section) verified that the state trajectory in the pitch mode was almost always in the pitch basin. For the animal that freely moved laterally and yawed, the average landscape was a much poorer approximation of the actual landscape for each trial.

\subsection{Percentage of trials in which system is attracted to basin of observed mode on actual landscape}
Because the average landscape did not account for trial-to-trial variation, to better quantify how well the potential energy landscape explained the observed locomotor modes, for both the robot and animal, we further calculated the actual (not averaged) potential energy landscape for each trial using the measured position ($x$, $y$) and body yaw of that trial. We then counted the number of trials in which the system state either stayed in the pitch basin or transitioned to the roll basin, in accord with the locomotor mode observed, and we calculated the percentage of trajectories attracted to the corresponding basin.

\subsection{Energy barrier to escape from pitch local minimum}
We measured the potential energy barrier that must be overcome to escape from the pitch local minimum. First, at each body forward position $x$, we considered imaginary straight paths away from the pitch local minimum (Figure \ref{fig:p13_2}B, $iii$, blue dot) in the full pitch-roll space ([$-180\degree$, 180$\degree$]), parameterized by an angle $\Psi$ relative to the negative pitch direction (body pitched up). Along each imaginary straight path, we obtained a cross section of the potential energy landscape (Figure \ref{fig:p13_2}B, $iii$, inset). Then, we measured and defined the maximal increase in potential energy in the cross section as the escape barrier along this imaginary straight path, which was a function of $\Psi$, as shown by a polar plot (Figure \ref{fig:p15_4}B). Then, we calculated how escape barrier along different directions away from pitch local minimum changed as traversal progressed (increasing $x$). We defined pitch-to-roll transition barrier as the lowest escape barrier, which occurred at the saddle point between pitch and roll basins. We measured how pitch-to-roll transition barrier and the location of saddle point in the pitch-roll space changed as $x$ increased. For the robot, we calculated pitch-to-roll transition barrier using the average landscape at each \textit{K}. For the animal, we used the average landscape with zero average lateral position and body yaw for simplicity, considering its large trial-to-trial variation in lateral position and body yaw.

\subsection{Robot system state velocity directions}
To measure the direction towards which the robot state trajectory was moving in the pitch-roll space during transition, for each trial, we calculated the velocity vector of the state trajectory in the pitch-roll space from the measured body roll and pitch, low-pass filtered data using a sixth order Butterworth filter. Then, we calculated the polar angle of this velocity vector relative to the pitch-roll axes of the landscape. To focus on the transition, for each trial in which pitch-to-roll transition occurred, we only considered the portion of the trial occurring over the $x$ range from start of beam contact to the onset of transition (Figure \ref{fig:p9_s4}, vertical dashed lines ($i$)-($ii$)). For trials in which pitch-to-roll transition did not occur, we considered the portion of the trial within the average $x$ range where transition was observed at higher \textit{K} ($x$ = [$-69$, $-39$] mm). For each \textit{K}, we pooled data of trials in which the system was trapped in the pitch mode and those in which the system transitioned to the roll mode to calculate their respective distribution (polar histogram) of velocity directions (Figure \ref{fig:p15_4}D, blue and red). We also measured the directions of the saddle point between the pitch and roll basins and the local maximum along the pitch-up and pitch-down directions, averaged over the $x$ range in which transition was observed (Figure \ref{fig:p15_4}D, yellow and gray dashed lines).

\subsection{Animal active body and limb adjustments}
We observed high speed videos of animal experiments to search for evidence of the animal using active adjustments to make transition. For each \textit{K}, we counted the percentage of trials in which the animal repeatedly flexed its head relative to the body, differentially used its hind legs, or did both \citep{wang2021a}.

\clearpage
\clearpage
\section{Results}
\begin{figure}
    \centering
    \includegraphics[width=1.0\linewidth]{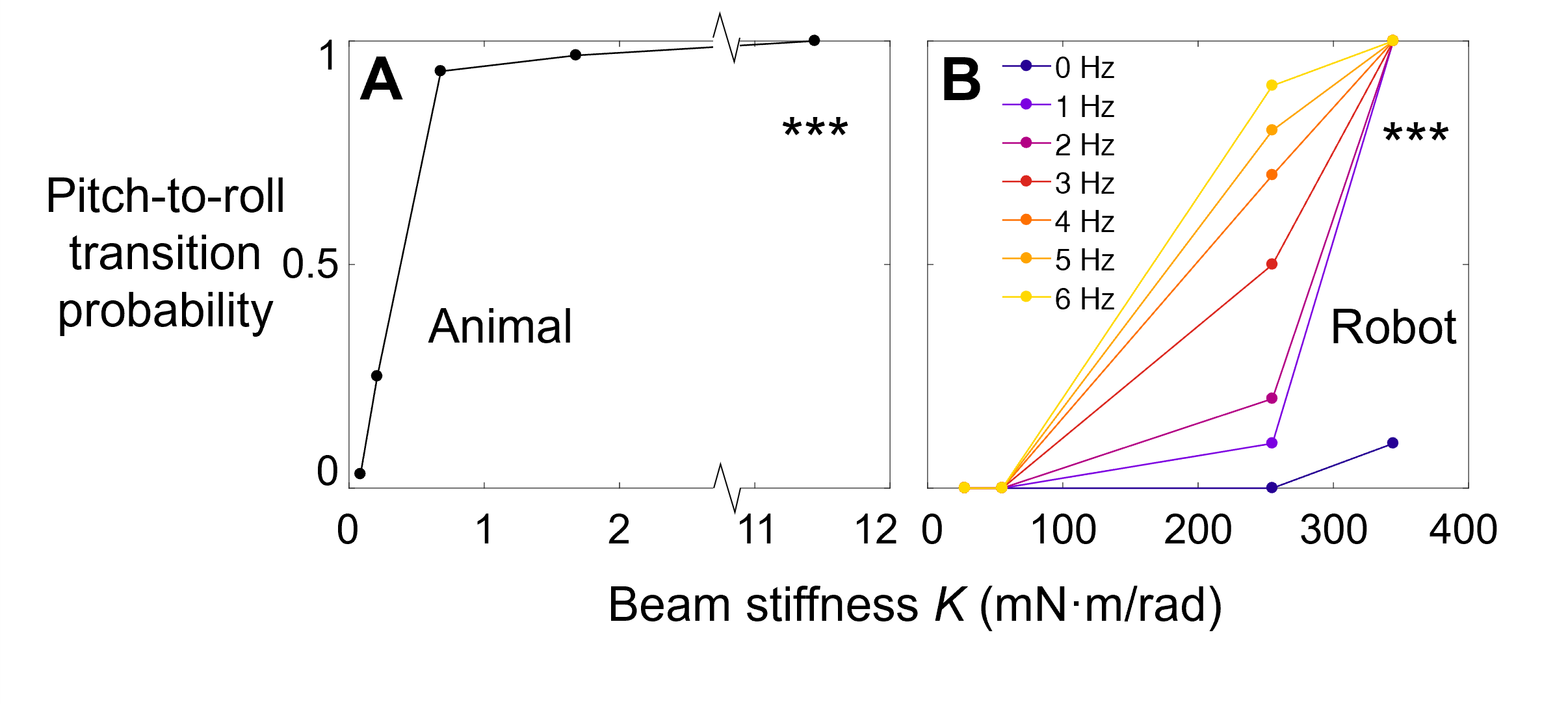}
    \caption[Pitch-to-roll transition probability of animal and robot as a function of beam stiffness \textit{K}] %
    {Pitch-to-roll transition probability of animal (A) and robot (B) as a function of beam stiffness \textit{K} . For robot, we varied oscillation frequency \textit{f}  to vary kinetic energy fluctuation. *** indicates a significant dependence on \textit{K}  (animal: mixed-effects chi-squared test, \textit{P} < 0.0001, $\chi^2$ = 297.4; robot: chi-squared test, P < 0.0001,  $\chi^2$ = 247.1). \textit{n} = 64, 60, 60, 62, 64 trials for animal and \textit{n} = 70 trials at each \textit{K} for robot.}
    \label{fig:p12_1_ef}
\end{figure}

Before encountering the beams, both the robot and animal moved forward with a near horizontal body posture. After beam contact, both the robot and animal started traversing by pushing against the beams, with the body pitched up. As beam stiffness \textit{K} increased, pitch-to-roll transition probability increased for both the robot and animal (Figure \ref{fig:p12_1_ef}; \textit{P} < 0.0001, mixed-design chi-squared test). At low K, neither transitioned to the roll mode even with body oscillation. At the highest K, both always transitioned, except for the robot without oscillation. In addition, for the robot at high \textit{K} (255 mN$\cdot$m/rad), pitch-to-roll transition probability increased with oscillation frequency \textit{f} (Figure \ref{fig:p12_1_ef}B) and thus with kinetic energy fluctuation (Figure \ref{fig:p10_s5}A). At the highest \textit{K} tested (344 mN$\cdot$m/rad), pitch-to-roll transition probability reached one for all \textit{f} > 0 tested. For simplicity, below we first describe robot results followed by animal results.

\begin{figure}
    \centering
    \includegraphics[width=1.0\linewidth]{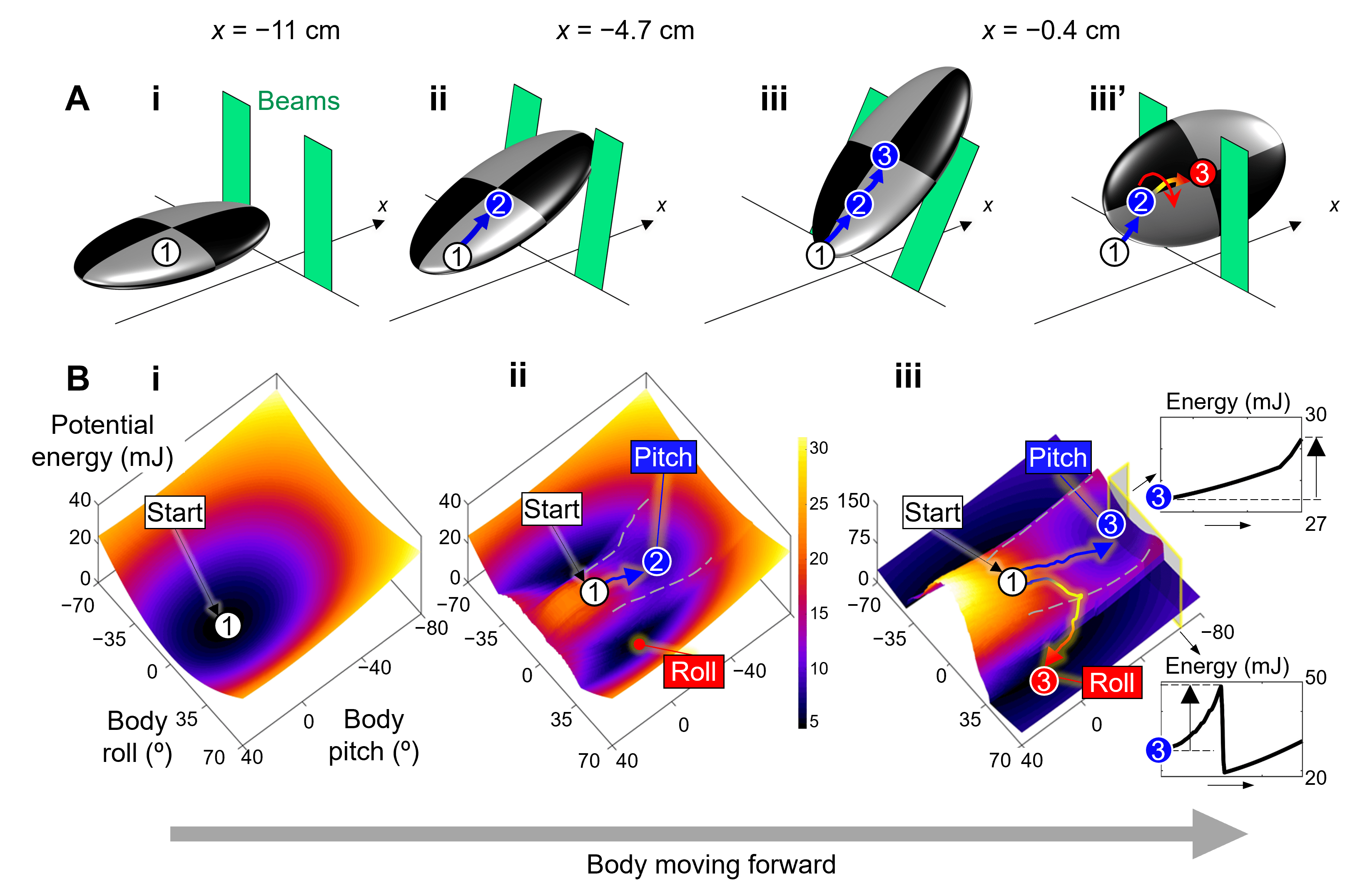}
    \caption[Robot locomotor transitions on a potential energy landscape] %
    {Robot locomotor transitions on a potential energy landscape. Results are shown at \textit{K} = 255 mN$\cdot$m/rad. (A) Snapshots of body before and during interaction with two beams in pitch (i, ii, iii) and roll (iii’) modes. (B) Snapshots of landscape over body pitch-roll space before (i) and during (ii, iii) interaction. Representative system state trajectories are shown for being trapped in pitch basin (blue) and transitioning to roll basin (red). Insets in (iii) define potential energy barriers to escape from pitch local minimum in pitch-up and positive roll directions. Dashed gray curves on landscape show boundaries between pitch and roll basins. Note that landscape evolves as body moves forward (increasing \textit{x}), and only part of landscape over pitch-roll space is shown to focus on pitch and roll basins.}
    \label{fig:p13_2}
\end{figure}

We tested the first hypothesis by reconstructing the robot’s potential energy landscape and evaluating how its system state behaved on the landscape (Figure \ref{fig:p13_2}). Using the measured physical and geometric parameters of the body and beams, we calculated the robot’s system potential energy (sum of body and beam gravitational energy and beam elastic energy) as a function of body pitch, roll, and forward position x relative to the beams. For simplicity, we first examine results at \textit{K} = 255 N$\cdot$m/rad. Before the body contacted the beams (Figure \ref{fig:p13_2}A, i), pitching or rolling increased body gravitational energy (because body center of mass was below rotation axes, Figure \ref{fig:p11_s6}). Thus, the potential energy landscape over body pitch-roll space had a global minimum at zero pitch and zero roll, i.e., when the body was horizontal (Figure \ref{fig:p13_2}B, i). As the body moved closer and interacted with the beams (Figure \ref{fig:p13_2}A, ii, iii), the global minimum evolved into a “pitch” local minimum at a finite pitch and zero roll (Figure \ref{fig:p13_2}B, ii, iii, blue). Meanwhile, two “roll” local minima emerged at near zero pitch and a finite positive or negative roll (Figure \ref{fig:p13_2}B, ii, iii, red, for rolling right or left), whose energies were lower than the pitch local minimum. Hereafter, we refer to these local minimum basins as pitch and roll basins \footnote{A fourth basin also emerged with its local minimum at a finite positive pitch and zero roll, corresponding to the body pitching down against the beams. However, such a configuration was never observed in the robot or animal.}.

\clearpage

We discovered that the robot’s system state during the observed pitch and roll modes were attracted to the pitch and roll basins, respectively. When the body was far away from the beams, the system state in pitch and roll space settled to the global minimum of the landscape (Figure \ref{fig:p13_2}B, i). During beam interaction, without oscillation, the system state was trapped in the pitch basin, leading to the body pushing across the beams in a pitched-up orientation with little roll (Figures \ref{fig:p13_2}A, B, ii, iii). With oscillation, the system stochastically escaped from the pitch basin and crossed a potential energy barrier to reach the roll basin (Figure \ref{fig:p13_2}B, iii), thereby transitioning from the pitch to the roll mode (Figure \ref{fig:p13_2}B, ii, iii’). We examined system state trajectory on the landscape reconstructed for each trial. Whether the robot was trapped in the pitch mode (blue trajectories) or transitioned to the roll mode (red trajectories), its system state was attracted to the corresponding basin in nearly all trials (99\%, not significantly different from 1, \textit{P} > 0.15, Student’s t-test, Figure \ref{fig:p15_4}A, iii). Because of this strong attraction, the measured system potential energy closely matched the observed mode basin’s local minimum energy throughout traversal (Figure \ref{fig:p17_5}iii, solid vs. dashed curves). All these findings held true at other \textit{K}  (near 100\%, Figures \ref{fig:p15_4}A, \ref{fig:p17_5}). Together, these robot results supported our first hypothesis.

\begin{figure}
    \centering
    \includegraphics[width=1.0\linewidth]{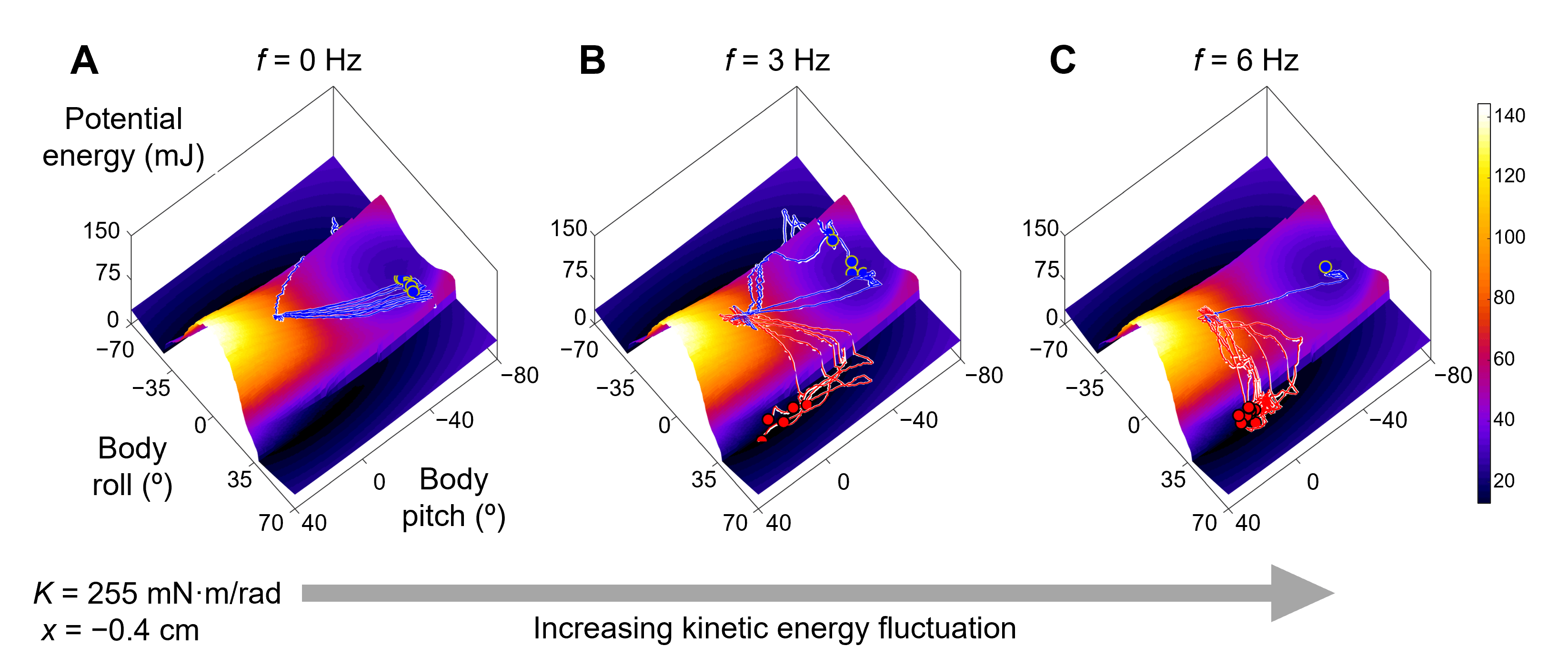}
    \caption[Robot locomotor transitions are stochastic and become more likely as kinetic energy fluctuation increases] %
    {Robot locomotor transitions are stochastic and become more likely as kinetic energy fluctuation increases. Comparison of state trajectory ensemble on average landscape (snapshot at \textit{x} = -0.4 cm) across oscillation frequencies: (A) \textit{f} = 0 Hz; (B) \textit{f} = 3 Hz; (C) \textit{f} = 6 Hz. Results are shown at \textit{K} = 255 mN$\cdot$m/rad. Blue and red curves show trials trapped in pitch basin and transitioning to roll basin, respectively. Trials in which body rolls left are flipped to rolling right considering lateral symmetry. \textit{n} = 10 trials at each \textit{f}. Only part of landscape over pitch-roll space is shown to focus on pitch and roll basins. Blue trajectories exiting pitch basin is an artifact of landscape averaging.}
    \label{fig:p14_3}
\end{figure}

\clearpage
Next, we tested the second hypothesis. We first observed how kinetic energy fluctuation affected the robot’s escape from a basin. Again, we examine results at \textit{K} = 255 mN$\cdot$m/rad first for simplicity. As \textit{f}  increased (which increased kinetic energy fluctuation), the system was more likely to escape from the pitch basin it was initially attracted to and reach the roll basin (Figure \ref{fig:p14_3}), resulting in more likely pitch-to-roll transitions (Figure \ref{fig:p12_1_ef}B, \textit{K} = 255 mN$\cdot$m/rad). 

\begin{figure}[p]
    \centering
    
    \caption[Robot tends to transition to roll basin when kinetic energy fluctuation is comparable to potential energy barrier to escape pitch local minimum and towards direction of lower barrier] %
    {Robot tends to transition to roll basin when kinetic energy fluctuation is comparable to potential energy barrier to escape pitch local minimum and towards direction of lower barrier. (A) Average potential energy landscape over pitch-roll space (snapshot at \textit{x} = 8 mm) with ensemble of state trajectories. Blue and red curves show trials trapped in pitch basin and transitioning to roll basin, respectively. Note that landscape evolves as body moves forward (increasing \textit{x}) and only part of the landscape over pitch-roll space is shown to focus on the pitch and roll basins. Top right number on each landscape shows percentage of trials in which system state is attracted to pitch/roll basin corresponding to observed mode. Blue trajectories exiting pitch basin is an artifact of landscape averaging. (B) Polar plot of potential energy barrier to escape from pitch local minimum (blue dot) along all directions in pitch-roll space (snapshot at \textit{x} = -53 mm). Pitch-to-roll transition barrier is defined as minimal escape barrier (arrows in iv), which occurs at saddle point between pitch and roll basins (yellow dot). (C) Pitch-to-roll transition barrier as a function of \textit{x}. Gray band shows \textit{x} range in which pitch-to-roll transition is observed (mean $\pm$ s.d.). Green circle/line in B, D shows measured average kinetic energy fluctuation of 2.3 mJ at highest \textit{f} = 6 Hz tested (Figure \ref{fig:p10_s5}). (D) Probability distribution of state velocity directions in pitch-roll space in the x range where transition is observed (gray band in C). Blue and red are data from trials trapped in pitch basin and transitioning to roll basin, respectively. Trials in which body rolls left are flipped to rolling right considering lateral symmetry. Black dashed lines and gray shaded sectors show angular direction of maximal escape barriers (mean $\pm$ s.d) along pitch up and down directions. Yellow dashed line and shaded sector show angular direction of minimal escape barrier (mean $\pm$ s.d), which occurs at saddle point. Columns i-iv are at \textit{K} = 28, 55, 255, and 344 mN$\cdot$m/rad. Data shown in A, C and D are for all \textit{f} tested (\textit{n} = 70 trials) at each \textit{K}.}
    \newpage
    \label{fig:p15_4}
\end{figure}

\clearpage
\begin{center}
   \includegraphics[width=1.0\linewidth]{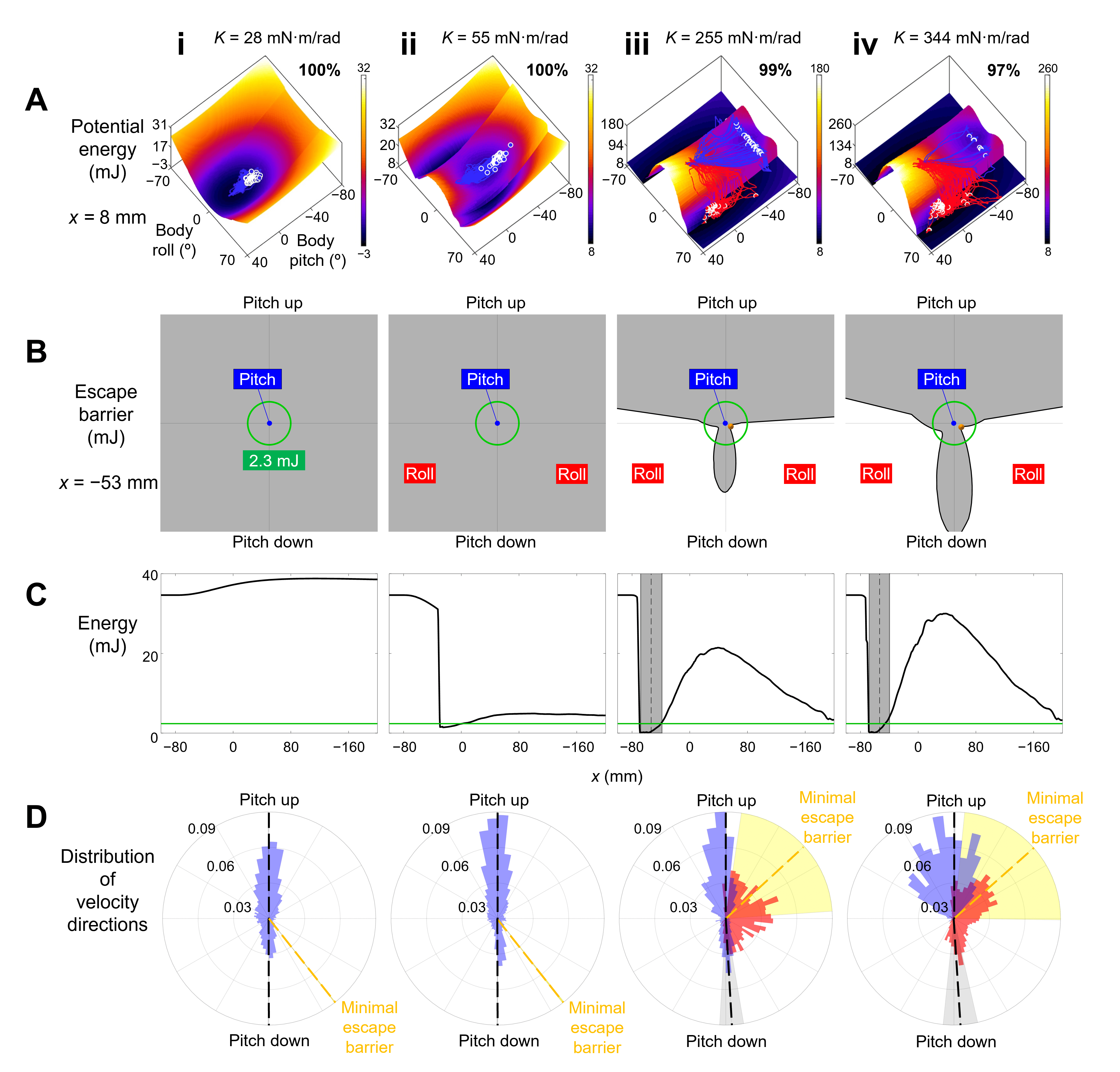}
\end{center}
\clearpage

Then, we compared the minimal potential energy barrier to escape from the pitch local minimum with the average kinetic energy fluctuation at $f$ =  6 Hz (Figure \ref{fig:p15_4}C, iii). The escape barrier depended on both towards which direction the system moved in the pitch-roll space (Figure \ref{fig:p13_2}B, iii, insets, Figure \ref{fig:p15_4}B, iii) and body forward position x relative to the beams (Figure \ref{fig:p15_4}C, iii). Minimal escape barrier occurred at the saddle point between the pitch and roll basins (Figure \ref{fig:p15_4}C, yellow dot), which we defined as pitch-to-roll transition barrier. Only within a small range of x was average kinetic energy fluctuation at $f$ =  6 Hz (Figure \ref{fig:p15_4}C, iii, green) sufficient for overcoming pitch-to-roll transition barrier (Figure \ref{fig:p15_4}C, iii, black). This range matched remarkably well with the x range over which pitch-to-roll transition was observed with increasing likelihood with \textit{f}  (gray band showing mean $\pm$ s.d. from all trials across \textit{f}). All these findings held true at \textit{K} = 344 N$\cdot$m/rad. At \textit{K} = 28 N$\cdot$m/rad, minimal escape barrier far exceeded kinetic energy fluctuation, consistent with the absence of transition. Together, these robot results supported our second hypothesis.

\clearpage
Finally, we tested the third hypothesis by examining the direction towards which the robot’s system state moved during interaction. At each \textit{K}, when the body was not in contact with the beams, the escape barrier was large along all directions in the pitch-roll space (e.g., \textit{x} = -80 mm). As the body moved forward (increasing \textit{x}), the escape barrier towards the direction of roll basins reduced drastically, becoming comparable to or even smaller than average kinetic energy fluctuation at $f$ =  6 Hz (green circle) at the saddle point (yellow dot). By contrast, escape barrier in the direction of pitching up or down was always greater than average kinetic energy fluctuation (Figure \ref{fig:p15_4}B). Examination of how the system state moved on the landscape and probability distribution of system state velocity directions in the pitch-roll space (Figure \ref{fig:p15_4}D) showed that escape was more aligned with the direction of the saddle point between pitch and roll basins, i.e., escape was more likely towards the direction of lower barrier. This is intuitive because in other directions escape barrier was higher and often exceeded kinetic energy fluctuation. Together, these robot observations supported our third hypothesis.

Comparison of robot observations across \textit{K}  further suggested a concept of favorability for locomotor transitions. As \textit{K}  increased, pitch-to-roll transition became more likely (Figure \ref{fig:p15_4}A), saturating at one for all \textit{f} > 0 tested at the highest \textit{K}  (Figure \ref{fig:p12_1_ef}B). Intuitively, when the beams were flimsy, the body pushed across (trapped in the pitch mode) as if nothing were there; when the beams were rigid, the body could not push across and must roll. Thus, the likelihood of pitch-to-roll transition is positively correlated with how favorable transitioning to the roll mode is relative to staying in the pitch mode.

\begin{figure}
    \centering
    \includegraphics[width=1.0\linewidth]{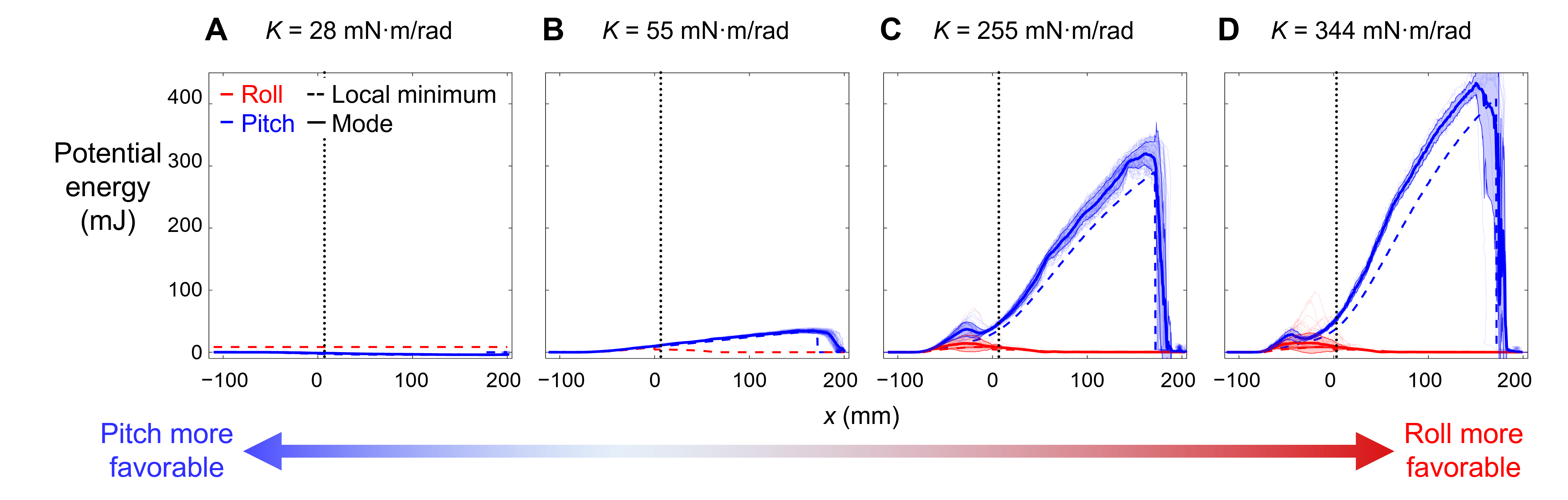}
    \caption[Favorability measure for robot] %
    {Favorability measure for robot. Potential energy of measured pitch and roll modes (solid, mean $\pm$ s.d.) and of pitch and roll local minima  (dashed) as a function of \textit{x}. Measured data are for all \textit{f} tested (\textit{n} = 70 trials) at each \textit{K}. Blue and red show trials trapped in pitch basin and transitioning to roll basin, respectively. Columns i-iv are at \textit{K} = 28, 55, 255, and 344 mN$\cdot$m/rad. Dotted line at \textit{x} = 8 mm shows location of snapshots in Figure \ref{fig:p15_4}A.}
    \label{fig:p17_5}
\end{figure}

To provide a measure of favorability, we compared whether the pitch or roll basin was lower during traversal, measured at their respective local minimum (Figure \ref{fig:p17_5}). At low \textit{K}  (28 mN$\cdot$m/rad), the pitch basin remained the global minimum basin throughout traversal (Figure \ref{fig:p17_5}i), indicating that the pitch mode was more favorable. As \textit{K}  increased, the pitch basin became increasingly higher than the roll basin (Figure \ref{fig:p17_5}ii-iv), indicating that the roll mode became increasingly more favorable. At small \textit{K} = 55 mN$\cdot$m/rad for \textit{x} > 0, although the roll mode was more favorable (Figure \ref{fig:p17_5}ii), kinetic energy fluctuation was smaller than the transition barrier (Figure \ref{fig:p15_4}C, ii); thus, transition did not occur (Figure \ref{fig:p15_4}A, ii). We emphasize that the negative correlation between the probability of staying in or transitioning to a mode and its relative basin height is only an emergent outcome of the transition physics. The passive robot does not directly feel how high or how low an adjacent basin is; whether it escapes and makes a transition only depends on the basin in which it currently resides. Exactly how favorability difference between basins emerges from the local dynamics of escaping from a basin remains to be understood.

\begin{figure}
    \centering
    \includegraphics[width=1.0\linewidth]{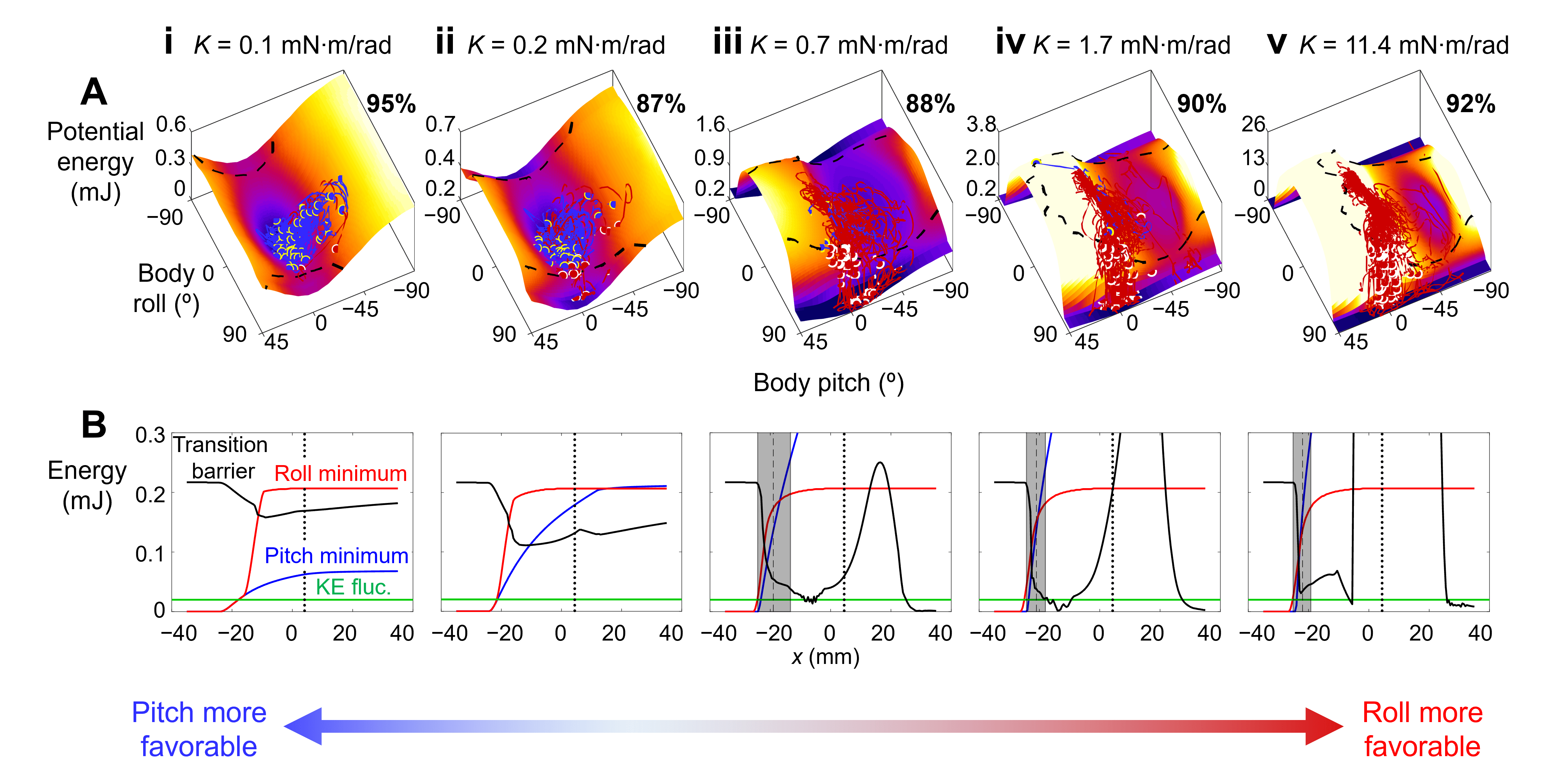}
    \caption[Animal tends to transition to roll basin when kinetic energy fluctuation is comparable to potential energy barrier to escape pitch basin and when roll basin is more favorable.] %
    {Animal tends to transition to roll basin when kinetic energy fluctuation is comparable to potential energy barrier to escape pitch basin and when roll basin is more favorable. (A) Potential energy landscape over pitch-roll space (snapshot at \textit{x} = 4 mm, dotted lines in B) with ensemble of state trajectories. Dashed black curves on landscape show boundary of pitch basin. Note that landscape evolves as body moves forward (increasing \textit{x}) and only part of landscape over pitch-roll space is shown to focus on pitch and roll basins. We set color map scale to saturate at high energy to highlight landscape basins. (B) Potential energy of pitch and roll local minima and pitch-to-roll transition barrier as a function of \textit{x}. Green line is measured average kinetic energy fluctuation of 0.02 mJ. Columns i-v are at \textit{K} = 0.1, 0.2, 0.7, 1.7, and 11.4 mN$\cdot$m/rad (\textit{n} = 64, 60, 60, 62, and 64 trials).}
    \label{fig:p18_6}
\end{figure}

Similar to the feedforward-controlled robot, the animal’s system state during the observed pitch or roll mode was attracted to the corresponding basin of the potential energy landscape (Figure \ref{fig:p18_6}A, ~90\% of trials at all \textit{K}). In addition, pitch-to-roll transition mostly occurred when both average kinetic energy fluctuation became comparable to transition barrier and the roll mode became more favorable than the pitch mode (Figure \ref{fig:p18_6}B). These similar observations were remarkable because, for the animal that displayed larger lateral motion and yawing, leg motion, and individual variation, the landscape (which was averaged from all trials) provided a much coarser approximation of the system than for the simpler, well-controlled robot. These animal results supported our first and second hypotheses. We did not test the third hypothesis in the animal, considering that the measured system state velocity was noisy and the animal had higher lateral and yaw motion during traversal.

These results showed that physical interaction with the terrain also played a major role in the animal’s probabilistic locomotor transitions, even when active behavior was likely at play. In some trials, the animal transitioned even when its average kinetic energy fluctuation was smaller than transition barrier (Figure \ref{fig:p18_6}B). In addition, the animal occasionally transitioned to the less favorable roll mode at low \textit{K} (Figure \ref{fig:p18_6}A, i, ii, red trajectories). Further, the animal often flexed its head relative to the body and used the two hind legs differentially \citep{wang2021a} during beam interaction (23\%, 63\%, 89\%, 79\%, and 85\% of the trials at the five \textit{K}’s). All these were evidence that the animal’s transition involved active behavior (see discussion). Unlike the robot that was pulled forward at a constant speed (pulling force always exceeded beam resistive force), the animal had a finite ability to push forward and may rely more on such active behavior to facilitate transition \citep{wang2021a}.

\clearpage
\section{Discussion}
In summary, using a transition between two representative modes in a model system, we demonstrated that an energy landscape approach helps understand how stochastic transitions of animals and robots across locomotor modes statistically emerge from physical interaction with complex 3-D terrain. We discovered that kinetic energy fluctuation from oscillatory self-propulsion helps the system cross barriers on a potential energy landscape to make locomotor transitions. This provided compelling evidence about why variation in movement can lead to stochastic outcome \citep{stephens2011a} and can be advantageous when locomotor behavior is separated into distinct modes. This also explained early observations of surprising ability to traverse unstructured terrain of bandwidth-limited, rapid-running insects \citep{sponberg2008a} and feedforward-controlled legged robots \citep{altendorfer2001a}, as both have substantial body oscillation during locomotion. However, we view this way of “vibrate like a particle” as only one of a suite of transition strategies. Animals and robots may use other strategies to make transitions, such as plan anticipatory actions \citep{gart2018a} and use random search \citep{xuan2020a} to overcome barriers, use sensory feedback adjustments to move towards lower barriers or reduce barriers \citep{wang2021a}, or even change morphology to modify landscape topology to introduce or eliminate certain modes \citep{han2021a}.

We posit that there is an “energy landscape dominated” regime of locomotion, where along certain directions there exist large potential energy barriers that are comparable to or exceed kinetic energy and/or mechanical work generated by each propulsive cycle or motion. This may happen when propulsive forces are either limited by physiological, morphological, and environmental (e.g., low friction) constraints or do not well align with directions along which large barriers occur. In complex terrain with many large obstacles \citep{gart2018a,han2021a,li2015a,wang2021a} and even during strenuous maneuvers \citep{li2019a,othayoth2021a,xuan2020a,xuan2020b}, these situations are frequent. In this regime, not only does energy landscape modeling provide a useful statistical physics approach for understanding locomotor transitions across modes, but it may also allow comparison across systems (different animal species, robots, terrain, and modes) to discover general physical principles. Outside of this regime, energy landscape modeling is not useful—for example, not for ballistic jumping over small obstacles with kinetic energy far exceeding potential energy barriers.

We discovered that distinct attractive basins of the potential energy landscape can lead to stereotyped locomotor modes and transitions in both the animal and feedforward-controlled robot. Because our potential energy landscape is directly derived from first principles (as opposed to fitting a model to behavioral data \citep{mearns2020a,stephens2008a,wiltschko2015a}, this result provided compelling evidence that behavioral stereotypy of animals emerges from their neural and mechanical systems directly interacting with the physical environment \citep{berman2018a,brown2018a}. In addition, our approach should inform how direct physical interaction with the environment constrains behavioral hierarchy \citep{berman2018a,brown2018a}. For example, for grass-like obstacle traversal, starting with our coarse-grained landscape here resulting from a rigid body interacting with rigid “beams” on torsional springs, we can add degrees of freedom describing head flexion \citep{wang2021a}, body bending and twisting, articulated leg motions, and more realistic beam obstacles with cantilever bending and spatial heterogeneity. This will reveal more nuanced pathways of transitioning between fine-grained locomotor modes that have a variety of body and appendage configuration and terrain responses (e.g., flexing the head and tucking the legs to roll into the gap \citep{wang2021a}, separating beams laterally, etc.). Analyzing the disconnectivity \citep{wales2003a} of basins of such a more complete, high-dimensional energy landscape will reveal the hierarchy (“treeness” \citep{berman2016a}) of locomotor modes in complex terrain.

\subsection{Towards understanding emergent behaviour in natural environments }
More broadly, these considerations suggest that our energy landscape approach provides a means towards first-principle, physical understanding of the organization of locomotor behavior, filling a critical knowledge gap. The field of movement ecology \citep{nathan2008a} makes field observations of trajectories of animals as a point mass moving and making behavioral transitions in natural environments (e.g., \citep{suraci2019a}), whose physical interactions are difficult to measure. Recent progress in quantitative ethology  has advanced understanding of the organization of behavior \citep{berman2016a,berman2018a,brown2018a,mearns2020a,wiltschko2015a}, often by quantifying kinematics in homogeneous, near featureless laboratory environments \citep{berman2016a,cande2018a,mearns2020a,stephens2008a}. Our work highlights the importance and feasibility of, and opens new avenues for, studying how the organization of behavior is constrained by an animal’s direct physical interaction with realistic environments \citep{li2013a}. Doing so will help inform how animal behavior evolves in nature; it will also simplify robot design, control, and planning to generate robust locomotor transitions in complex terrain \citep{Mi2022}, which may be otherwise intractable in the large locomotor-terrain parameter space. This is analogous to rugged free energy landscapes allowing divide-and-conquer in protein folding \citep{dill2012a}.

\begin{figure}
    \centering
    \includegraphics[width=1.0\linewidth]{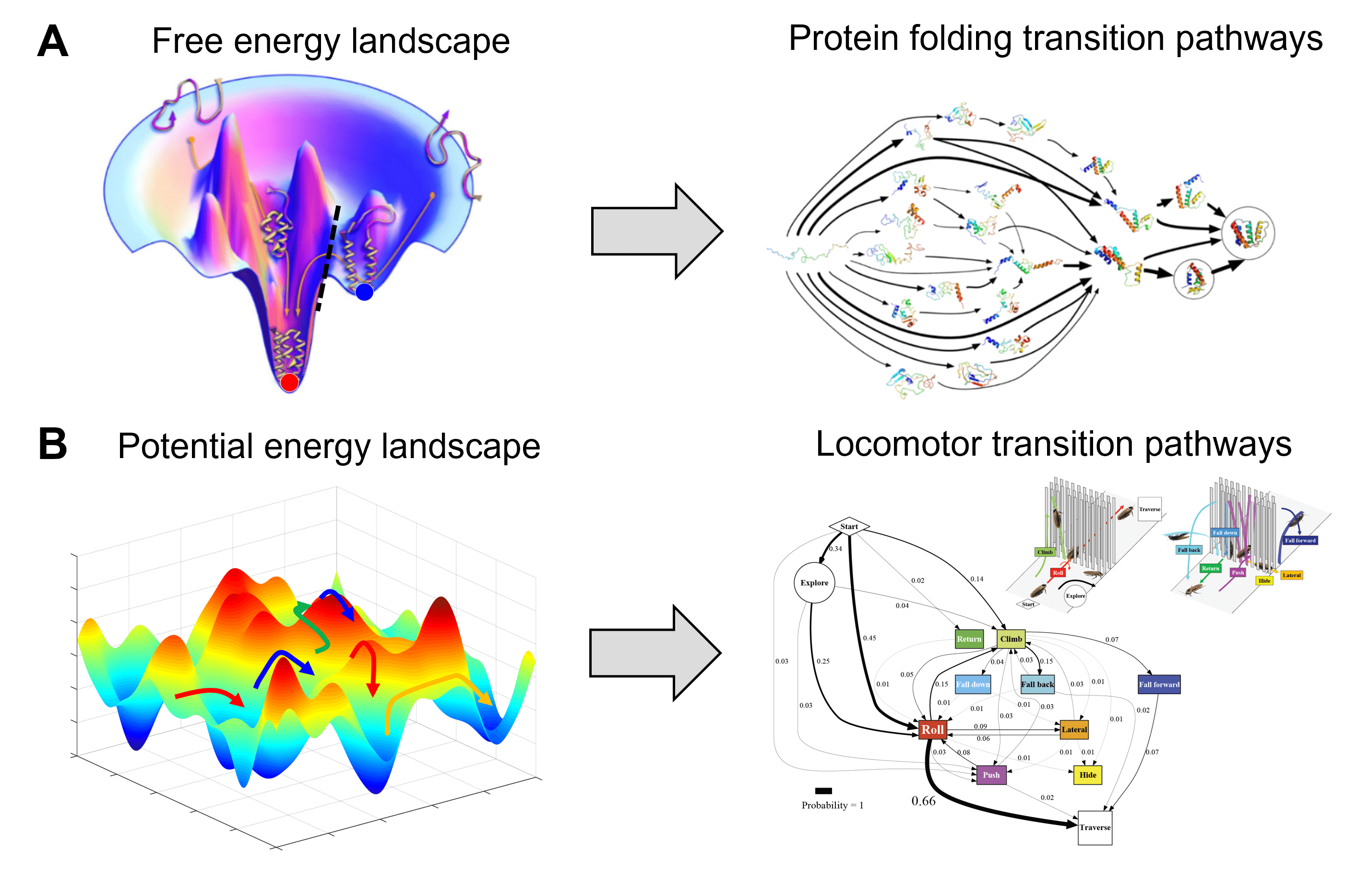}
    \caption[Comparison of energy landscape between protein-folding transitions and locomotor transition.] %
    {Comparison of energy landscape between protein-folding transitions and locomotor transition. (A) Energy landscape theories help understand physical principles and predict global structures and emergent properties of probabilistic protein folding transitions via multiple pathways. Image credits: A, left panel Reproduced from \citep{dill2012a}; A, right panel reproduced from \citep{voelz2012a}. (B) We envision energy landscape modeling as a beginning of a statistical physics approach for understanding and predicting probabilistic, multi-pathway locomotor transitions in complex terrain \citep{li2015a}.}
    \label{fig:p19_s7}
\end{figure}

Our empirically discovered physical principles of locomotor transitions are surprisingly similar to those of microscopic systems (Figure \ref{fig:p19_s7}), especially multi-pathway protein folding transitions where predictive energy landscape theories have been very successful \citep{dill2008a,onuchic2004a,wales2003a}. Thus, we envision our energy landscape as the beginning of a statistical physics theory that will quantitatively predict global structures and emergent dynamics of multi-pathway locomotor transitions in the energy landscape dominated regime. An immediate next step towards this is to model conservative forces using potential energy landscape gradients and add stochastic, non-conservative propulsive and dissipative forces that perturb the system to “diffuse” across landscape barriers (analogous to \citep{bryngelson1989a,socci1996a}). Doing this will also elucidate how escape dynamics from a basin locally leads to emergent favorability difference between basins. These physical principles will help reveal how animals, and how robots should, use local force sensing to control motion to facilitate locomotor transitions on the landscape. Further, although it seems obvious that near-equilibrium statistical thermodynamics does not directly apply here, an energy landscape approach to locomotor transitions in complex terrain provides opportunities to test and develop new theories of few-body active matter \citep{savoie2019a}.

Finally, our energy landscape approach provides a conceptual way of thinking about locomotor modes beyond near-steady-state, limit-cycle-like behavior (e.g., walk, run, climb \citep{blickhan1993a,goldman2006a,kuo2007a}) by adding metastable behavior \citep{byl2009a} locally attracted to landscape basins (e.g., pitch and roll modes here, which are far-from-steady maneuvers). We foresee the creation of new dynamical systems theories of terrestrial locomotion \citep{holmes2006a} that produce transitions across locally attractive landscape basins as well as between limit-cycle attractors \citep{diederich2002a,geyer2006a}. They will enable using physical interaction to design, control, and plan basins funneled into one another to compose \citep{burridge1999a} locomotor transitions to perform high-level tasks in the real world. Terradynamics of locomotor-terrain interaction starting from first principles \citep{li2013a} such as illustrated here will facilitate this progress.

\cleardoublepage

\chapter{Propelling and perturbing appendages together facilitate strenuous ground self-righting}
\label{chap:elife}

\let\thefootnote\relax\footnotetext{This chapter is a published paper by Ratan Othayoth and Chen Li in \textit{eLife} (2021) (\cite{othayoth2021a})}

\section{Summary}
Terrestrial animals must self-right when overturned on the ground, but this locomotor task is strenuous. To do so, the discoid cockroach often pushes its wings against the ground to begin a somersault which rarely succeeds. As it repeatedly attempts this, the animal probabilistically rolls to the side to self-right. During winged self-righting, the animal flails its legs vigorously. Here, we studied whether wing opening and leg flailing together facilitate strenuous ground self-righting. Adding mass to increase hind leg flailing kinetic energy increased the animal’s self-righting probability. We then developed a robot with similar strenuous self-righting behavior and used it as a physical model for systematic experiments. The robot’s self-righting probability increased with wing opening and leg flailing amplitudes. A potential energy landscape model revealed that, although wing opening did not generate sufficient kinetic energy to overcome the high pitch potential energy barrier to somersault, it reduced the barrier for rolling, facilitating the small kinetic energy from leg flailing to probabilistically overcome it to self-right. The model also revealed that the stereotyped body motion during self-righting emerged from physical interaction of the body and appendages with the ground. Our work demonstrated the usefulness of potential energy landscape for modeling self-righting transitions.

\section{Significance Statement}
Animals can, and robots should, use different types of appendages together to propel and perturb themselves to self-right when overturned, a strenuous yet crucial locomotor task.

\section{Author contributions}
Ratan Othayoth designed robotic physical model, performed animal and robot experiments, analyzed and validate results, and created visualizations. Chen Li conceived and designed study, supervised the project, obtained funding and reviewed and edited the manuscript for publication, 

\clearpage
\section{Introduction}
Ground self-righting is a critical locomotor capability that animals must have to survive (for a review, see \cite{li2019a}). The longer an animal is flipped over and stranded, the more susceptible it is to risks like predation, starvation, desiccation \citep{steyermark2001a}, and limited mating success \citep{penn1995a}. Thus, it is crucial for animals to be able to self-right at a high probability because it can mean the difference between life or death. Similarly, ground self-righting is critical for the continuous operation of mobile robots (for a review, see \cite{li2017a}).
	
Ground self-righting is a strenuous task. For example, to self-right, cockroaches must overcome potential energy barriers seven times greater than the mechanical energy required per stride for steady-state, medium speed running (8 body lengths s$^{-1}$) \citep{kram1997a} or, exert ground reaction forces eight times greater than that during steady-state medium speed running (5 body lengths$^{-1}$) \citep{full1995a}. Often, animals struggle to self-right quickly and needs multiple attempts \citep{brackenbury1990a,domokos2008a,hoffman1980a,kopp1927a,li2019a,silvey1973a} to self-right due to constraints from morphology, actuation, and the terrain \citep{domokos2008a,faisal2001a,golubovi2017a,li2019a,steyermark2001a}.

Ground self-righting has been studied in a diversity of animals, including insects \citep{brackenbury1990a,delcomyn1987a,faisal2001a,frantsevich1980a,li2019a,sherman1977a,zill1986a}, crustaceans \citep{davis1968a,silvey1973a} , mollusks \citep{hoffman1980a,weldon1979a,zhang2020a}, and vertebrates \citep{ashe1970a,bartholomew1951a,creery1980a,domokos2008a,golubovi2015a,kopp1927a,malashichev2016a,pellis1991a,robins1998a,vince1986a,winters1986a}. A diversity of strategies have been described, including using appendages such as legs, wings, tail, and neck and deforming the body substantially. Often, rather than using a single type of appendages or just deforming the body without using appendages, animals use them together to propel and perturb the body to destabilize from the upside-down state (\citep{brackenbury1990a}; \citep{davis1968a}; \citep{domokos2008a}; \citep{faisal2001a}; \citep{hoffman1980a}; \citep{li2019a}). In particular, vigorous appendage flailing is a ubiquitous behavior observed across a diversity of species (\citep{ashe1970a}; \citep{brackenbury1990a}; \citep{davis1968a}; \citep{delcomyn1987a}; \citep{domokos2008a}; \citep{faisal2001a}; \citep{hoffman1980a}; \citep{kleitman1926a}; \citep{kopp1927a}; \citep{li2019a}; \citep{zill1986a}). Some of these animals also use other appendages or the body to propel against the ground (\citep{brackenbury1990a}; \citep{davis1968a}; \citep{domokos2008a}; \citep{faisal2001a}; \citep{hoffman1980a}; \citep{li2019a}), and such vigorous appendage flailing appears to be a desperate, wasteful struggle.

\begin{figure}[p]
    \centering
    \includegraphics[width=1.0\linewidth]{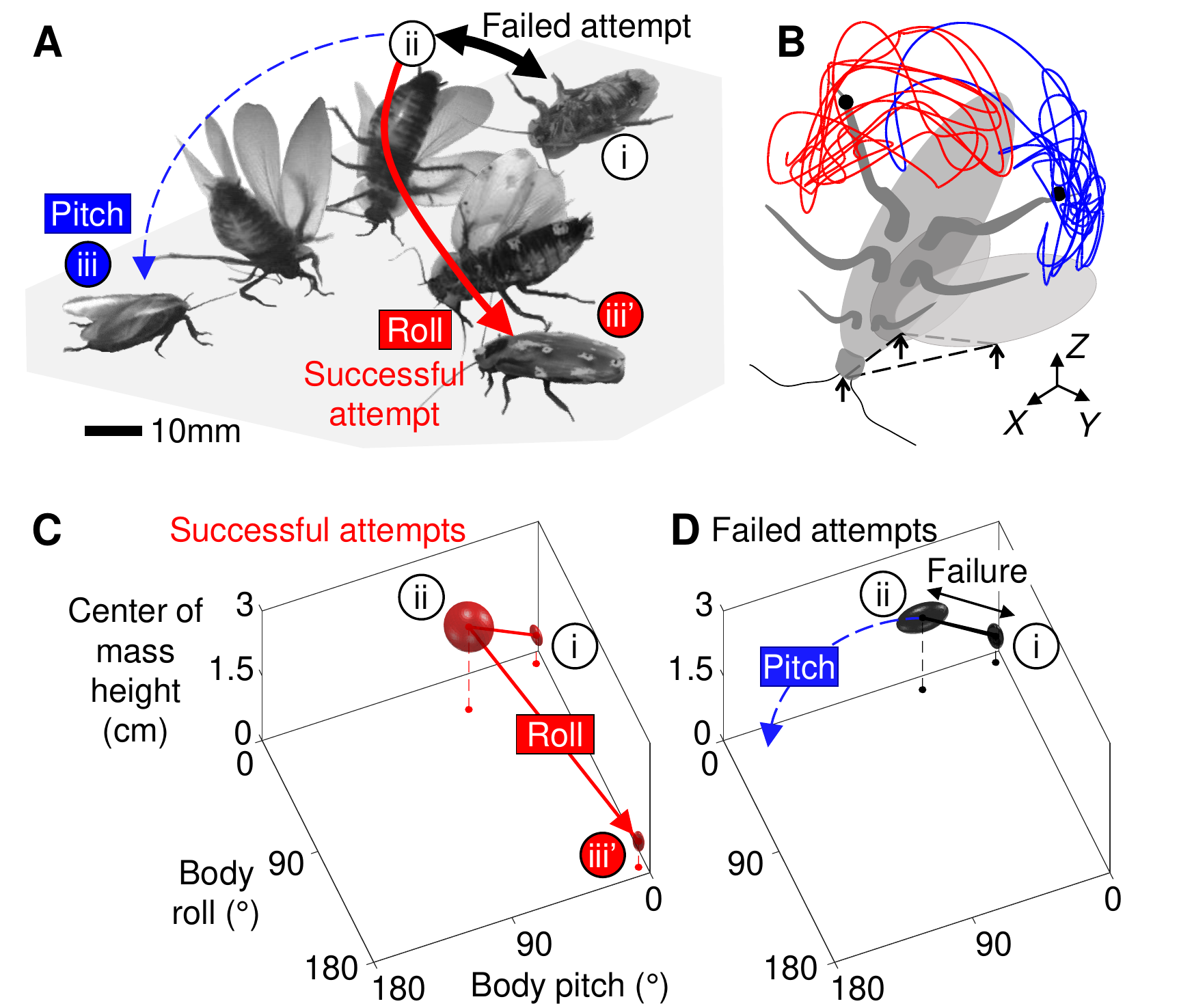}
    \caption[Strenuous, leg-assisted, winged ground self-righting of discoid cockroach]{Strenuous, leg-assisted, winged ground self-righting of discoid cockroach. (A) Representative snapshots of animal successfully self-righting by pitch (blue) and roll (red) modes after multiple failed attempts (black arrow). See Figure 1—video 1 for a typical trial, in which the animal makes multiple failed attempts to pitch over the head and eventually rolls to self-right. (B) Schematic of metastable state with a triangular base of support (dashed triangle) formed by ground contacts of head and two wing wedges, with vigorous leg flailing. Red and blue curves show representative trajectories of left and right hind leg tips from a trial. \textit{x}-\textit{y}-\textit{z} is lab frame. (C, D) Stereotyped body motion during successful (C) and failed (D) self-righting attempts in body pitch, body roll, and center of mass height space. i, ii, and iii in A, C, and D show upside-down (i), metastable (ii), and upright (iii, iii’) states. Ellipsoids show means (center of ellipsoid) $\pm$ s.d. (principal semi-axis lengths of ellipsoid) of body pitch, body roll, and center of mass height at the beginning, highest center of mass height, and end of the attempt. For failed attempts, the upside-down state at the end of the attempt is not shown because it overlaps with the upside-down state at the start of the attempts (i). Data from \cite{li2019a}}
    \label{fig:elife_f1}
\end{figure}

Here, we study how propulsive and perturbing appendages together contribute to successful strenuous ground self-righting. Our model system is the discoid cockroach’s strenuous ground self-righting using wings \citep{li2019a}) (Figure \ref{fig:elife_f1}). The overturned animal opens and pushes its wings against the ground in an attempt to self-right, resulting in its body pitching forward (Figure \ref{fig:elife_f1}Ai). Because the two opened wings and head form a triangular base of support, in which the center of mass projection falls (Figure \ref{fig:elife_f1}Aii, B), this intermediate state is metastable. However, wing pushing rarely pitches the animal all the way over its head to self-right (the pitch mode, Figure \ref{fig:elife_f1}Aii-iv, blue). Thus, the animal often opens and closes its wings (hereafter referred to as an attempt ) multiple times, resulting in its body repeatedly pitching up and down, but it fails to self-right (Figure \ref{fig:elife_f1}A, black arrows). Eventually, the animal almost always self-rights by rolling sideways over one of the wings (the roll mode; Figure \ref{fig:elife_f1}Aiii’-iv’, red). Although wings are the primary propulsive appendages in this self-righting strategy, the animal also vigorously flails its legs mediolaterally, even when body pitching nearly prevents them from reaching the ground (Figure \ref{fig:elife_f1}B, dashed curves). The legs occasionally scrape the ground, the abdomen occasionally flexes and twists, and the wings often deform passively under load \citep{li2019a}. For simplicity, we focused on the perturbing effects of the more frequent leg flailing (but see discussion of these other perturbing motions). Another curious observation is that, although the animal can in principle rotate its body in arbitrary trajectories to self-right, the observed body motion is stereotyped (Figure \ref{fig:elife_f1}) \citep{li2019a}.

A recent potential energy landscape approach to locomotor transitions \citep{othayoth2020a,othayoth2021b} provides a modeling framework to understand how propelling and perturbing appendages together contribute to strenuous ground self-righting. A previous study modeling ground self-righting of turtles in two dimensions (the transverse plane in which the body rolls) suggested that, when trapped in a gravitational potential energy well, modest kinetic energy from perturbing appendages (legs and neck) helps overcome the small potential energy barriers \citep{domokos2008a}. A recent study of cockroaches took an initial step in expanding potential energy landscape modeling of ground self-righting to three dimensions \citep{li2019a}. However, due to frequent camera occlusions, this study was unable to measure the complex 3-D motions of appendages and only modeled the animal as a rigid body. For turtles with a rigid shell interacting with the ground, modeling self-righting with a rigid body is a good first-order approximation. However, this approximation is no longer good for modeling winged self-righting of the discoid cockroach because wing opening will change potential energy landscape.

\begin{figure}[t]
    \centering
    \includegraphics[width=1.0\linewidth]{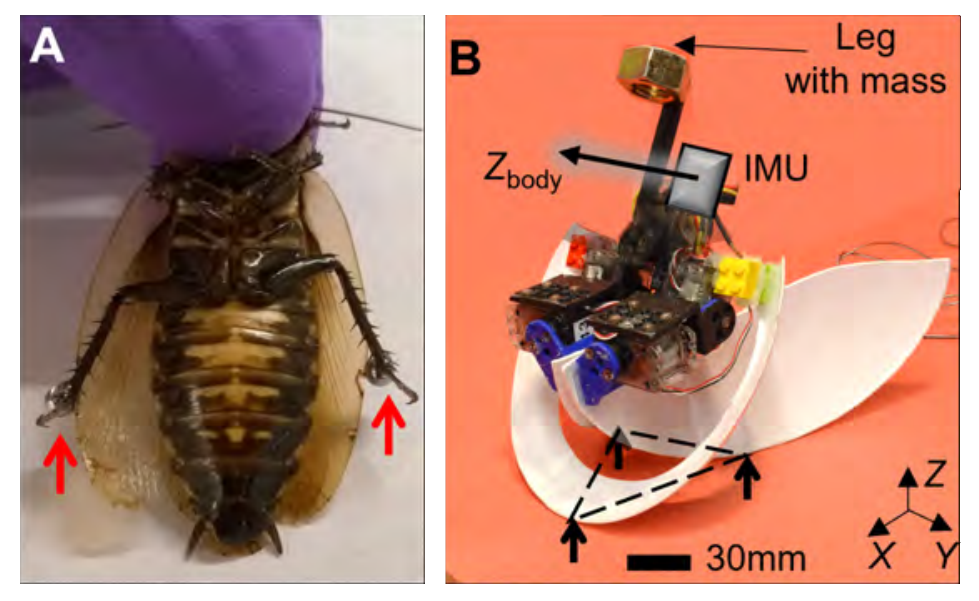}
    \caption[Animal leg modification and robotic physical model]{Animal leg modification and robotic physical model. (A) Discoid cockroach with modified hind legs with stainless steel spheres attached. (B) Robotic physical model in metastable state with a triangular base of support (dashed triangle), formed by ground contacts of head and two wing edges. Black arrow shows body \textit{Z}-axis, $\textit{Z}_{body}$.}
    \label{fig:elife_f2}
\end{figure}
\begin{figure}[t]
    \centering
    \includegraphics[width=1.0\linewidth]{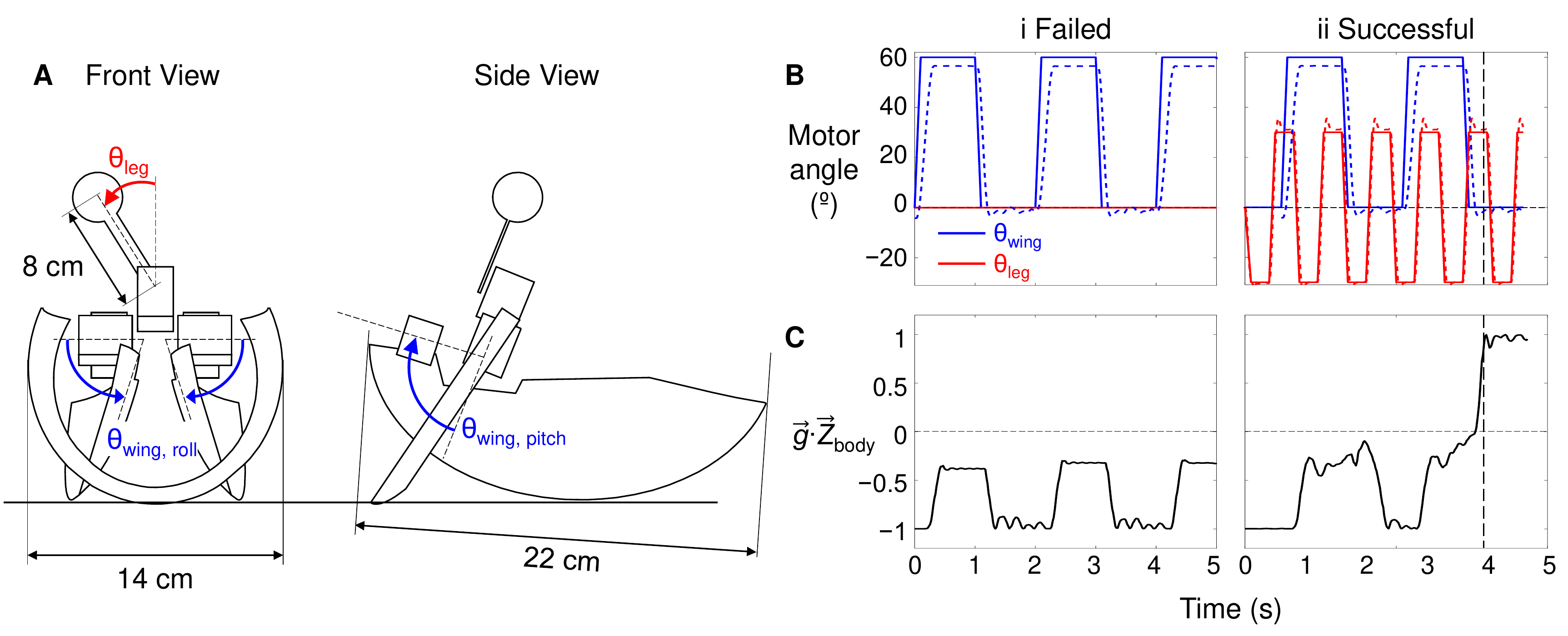}
    \caption[Robot wing and leg actuation and body orientation measurement.]{Robot wing and leg actuation and body orientation measurement. (A) Schematic of leg-assisted, winged self-righting robot from front and side views with geometric dimensions. Front view illustrates wing rolling and leg oscillation and side view illustrates wing pitching. Wing pitching and rolling are by the same angle, synchronized, and together compose wing opening. (B) Motor angles of wings (blue) and leg (red) as a function of time. Solid and dashed curves are commanded and measured motor actuation profiles, respectively. (C) Projection of gravitational acceleration vector "g"onto body Z-axis "Z" body as a function of time, measured using onboard IMU. Vertical dashed line shows the instant when the robot self-righted. In B, C, columns i and ii are for a representative failed and successful trial, respectively.}
    \label{fig:elife_f2_fs1}
\end{figure}

Inspired by these insights and limitations, we hypothesized that the discoid cockroach’s wing opening reduces the barriers to be sufficiently low for small kinetic energy from leg flailing to overcome. This hypothesis predicted that the greater the wing opening and leg flailing are, the more likely self-righting is to occur. We first tested this prediction in the animal, by directly modifying the hind leg inertia  to increase kinetic energy from leg flailing (Figure \ref{fig:elife_f2}A) and studying how it impacted self-righting probability. Then, we developed a robotic physical model (Figure \ref{fig:elife_f2}B) to systematically test the prediction using repeatable experiments over a wide range of wing opening and leg oscillation amplitudes. In addition, we modeled the escape from the metastable state to self-right as a probabilistic barrier-crossing transition on an evolving potential energy landscape of the self-deforming robot/animal, facilitated by kinetic energy. The landscape is the gravitational potential energy of the robot in its body pitch-roll space. Because self-righting could in principle occur via both roll and pitch modes, we analyzed the potential energy barriers on landscape and the kinetic energy from wing opening (primary propulsion) and leg flailing (secondary perturbation) along roll and pitch directions. Considering the effects of wing opening and leg flailing separately gave new insight into the physical mechanism of self-righting. Finally, we examined whether the observed stereotypy of the animal’s body motion can be explained by the potential energy landscape.

We designed and controlled our robotic physical model to achieve similar, strenuous self-righting behavior as the animal’s, where both wing and leg use are crucial (see discussion). The robot consisted of a head, two wings, a leg, and motors to actuate the wings and leg (Figure 2B). To emulate the animal’s wing opening, both robot wings opened by rolling and pitching about the body by the same angle (defined as wing opening amplitude, $\theta_{wing}$; Figures \ref{fig:elife_f2}B, \ref{fig:elife_f2_fs1}). To simplify leg flailing of the animal, the robot used a pendulum leg which oscillated in the coronal plane by the same angle to both sides (defined as leg oscillation amplitude, $\theta_{leg}$; Figures \ref{fig:elife_f2}B, \ref{fig:elife_f2_fs1} ). We opened and closed the robot’s wings (hereafter referred to as an attempt) repeatedly while oscillating its legs to generate repeated attempts observed in the animal. The robot’s leg oscillation was feedforward-controlled, considering that the animal’s leg flailing motion did not correlate with wing opening motion (see Materials \& Methods for detail). Sufficiently large or sufficiently asymmetric wing opening alone guarantees self-righting \citep{li2016a,li2017a}). Here, to study the effect of using both wings and legs under the most strenuous condition, we chose to open both wings symmetrically and only used sufficiently small $\theta_{wing}$ with which the robot did not always self-right with wing opening alone. We emphasize that our goal was not to simply achieve successful self-righting in a robot.

We chose to focus potential energy landscape modeling on the robotic physical model because it offers two advantages. First, the animal’s complex 3-D motion with many degrees of freedom is difficult to quantify. It would take ~540 hours (~12 working weeks) to track our animal dataset (~5 seconds per trial at 200 frames/s, with 3 markers on the body, each wing, and each leg) to quantify 3-D motion required for calculating the potential energy landscape. In addition, wing motion is often impossible to quantify due to occlusion under the body. By contrast, the robot’s simpler mechanical design, controlled actuation, and an onboard inertial measurement unit (IMU) sensor allowed easier reconstruction of its 3-D motion. Second, the animal’s wing opening and leg flailing are highly variable (Xuan and Li, 2020a) and cannot be controlled. This results in the potential energy landscape varying substantially from trial to trial and makes it difficult to evaluate how the system behaved probabilistically on the landscape. By contrast, the robot’s controlled variation of wing opening and leg flailing allowed us to do so. Considering that body rolling is induced by centrifugal force from leg flailing, we compared the ratio of leg centrifugal force to leg gravitational force between the animal and robot and verified they are dynamically similar (see Materials and Methods for detail). In addition, because the animal and robot are geometrically similar, their potential energy barriers also scale as expected (Table 2). Thus, the physical principles discovered for the robot are applicable to the animal.

\clearpage

\section{Methods}
\section*{\textit{Animal experiments}}
\section{Animals}
We used 30 adult male \textit{Blaberus discoidalis} cockroaches (Figure \ref{fig:elife_f2}A) (Pinellas County Reptiles, St Petersburg, FL, USA), as females were often gravid and under different load-bearing conditions. Prior to experiments, we kept the animals in individual plastic containers at room temperature (24 $\degree$C) on a 12h:12h light: dark cycle and provided water and food (rabbit pellets) ad libitum. Animals weighed 2.6 $\pm$ 0.2 g and measured 5.3 $\pm$ 0.2 cm in length, 2.3 $\pm$ 0.1 cm in width, and 0.8 $\pm$ 0.1 cm in thickness. All data are reported as mean $\pm$ s.d. unless otherwise specified.
\subsection{Leg modification}
To study the effect of leg flailing, we directly modified both hind legs of the animal. We attached stainless steel spheres of diameter 0.32 cm and mass 0.14 g (5\% of body weight, 180\% of leg weight (McMaster-Carr, Elmhurst, IL, USA) to the tibia-tarsus joint of both hind legs (Figure \ref{fig:elife_f2}A) using ultraviolet curing glue (BONDIC, Ontario, Canada). We verified that the added mass increased the average kinetic energy during leg flailing (Figure \ref{fig:elife_f3_fs2}, see section ‘Kinetic energy measurement’).

\subsection{Experiment protocol}
We used a flat, wooden surface (60 cm $\times$ 60 cm) covered with cardstock and walled with transparent acrylic sheets as the righting arena. Four 500 W work lights (Coleman Cable, Waukegan, IL, USA) illuminated the arena for high-speed imaging. We maintained the arena at an ambient temperature of $40 \pm 2 $C during experiment. We used two synchronized cameras (Fastec IL5, Fastec Imaging, San Diego, CA, USA) at 200 frames s$^{-1}$ and 200 $\mu$s shutter time to record the self-righting maneuver from top (1200 $\times$ 1080 pixels) and side (1200 $\times$ 400 pixels) views, with a small lens aperture to maximize the focal depth of field.

For each trial, we first started video recording, held the animal upside-down by its pronotum, and gently released it from a height of $\approx$ 1 cm above the center of the righting arena. The small drop was to ensure that the animal did not begin leg searching, a common strategy used to self-right \citep{camhi1977a}, before it was released. The animal was given 10 seconds to attempt to self-right during each trial. After it self-righted or 10 seconds elapsed, the animal was picked up, and video recording was stopped. After each trial, we returned the animal to its container and continued testing a different animal. This way, each animal was allowed to rest for $\approx$ 30 minutes before its next trial to minimize the effects of fatigue \citep{camhi1977a}.

We tested 30 animals, each with five trials with its hind legs intact and then modified, resulting in a total of 300 accepted trials (\textit{N} = 30 animals, \textit{n} = 150 trials for each leg treatment). We excluded trials in which the animal collided with the walls of the righting arena or moved out of both camera views.

\subsection{Self-righting performance} 
For each animal trial, we watched the videos to determine whether the animal self-righted. Because the animal did not always immediately begin to self-right when placed on the arena \citep{camhi1977a,li2019a}, we defined the beginning of the self-righting attempt as the instant when the animal began moving its body or appendages to self-right. We defined the animal to have successfully self-righted if it attained an upright orientation with all six legs on the ground within 10 s of starting its attempt. We identified the trials in which animal succeeded in self-righting using the leg-assisted, winged strategy. For each animal and each leg treatment, we defined and measured self-righting probability as the number of trials that self-righted using winged attempts divided by the total number of trials. We counted the trials that used the legged strategy as failed. We then calculated average self-righting probability for each leg treatment by averaging across all animals.

\subsection{Preference of self-righting strategies}
We verified that the animal’s preference of winged and legged self-righting strategies (see Footnote 1) did not change with leg modification. To compare the animal’s preference of winged and legged self-righting strategies before and after leg modification, for each trial, we examined the videos to identify winged and legged self-righting attempts and measured the percentage of time spent on each strategy. Then, for each leg treatment and each animal, we averaged it across all the trials from that animal. For each treatment, we then averaged across each animal to calculate the average percentage of time spent on each strategy (Figure \ref{fig:elife_f3_fs1}).

\subsection{Kinetic energy measurement}
To measure the animal’s pitch and roll kinetic energy during self-righting, in a separate experiment, we used three high speed cameras (Photron FASTCAM Mini UX-100) to record the animal self-righting at 2000 frames $s^{-1}$ and a resolution of 1280 $\times$ 1024 pixels, first with its hind legs intact (\textit{N} = 2 animals, \textit{n} = 2 trials) and then modified (\textit{N} = 2 animals, \textit{n} = 2 trials).

We used DeepLabCut (Mathis et al., 2018) to track the tip and femur-tibia joint of both hind legs, head anterior tip, abdomen posterior tip, and body midpoint (Figure \ref{fig:elife_f3_fs2}A, B). We then used Direct Linear Transformation software DLTdv5 (Hedrick, 2008b) to reconstruct 3-D motion of the tracked points and used a sixth order Butterworth filter with a cut-off frequency of 25 Hz to filter their 3D positions.
To calculate kinetic energy, we approximated the animal body as an ellipsoid cut into two parts at 38\% of total length from the anterior end, connected by a hinge joint (thorax-abdomen joint, Figure \ref{fig:elife_f3_fs2}A). The smaller part represented the animal’s head and thorax, and the larger part represented its abdomen. We assumed uniform mass distribution for both parts. We used the geometric center of the body parts when their fore-aft axes are aligned to approximate body center of mass . For both hind legs, we approximated the coxa-femur and tibia-tarsus segment as rigid rods. One end of the rod representing coxa-femur segment was connected to the body at the midpoint of thorax-abdomen joint, and the other end connected to the rod representing tibia-tarsus segment, both via spherical joints (Figure \ref{fig:elife_f3_fs2}A, B, thick black lines connected by blue dots). For modified hind legs, we approximated the stainless steel spheres at the leg tip as a point mass attached to the free end of the tibia-tarsus rod (Figure \ref{fig:elife_f3_fs2}A, B).
We defined pitch and roll kinetic energy as the sum of kinetic energy from translational and rotational velocity components from all body parts that contribute to pitching and rolling motion, respectively. We obtained pitch and roll kinetic energy by summing contributions from the body ellipsoid parts and the hind leg segments. For each part, we measured its rotational velocity components about the animal’s body fore-aft ($X_{body}$) and lateral ($Y_{body}$) principal axes, and we measured the translational velocity components of its center of mass along the fore-aft and lateral directions (Figure \ref{fig:elife_f3_fs2}B, red vs. blue arrows). For the sphere attached to modified leg, we measured its translational velocities. Because vertical translational velocity and yaw angular velocity did not contribute to motion along the pitch or roll direction, we did not consider them. 
For each of the ellipsoid parts and rigid rods, we calculated its pitch and roll kinetic energy as follows:
\begin{equation}
    KE_{pitch,j} = \frac{1}{2}I_{yy,j}\omega^{2}_{y,j} + \frac{1}{2}m_{j}v^{2}_{x,j}
\end{equation}
\begin{equation}
    KE_{roll,j} = \frac{1}{2}I_{xx,j}\omega^{2}_{x,j} + \frac{1}{2}m_{j}v^{2}_{y,j}
\end{equation}
where $I_{xx,j}$ and $I_{yy,j}$ are the moments of inertia the $j^{th}$ object measured about the animal’s body fore-aft ($X_{body}$) and lateral ($Y_{body}$) principal axes, respectively, $m_j$ is the mass of $j^{th}$ object, $\omega_{x,j}$ and $\omega_{y,j}$ are the rotational velocities of the $j^{th}$ object about body fore-aft and lateral principal axes, and $v_{x, j}$ and $v{y, j}$ are the translational velocity of the center of mass of the $j^{th}$ object along fore-aft and lateral directions, respectively (Figure \ref{fig:elife_f3_fs2}B). For both hind leg segments, we used the mass reported in  (\citep{kram1997a}) (0.07 g for coxa-femur segments and 0.01 g for tibia-tarsus segment). To calculate the mass of the two body parts, we assumed body density to be  uniform.

We calculated the pitch and roll kinetic energy of the added spherical mass as follows:
\begin{equation}
    KE_{pitch,sphere} = \frac{1}{2}m_{sphere}v^{2}_{x,sphere}
\end{equation}
\begin{equation}
    KE_{roll,sphere} = \frac{1}{2}m_{sphere}v^{2}_{y,sphere}
\end{equation}
where $m_{sphere}$ is the added spherical mass, and $v_x$, sphere and $v_y$, sphere are the translational velocity components of the sphere along fore-aft and lateral directions, respectively. We considered kinetic energy from the added spherical mass only for animal with modified legs. 

We obtained the pitch and roll kinetic energy of the intact animal from equations (1) and (2) respectively. For the modified animal, we added equations (1) and (3) to obtain total pitch kinetic energy and added (2) and (4) to obtain total roll kinetic energy. For each trial, we first averaged the measured kinetic energy along pitch and roll directions over the recorded interval (2.5 s) for each trial. Then for each leg treatment, we further averaged it across all the trials of that treatment (intact: \textit{N} = 2 animals, \textit{n} = 2 trials; modified: \textit{N} = 2 animals, \textit{n} = 2 trials).
    
\subsection{Relationship between wing opening and leg flailing}
We examined whether the animal’s leg flailing during self-righting was more feedforward-driven or more towards a feedback-controlled reflex coordinated with wing opening. To do so, we measured the correlation between wing opening and leg flailing motions as well as their self-correlations (Figure \ref{fig:elife_f3_fs3}). Because wing opening was difficult to measure due to occlusion of wings by the body during self-righting, we used abdomen tip height as a proxy for wing opening, considering that abdomen tip height typically increased as wings opened. For each hind leg, we used its leg tip height as a proxy of the flailing motion (Figure \ref{fig:elife_f3_fs2}). To check whether the height of abdomen tip and hind leg tips were correlated to each other and to themselves, we measured the normalized cross-correlations between each pair of these variables and the normalized autocorrelation of each of them (Figure \ref{fig:elife_f3_fs3}). Normalized cross correlation h between two signals f (t) and g(t) is defined as
\begin{equation}
    h(t) = \frac{\int_{-\infty}^{\infty}f^{*}(\tau-t)g(\tau)d\tau}{\sqrt{\int_{-\infty}^{\infty}|f(\tau)|^{2}d\tau.\int_{-\infty}^{\infty}|g(\tau)|^{2}d\tau}}
\end{equation}
where $t$ is the time lag between $f(t)$ and $g(t)$ and is a variable, $\tau$ is the variable of integration, and $f*(t)$ is the complex conjugate of $f(t)$. When $f(t) = g(t)$, $h(t)$ is the normalized autocorrelation.

All normalized cross correlations plots lacked a prominent peak whose value was close to 1, and all normalized autocorrelations plots had a prominent peak only at zero lag. This showed that abdomen tip height did not correlate with itself or with either of the two hind leg tips heights (Figure \ref{fig:elife_f3_fs3}A, B, F). This meant that wing opening and leg flailing motions were not correlated to each other during self-righting. However, the normalized cross correlation between both hind legs had recurring oscillations as the lag increased in magnitude (Figure \ref{fig:elife_f3_fs3}). This suggested that leg flailing had some rhythm, despite a large temporal variation and difference between the two hind legs (\citep{delcomyn1987a}; \citep{sherman1977a}; \citep{zill1986a}). 

\section*{Robotic physical modelling}
\subsection{Design and actuation}
The robot consisted of a head, two wings, a leg, and four motors to actuate the wings and one to actuate the leg (Table 1, Figures \ref{fig:elife_f2}B, Figure \ref{fig:elife_f2_fs1}). The head and wings were cut from two halves of a thin ellipsoidal shell thermo-formed (Formech 508FS, Middleton, WI, USA) from 0.16 cm thick polystyrene plastic sheet (McMaster-Carr, Elmhurst, IL, USA). We connected different parts using joints 3-D printed using PLA (Ultimaker 2+, Geldermalsen, Netherlands) (Figure \ref{fig:elife_f2}B). We used DC servo motors (Dynamixel XL-320, ROBOTIS, Lake Forest, CA, USA) to actuate both the wings and the leg.

To measure the robot’s 3-D orientation (roll, pitch, and yaw angles), we attached an inertial measurement unit (IMU) (BNO055, Adafruit, New York, NY, USA) near its center of mass determined from the robot CAD model. We used the Robot Operating System (Version: melodic) (\citep{quigley2009a}) to send actuation signals for the wing and leg motors and record IMU data. To ensure a constant voltage for repeatable experiments, we used an external 8 V voltage source (TP3005DM, TEK Power, Montclair, CA, USA) to power the robot. We used fine flexible wires (30 AWG, 330-DFV, Vishay Sensor, USA) for powering robot and sending/acquiring signals and ensured that they were loose and did not interfere with robot motion.

\begin{table}[]
\caption{\label{tab:elife_table1} Mass distribution of the robot.}
\centering
\begin{tabular}{|l|l|}

\hline
\textbf{Component}    & \textbf{Mass (g)} \\ \hline
Head                  & 13.4              \\ \hline
Leg rod               & 4.3               \\ \hline
Leg added mass        & 51.5              \\ \hline
Leg motor             & 28.6              \\ \hline
Two wings             & 57.4              \\ \hline
Two wing pitch motors & 56.0              \\ \hline
Two wing roll motors  & 48.8              \\ \hline
Total                 & 260.0             \\ \hline
\end{tabular}
\label{elife_table1}

\end{table}

\subsection{Similarity to animal}
To examine whether the robotic physical model was similar to the animal and reasonably approximated its self-righting motion, we examined how well they were geometrically similar and their leg flailing motions were dynamically similar. To evaluate geometric similarity, we compared their dimensions. For geometrically similar objects, length $l$ should scale with mass $m$ and density $\rho$ as $l \propto (m/\rho)^{1/3}$ (Alexander, 2006). Following this, potential energy should scale as $E \propto m\cdot(m/\rho)^{1/3} \propto m^{4/3}\rho^{-1/3}$.The robot, which was 90 times as much heavy and 2.3 times as much dense as the animal with modified legs (Table 2) was expected have dimensions $(90/2.63)^{1/3} = 3.4$ times those of the animal. For the animal, $m$ includes the added mass from leg modification because we used the same for calculating the potential energy landscape. Because gravitational potential energy is proportional to mass and center of mass height, the potential energy barriers should scale by a factor of $90^{4/3}{\times}2.63^{-1/3} = 305$ (Table 2). We found that the robot’s length, thickness, and pitch potential energy barriers scaled up roughly as expected (Table 2). The larger scaling factor for robot’s width and roll potential energy barrier is due to the robot being designed wider to make self-righting via rolling more strenuous.

To evaluate dynamic similarity between the robot and animal, we calculated Froude Number for their leg flailing. Here, we used the following definition of Froude number \citep{biewener2003a}:

\begin{equation}
Fr = \frac{\text{Inertial force from leg flailing}}{\text{Gravitational force of leg}} = \frac{mv^{2}/r}{mg} = \frac{v^2}{rg}    
\end{equation}
where $m$ is the mass of the animal or robot leg(s), plus the added mass attached it for the modified animal, $v$ is the leg translational velocity along the body lateral principal axis, $g$ is gravitational acceleration, and $r$ is leg length.

\begin{table}[]
\caption{\label{tab:elife_table2} Comparison between animal and robot.}
\begin{tabular}{c c c c c}
\multicolumn{2}{l}{\textbf{Parameter}}                               & \textbf{Animal} & \textbf{Robot}        & \textbf{Ratio} \\
\multicolumn{2}{l}{Body length 2a (mm)}                              & 53              & 260                   & 4.9            \\
\multicolumn{2}{l}{Body width 2b (mm)}                               & 23              & 220                   & 9.6            \\
\multicolumn{2}{l}{Body thickness 2c (mm)}                           & 8               & 43                    & 5.4            \\
\multicolumn{2}{l}{Mass attached to leg (g)}                         & 0.14            & 51.5                  & 368            \\
\multicolumn{2}{l}{Total mass m* (g)}                                & 2.84            & 260                   & 90             \\
\multicolumn{2}{l}{Density $\rho$ ($\times 10^{-3} g \cdot mm^{-3}$)}                         & 0.88            & 2.05                  & 2.3            \\
\multicolumn{2}{l}{Expected length scale factor $(m/\rho)^{1/3}$}            & 1.47            & 5.06                  & 3.4            \\
\multicolumn{2}{l}{Expected potential energy scale factor $m^{4/3}/\rho^{1/3}$} & 4.28            & 1306                  & 305            \\
\multicolumn{2}{l}{Maximum pitch potential energy barrier (mJ)}      & 0.58            & 282                   & 486            \\
\multicolumn{2}{l}{Maximum roll potential energy barrier (mJ)}       & 0.19            & 244                   & 1284           \\
\multirow{2}{*}{Froude number for leg flailing  Fr}  & Intact legs   & 0.37   &\multirow{2}{*}{0.78} & 2.1\\
                                                     & Modified legs & 1.27            &                       & 0.61    \\      
\end{tabular}
\label{elife_table2}
 * Includes mass attached to the legs.
\end{table}
 	     
We found that the Froude numbers for the robot and both the intact and modified animals were similar (within a factor of two). This dynamic similarity demonstrated that the robot provided a good physical model for studying the animal’s self-righting.      
\subsection{Experiment protocol}
For robot experiments, we used a level, flat, rigid wooden surface (60 cm $\times$ 60 cm) covered with sandpaper as the righting arena. We used two synchronized webcams (Logitech C920, Logitech, Newark, CA, USA) to record the experiment from top and side views at 30 frames s-1 and a resolution of 960 $\times$ 720 pixels. Using the onboard IMU, we recorded the robot body orientation relative to the lab coordinate system ($X$-$Y$-$Z$ in Figure \ref{fig:elife_f2}B) at $\approx$ 56 Hz and synchronized them with the motor actuation timings angles (Figure \ref{fig:elife_f2_fs1}, bottom right).

Before each trial, we placed the robot upside-down (Figure \ref{fig:elife_f5}Ai) on the arena, with its wings closed and leg aligned with the body midline and started video recording. We then actuated the robot to repeatedly open and close its wings at 2 Hz and oscillate its legs at 2.5 Hz to self-right. Because the animal was likely to move its leg before wings at the start of self-righting (59\% of intact leg trials and 81\% of modified leg trials), for non-zero robot leg oscillation amplitudes, the first wing opening was started after completing one cycle of leg oscillation (0.4 s). If the robot did not self-right after five wing opening attempts (10 s), we powered down the robot, stopped video recording, and reset the robot for the next trial. We tested self-righting performance of the robot by systematically varying leg oscillation amplitude $\theta_{leg}$ (0$\degree$, 15$\degree$, 30$\degree$, 45$\degree$) and wing opening amplitude $\theta_{wing}$ (60$\degree$, 72, 83$\degree$). We collected five trials for each combination of $\theta_{wing}$ and $\theta_{leg}$. This resulted in a resulted in a total of 60 trials with 134 attempts.

To reconstruct the robot’s 3-D motion, in a separate experiment, we characterized how the wing and leg actuation angles changed over time during an attempt (Figure \ref{fig:elife_f2_fs1}). We attached BEETag markers (Crall et al., 2015) to the body frame and to each link actuated by the motors and tracked their positions using two calibrated high speed cameras (Fastec IL5, Fastec Imaging, San Diego, CA, USA) at 500 frame $s{-1}$ and a resolution of 1080 $\times$ 1080 pixels, as the robot actuated its wings and legs to self-right. We obtained 3-D kinematics of the markers using the Direct Linear Transformation method DLTdv5 (Hedrick, 2008b). We then measured the rotation of the link actuated by each motor about its rotation axis as a function of time during an attempt. Because the wings were controlled to roll and pitch by the same angle, we used the average measured wing actuation profile of all the four motors (two for wing pitching and two for wing rolling). The actual wing opening and leg oscillation angles were smaller than the commanded (solid blue and red) due to the inertia of robot body components attached to each motor.

\subsection{Self-righting performance}
We defined the beginning of the righting attempt as the instant when the robot first started opening its wings and measured this instance from the commanded motor actuation profile (Figure \ref{fig:elife_f2_fs1}Bii). We defined the robot to have successfully self-righted if it attained an upright orientation within 10 seconds (five attempts). We used the IMU to measure the projection of the gravity acceleration vector "g" onto the body Z-axis "Z" body as a function of time. This allowed us to determine when the robot became upright. We then counted the number of successful and failed attempts for each trial. For each trial, we defined self-righting probability as the ratio of the number of successful attempts to the total number of attempts of that trial. At each wing opening and leg oscillation amplitude, we then averaged it across all trials of that treatment to obtain its average self-righting probability. Among all the 134 attempts observed across all 60 trials, 44 attempts succeeded (12, 15, and 17 attempts at $\theta_{wing}$ = 60$\degree$, 72$\degree$, and 83$\degree$, respectively), and 90 attempts failed (46, 27, and 17 attempts at $\theta_{wing}$ = 60$\degree$, 72$\degree$, and 83$\degree$, respectively).

\subsection{Robot 3-D motion reconstruction} 
For each robot trial, we measured the robot’s 3-D orientation in the lab frame using Euler angles (yaw $\alpha$, pitch $\beta$, and roll $\gamma$, Z-Y’-X” Tait-Bryan convention). We divided each trial temporally into 0.01 s intervals and used the measured motor actuation angles and body 3-D  orientation (Figure \ref{fig:elife_f2_fs1} B, C) at each interval to reconstruct the robot’s body shape and 3-D orientation, respectively. Because the IMU measured only the 3-D orientation of the robot, we constrained the robot’s center of mass to translate only along the vertical direction (Figure \ref{fig:elife_f2}B, Z-axis of lab frame) while maintaining contact with the ground. We then used the reconstructed 3-D motion of the robot to obtain the translational and rotational velocity components of all robot parts.

\subsection{Kinetic energy measurements}
For each robot trial, we measured pitch and roll kinetic energy for all attempts. We defined pitch and roll kinetic energy as the kinetic energy of the entire robot due to translational and rotational velocities along body fore-aft and lateral directions, respectively. Because vertical translation and yawing do not contribute to body pitching or rolling towards self-righting, we did not consider vertical velocities or rotational velocities about the vertical axis.
Considering that the five motors, leg, and mass added to the leg could be approximated as regular, symmetric shapes with uniform mass distribution (motors and leg as solid cuboids and added mass as a solid sphere), the moment of inertia at the center of mass of each part could be directly calculated. Then, we calculated the total pitch and roll kinetic energy of the motors and leg with added mass as:
\begin{equation}
    KE_{pitch} = \sum_{j=1}^k(\frac{1}{2}I_{yy,j}\omega^{2}_{y,j} + \frac{1}{2}m_{j}v^{2}_{x,j})
\end{equation}
\begin{equation}
    KE_{roll} = \sum_{j=1}^k(\frac{1}{2}I_{xx,j}\omega^{2}_{x,j} + \frac{1}{2}m_{j}v^{2}_{y,j})
\end{equation}
where $j$ enumerates the five motors, leg, and mass added to the leg, $I_{xx,j}$ and $I_{yy,j}$ are the moments of inertia of object $j$ about the body fore-aft and lateral principal axes (measured at the part’s center of mass), $m_j$ is the mass of object $j$, and $v_{x,j}$ and $v_{y,j}$ are translational velocities of object j along fore-aft and lateral directions of robot, and $\omega_x$ and $\omega_y$ are rotational velocities of object $j$ about fore-aft and lateral directions of the robot, respectively. 
For both the wings and head with complex shapes, we imported their CAD model and approximated them with uniformly distributed point mass clouds and calculated the pitch and roll kinetic energy of each part as:
\begin{equation}
    KE_{pitch,cloud} = \frac{m}{2k}\sum_{j=1}^kv^{2}_{x,j}
\end{equation}
\begin{equation}
    KE_{roll,cloud} = \frac{m}{2k}\sum_{j=1}^kv^{2}_{y,j}
\end{equation}
where $m$ is the total mass of the wing or head, $k$ is the number of point masses in the point cloud, and $v_{x,i}$ and $v_{y,i}$ are the velocity components of the $i_{th}$ point mass along the body fore-aft and lateral principal axes. 

To obtain total pitch and roll kinetic energy, we summed the pitch and roll kinetic energy of all the parts. To compare pitch and roll kinetic energy at each combination of wing opening and leg oscillation amplitudes, we first averaged the total pitch and roll kinetic energy respectively over the phase when wings were fully open in the first attempt of each trial to avoid bias from the large rolling kinetic energy during successful self-righting in later attempts. We then averaged these temporal averages across the five trials at each combination of wing opening and leg oscillation amplitudes (Figure \ref{fig:elife_f4}A, B).

\clearpage
\section{Potential energy landscape modelling}
\subsection{Model definition}
The gravitational potential energy of the animal or robot is: 
\begin{equation}
  E = mgz_{CoM}  
\end{equation}
where $m$ is the total mass of the animal or robot, $g$ is gravitational acceleration, $z_{CoM}$ is center of mass height from the ground. To determine the robot’s center of mass, we used a CAD model of the robot (Figures \ref{fig:elife_f2}A, Figure \ref{fig:elife_f2_fs1}) and measured the 3-D positions and orientations of all robot body parts for a given body orientation and wing opening (see consideration of leg oscillation below). We approximated the animal body as a rigid ellipsoid, with the animal’s center of mass at the body geometric center, and its wings as slices of an ellipsoidal shell. Because the animal or robot did not lift off during self-righting, in the model we constrained the lowest point of the animal or robot to be always in contact with the ground.
The potential energy depended on body pitch and roll, wing opening angle, and leg oscillation angle. Because the effect of leg oscillation was modelled as a part of kinetic energy, for simplicity, we set the leg to be held fixed in the middle when calculating the potential energy landscape. We verified that potential energy landscape did not change considerably (roll barrier changed only up to 13\%) when the leg moved. Because we used Euler angles for 3-D rotations, change in body yaw did not affect center of mass height. Because the robot’s initial wing opening was negative (-6$\degree$) due to body weight, in our model calculations, we varied wing opening angle within the range [-10$\degree$, 90$\degree$] with a 0.5$\degree$ increment. For each wing opening angle, we then varied both body pitch and roll within the range  [-180$\degree$, 180$\degree$] with a 1$\degree$ increment and calculated zCoM to obtain the system potential energy (Figure \ref{fig:elife_f2_fs1}). Because the animal or robot did not pitch backward significantly, in the figures we do not show landscape for body pitch < -90$\degree$; the full landscape maybe visualized using data and code provided (Othayoth and Li, 2021a).

\subsection{System state trajectories on potential energy landscape}
To visualize how the robot’s measured system state behaved on the landscape, we first discretized each righting attempt into time intervals of 0.01 s. For each interval, we used the measured the wing opening angle (Figure \ref{fig:elife_f2_fs1}, dashed blue curves) to calculate the potential energy landscape. We then projected the measured body pitch and roll onto the landscape to obtain the system state trajectory over time. Note that only the end point of the trajectory, which represented the current state, showed the actual potential energy of the system at the corresponding wing opening angle. The rest of the visualized trajectory showed how body pitch and roll evolved but, for visualization purpose, was simply projected on the landscape surface. The exact system state trajectories are shown in Figure \ref{fig:elife_f6}.

\subsection{Potential energy barrier measurements}
We measured the potential energy barrier that must be overcome to escape from metastable basin to transition to an upright basin (Figure \ref{fig:elife_f5}C, \ref{fig:elife_f7}). For each wing opening angle (Figure \ref{fig:elife_f7}B, dashed blue), at each time interval, we considered imaginary straight paths away from the metastable local minimum (Figure \ref{fig:elife_f5}B, white dot) in the body pitch-roll space, parameterized by the polar angle $\psi$ from the positive pitch direction (body pitching up, Figure \ref{fig:elife_f5}Ci). Along each path, we obtained a cross section of the landscape. Then, we defined and measured the potential energy barrier along this path as the maximal increase in potential energy in this cross section. Finally, we plotted the potential energy barrier as a function of $\psi$ (Figure \ref{fig:elife_f5}C). We defined the roll barrier as the lowest potential energy barrier within $\psi$ = $\pm$ [45$\degree$, 135$\degree$], because both roll upright minima always lay in this angular range. We defined the pitch barrier as the potential energy barrier at $\psi$ = 0$\degree$ towards the pitch local minimum. Finally, we measured both pitch and roll barriers as a function of wing opening angle (Figure \ref{fig:elife_f7}, Figure \ref{fig:elife_f7_fs1}).

\subsection{Comparison of kinetic energy and potential energy barriers}
To understand how wing opening and leg oscillation together contribute to the robot’s self-righting, we compared the measured kinetic energy and potential energy barriers along both pitch and roll directions throughout each attempt. For each attempt, we measured kinetic energy minus potential energy barrier over time along both pitch and roll directions (Figure \ref{fig:elife_f7_fs2}, \ref{fig:elife_f7_fs3} A-C). We then examined whether there was a surplus or deficit of kinetic energy to overcome the potential energy barrier in both pitch and roll directions, comparing between successful and failed attempts (Figure Figure \ref{fig:elife_f7_fs2}D, \ref{fig:elife_f7_fs3}D). To examine how maximal surplus varied with wing opening and leg oscillation amplitudes, for each combination of the two, we recorded the maximal surplus when the wings are held fully open in each attempt and averaged it across all attempts (Figure \ref{fig:elife_f7_fs2}E vs. \ref{fig:elife_f7_fs3}E).
	
\subsection{Data analysis and statistics}
We tested whether the animal’s percentage of time spent on winged and legged self-righting attempts and self-righting probability changed with leg modification using a mixed-effects ANOVA, with leg treatment as the fixed factor and individual as a random factor to account for individual variability. We tested whether the animal’s pitch and roll kinetic energy depended on leg modification using ANOVA with leg treatment a fixed-factor. We tested whether the animal’s self-righting probability depended on leg treatment using a mixed-effect ANOVA with leg treatment as a fixed factor and individual as a random factor.

We tested whether the robot’s self-righting probability, number of attempts required to self-right, pitch and roll kinetic energy depended on leg oscillation amplitude at each wing opening amplitude using a chi-squared test for probability and an ANOVA for the rest, with wing opening magnitude as a fixed factor. We tested whether kinetic energy minus potential energy barrier along the pitch and roll directions depended on leg oscillation amplitude at each wing opening amplitude, using ANOVAs with leg oscillation amplitude as the fixed factor. We also tested whether they depended on wing opening amplitude at each leg oscillation amplitude, using ANOVAs with wing opening amplitude as the fixed factor. To test whether kinetic energy minus potential energy barrier differed between successful and failed attempts, we used an ANOVA with the attempt outcome (success or failure) as the fixed factor. All statistical tests were performed using JMP Pro 14 (SAS Institute Inc., NC, USA).

\clearpage
\section{Results}
\subsection{Leg flailing facilitates animal winged self-righting}
\begin{figure}[t]
   \centering
   \includegraphics[width=1.0\linewidth]{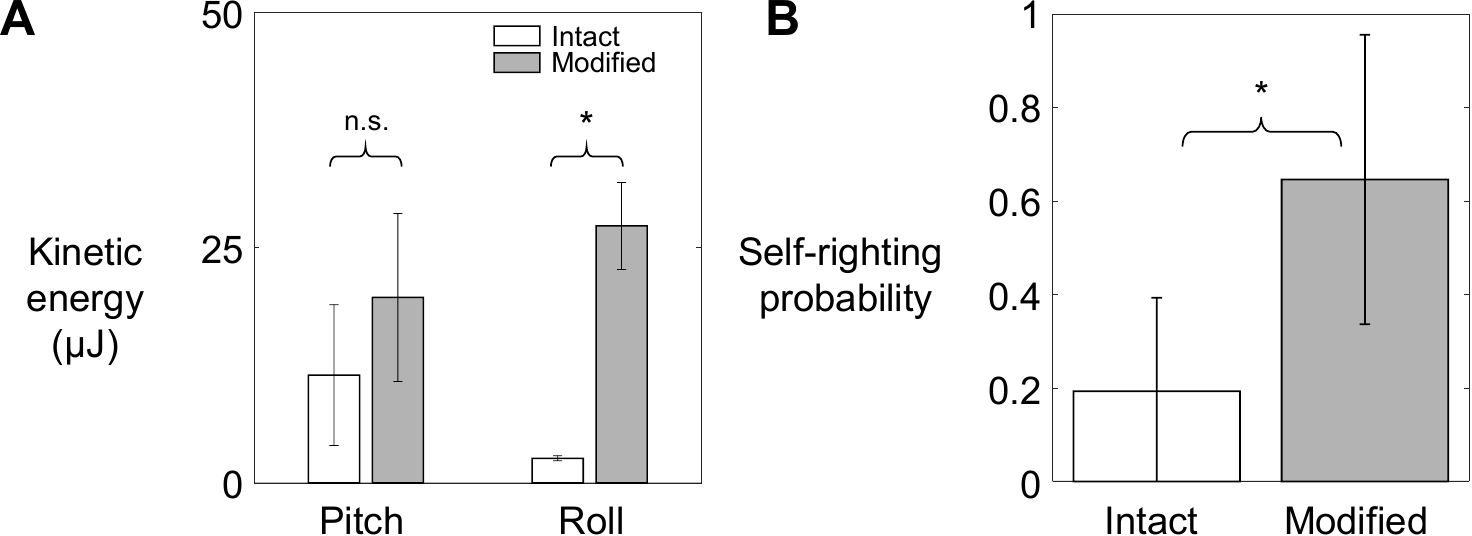}
   \caption[Animal’s kinetic energy and self-righting probability] %
   {Animal’s kinetic energy and self-righting probability. Comparison of (A) average pitch and roll kinetic energy and (B) self-righting probability between intact animals and animals with modified hind legs. Error bars in show $\pm$ s.d. Asterisk indicates a significant difference (P < 0.05) and n.s. indicates none. Statistical tests: Pitch kinetic energy: P = 0.34, F1, 1 = 1.53, ANOVA. Roll kinetic energy: P = 0.02, F1, 1 = 50.35, ANOVA. Probability: P < 0.0001, F1, 29 = 93.38, mixed-effect ANOVA. Sample size: (A) N = 2 animals, n = 2 trials. (B) Intact: N = 30 animals, n = 150 trials. Modified: N = 30 animals, n = 150 trials.}
   \label{fig:elife_f3}
\end{figure}

\begin{figure}[t]
   \centering
   \includegraphics[width=0.7\linewidth]{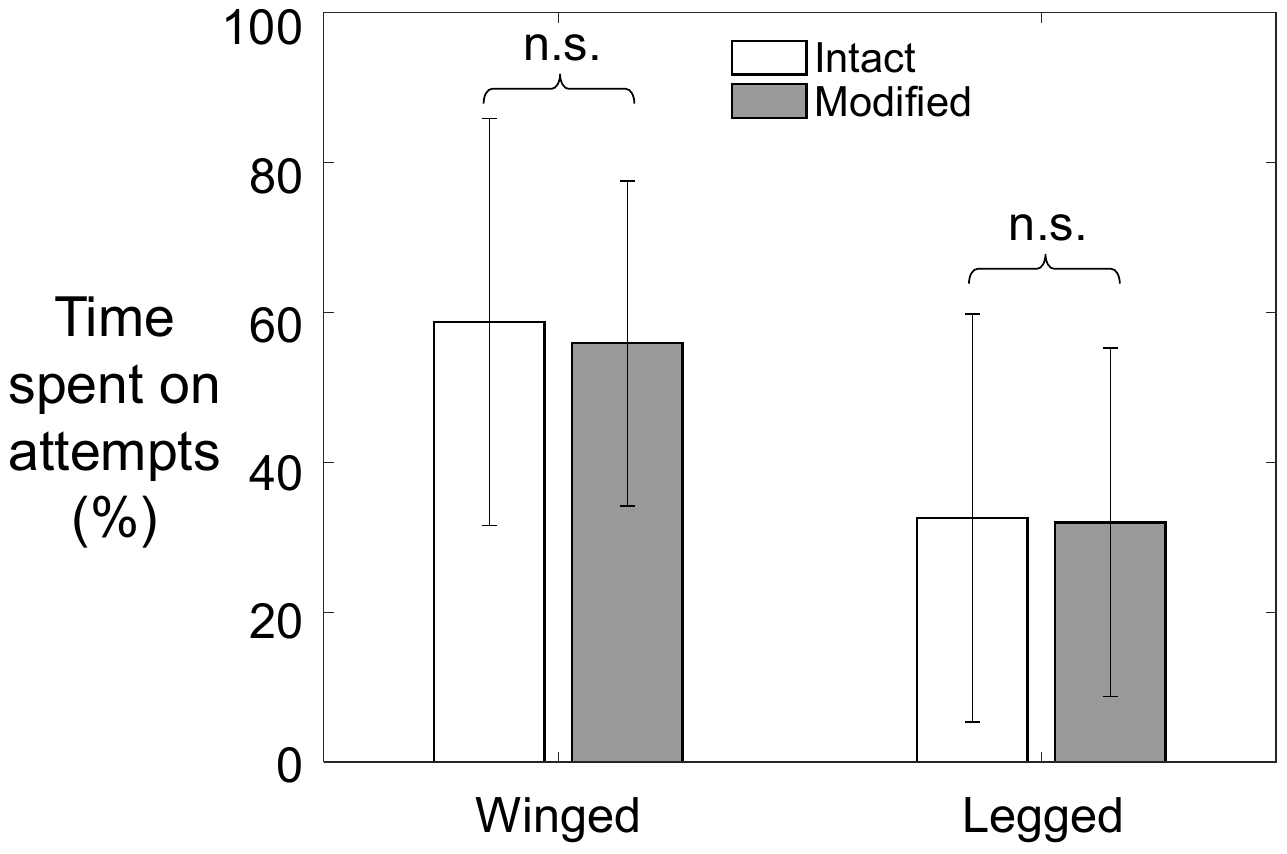}
   \caption[Comparison of average percentage of time spent on winged and legged self-righting attempts between animals with intact and modified legs] %
   {Comparison of average percentage of time spent on winged and legged\footnote{ADDFOOTNOTE} self-righting attempts between animals with intact and modified legs. Error bars show ± s.d. n.s. indicates no significant difference. Winged: P = 0.19, F1,269 =1.71; legged: P = 0.78, F1,269 = 0.07 mixed-effect ANOVA. Sample Size: N = 30 animals, n = 150 trials for each treatment. }
   \label{fig:elife_f3_fs1}
\end{figure}

\begin{figure}[h]
   \centering
   \includegraphics[width=1.0\linewidth]{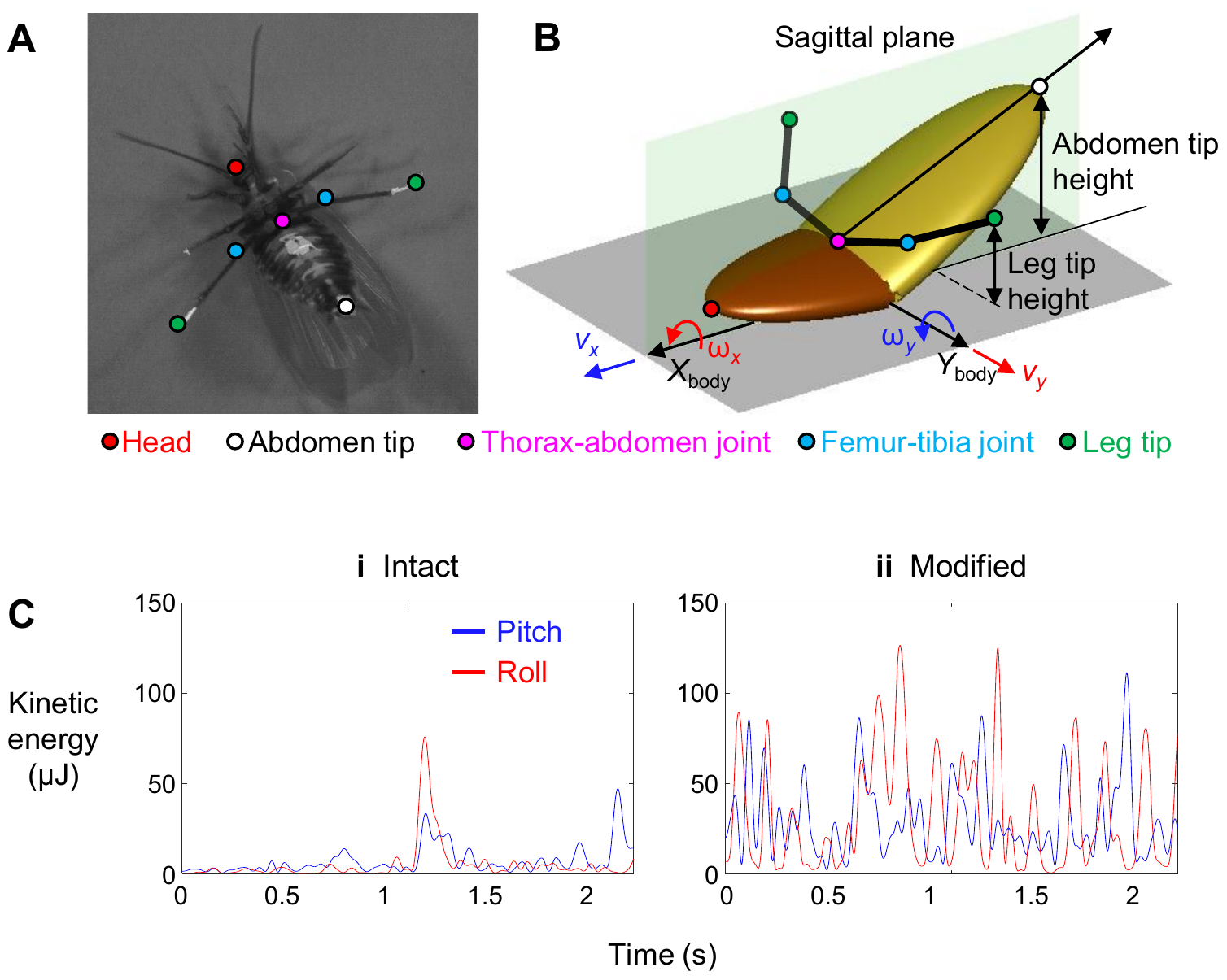}
   \caption[Animal kinetic energy calculation] %
   {Animal kinetic energy calculation. (A) Representative snapshot of body and appendage with definition of markers tracked. (B) Multi-body model of animal for calculating pitch and roll kinetic energy. Red and blue arrows show velocity components that contribute to pitch and roll kinetic energy, respectively. (C) Pitch and roll kinetic energy as a function of time for animal with (i) intact and (ii) modified hind legs from a representative trial.}
   \label{fig:elife_f3_fs2}
\end{figure}

\begin{figure}[h]
   \centering
   \includegraphics[width=1.0\linewidth]{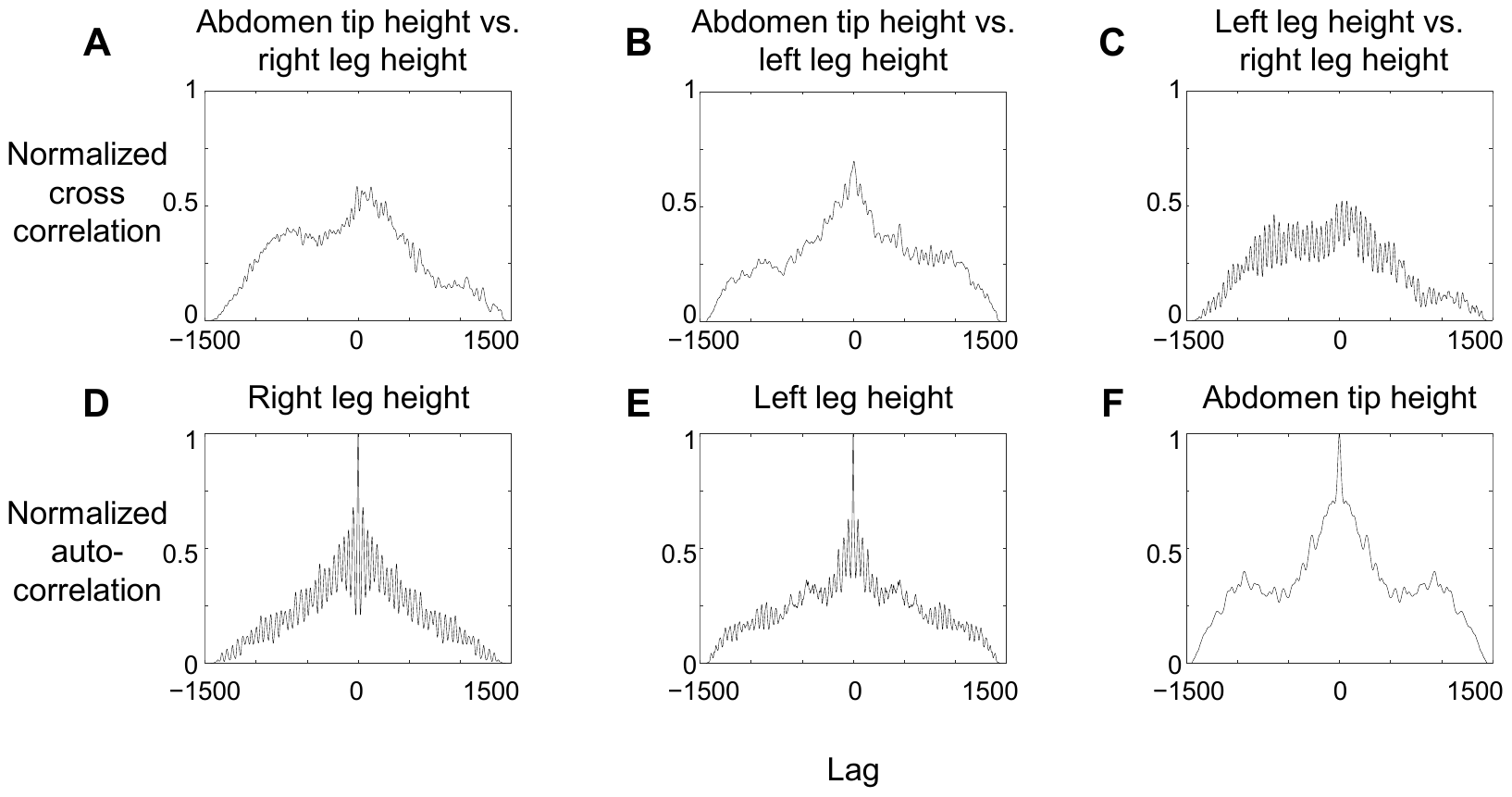}
   \caption[Correlation between animal’s body and leg motion] %
   {Correlation between animal’s body and leg motion. (A-C) Pair-wise normalized cross correlations between left hind leg tip height, right hind leg tip height, and abdomen tip height, as a function of lag between each pair of variables. (D-F) Normalized autocorrelation of left hind leg tip height, right hind leg tip height, and abdomen tip height as a function of lag between a variable and itself. N = 1 animal, n = 1 trial.}
   \label{fig:elife_f3_fs3}
\end{figure}

As leg modification increased the animal’s average kinetic energy in both pitch and roll directions (by 2 and 10 times, respectively; Figures \ref{fig:elife_f3}A, \ref{fig:elife_f3_fs2}; \textit{P} < 0.05, ANOVA), its probability of self-righting using wings increased (Figure \ref{fig:elife_f3}B; \textit{P} < 0.0001, mixed-effect ANOVA). These observations supported our hypothesis. Leg modification did not change the animal’s relative preference of using winged and legged self-righting strategies (Figure \ref{fig:elife_f3_fs1}). In addition, wing opening and leg flailing did not show temporal correlation. Furthermore, the approximate time period of leg flailing (100 ms) was comparable to combined sensory feedback (6-40 ms \citep{ritzmann2012a} and neuromuscular (45 ms \citep{sponberg2008a}) delays. These, combined with the fact that previous studies observed minimal proprioceptive sensory input from legs during flailing \citep{camhi1977a,delcomyn1987a,zill1986a}, indicate that leg flailing was more feedforward-driven than a feedback-controlled reflex coordinated with wing opening (Figure \ref{fig:elife_f3_fs2}). Moreover, large trial-to-trial variations in the number of attempts required to self-right showed that the animal’s self-righting was stochastic (Figure \ref{fig:elife_f3_fs3}).

\clearpage
\subsection{Wing opening and leg flailing together facilitate robot self-righting}
\begin{figure}[t]
   \centering
   \includegraphics[width=1.0\linewidth]{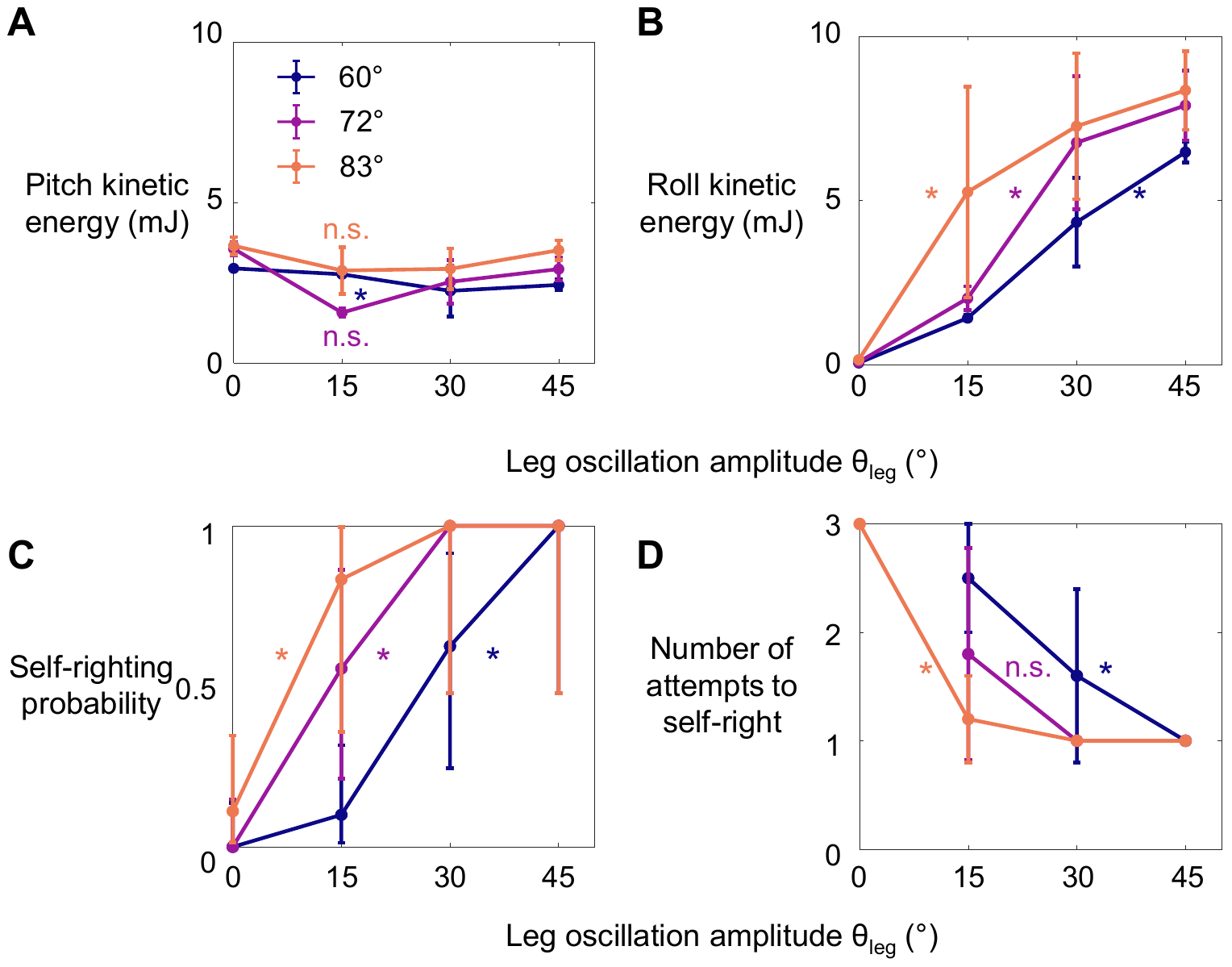}
   \caption[Robot’s kinetic energy and self-righting performance.] %
   {Robot’s kinetic energy and self-righting performance. (A, B) Average pitch and roll kinetic energy during self-righting as a function of leg oscillation amplitude $\theta_{leg}$ at different wing opening amplitudes $\theta_{wing}$. (C, D) Self-righting probability and average number of attempts required to self-right as a function of $\theta_{}$leg at different $\theta_{wing}$. Error bars in A, B, and D are $\pm$ s.d., and those in C are confidence intervals of 95\%. Asterisks indicate a significant dependence (\textit{P} < 0.05) on $\theta_{leg}$ at a given $\theta_{wing}$ and n.s. indicates none. See Figure 4 source data for detail of statistical tests. Sample size: Kinetic energy: \textit{n} = 20 attempts at each wing opening amplitude. Self-righting probability and number of attempts: \textit{n} = 58, 42, and 34 attempts at $\theta_{wing}$ = 60$\degree$, 72$\degree$, and 83$\degree$. For kinetic energy, only the first attempt from each trial is used to measure the average to avoid bias from large pitching or rolling motion during subsequent attempts that self-right.}
   \label{fig:elife_f4}
\end{figure}

The robot’s self-righting performance increased with both wing opening amplitude $\theta_{wing}$ and leg oscillation amplitude $\theta_{leg}$ (Figure \ref{fig:elife_f4}). Similar to the animal, the robot’s self-righting was stochastic, with large trial-to-trial variation in the number of attempts required to self-right and body pitching and rolling motions (Figures \ref{fig:elife_f6}, \ref{fig:elife_f6_fs1} ). For each $\theta_{wing}$ tested, as $\theta_{leg}$ increased, average roll kinetic energy increased (Figure \ref{fig:elife_f4}B; \textit{P} < 0.0001, ANOVA) and the robot’s self-righting probability increased (Figure \ref{fig:elife_f4}C; \textit{P} < 0.0001, Nominal logistic regression), reaching one at higher $\theta_{leg}$. Meanwhile, the number of attempts required for self-righting decreased (Figure \ref{fig:elife_f4}D; \textit{P} < 0.05, ANOVA). At the maximal $\theta_{leg}$ tested (45$\degree$), the robot always self-righted (\ref{fig:elife_f4}C) and always did so in the first wing opening attempt (Figure \ref{fig:elife_f4}D). Together, these results demonstrated that wing opening and leg flailing together facilitate the robot’s self-righting performance over the wide range of parameter space tested.

\begin{figure}
   \centering
   
   \caption[Robot’s self-righting motion and potential energy landscape.] %
   {Robot’s self-righting motion and potential energy landscape. (A) Snapshots of reconstructed robot upside down (i), in metastable state (ii), self-righting by pitch (iii) and roll (iii’) modes, and upright afterwards (iv, iv’). (B) Snapshots of potential energy landscape at different wing opening angles corresponding to (A) i, ii, iii. Dashed curves are boundary of upside-down/metastable basin. Green dots show saddles between metastable basin and the three upright basins. Gray curves show constant potential energy contours. Black, dashed blue, and red curves are representative trajectories of being attracted to and trapped in metastable basin, self-righting by pitch mode, and self-righting by roll mode, respectively. i, ii, iii in (A, B) show upside-down (1), metastable (2), and upright (3iii, iii’) states. (C) Polar plot of potential energy barrier to escape from upside-down or metastable local minimum along all directions in pitch-roll space. $\Psi$ is polar angle defining direction of escape in body pitch-roll space. Green arrow in (i) shows direction of upright minima at pitch = 180$\degree$ ($\Psi$ = 0$\degree$). Black circle shows scale of energy barrier (100 mJ). Blue and red arrows in (ii) define pitch and roll potential energy barriers. Blue and red error bars in (iii) show average maximal pitch and roll kinetic energy, respectively. }
   \label{fig:elife_f5}
\end{figure}
\clearpage
\begin{center}
    \includegraphics[width=1.0\linewidth]{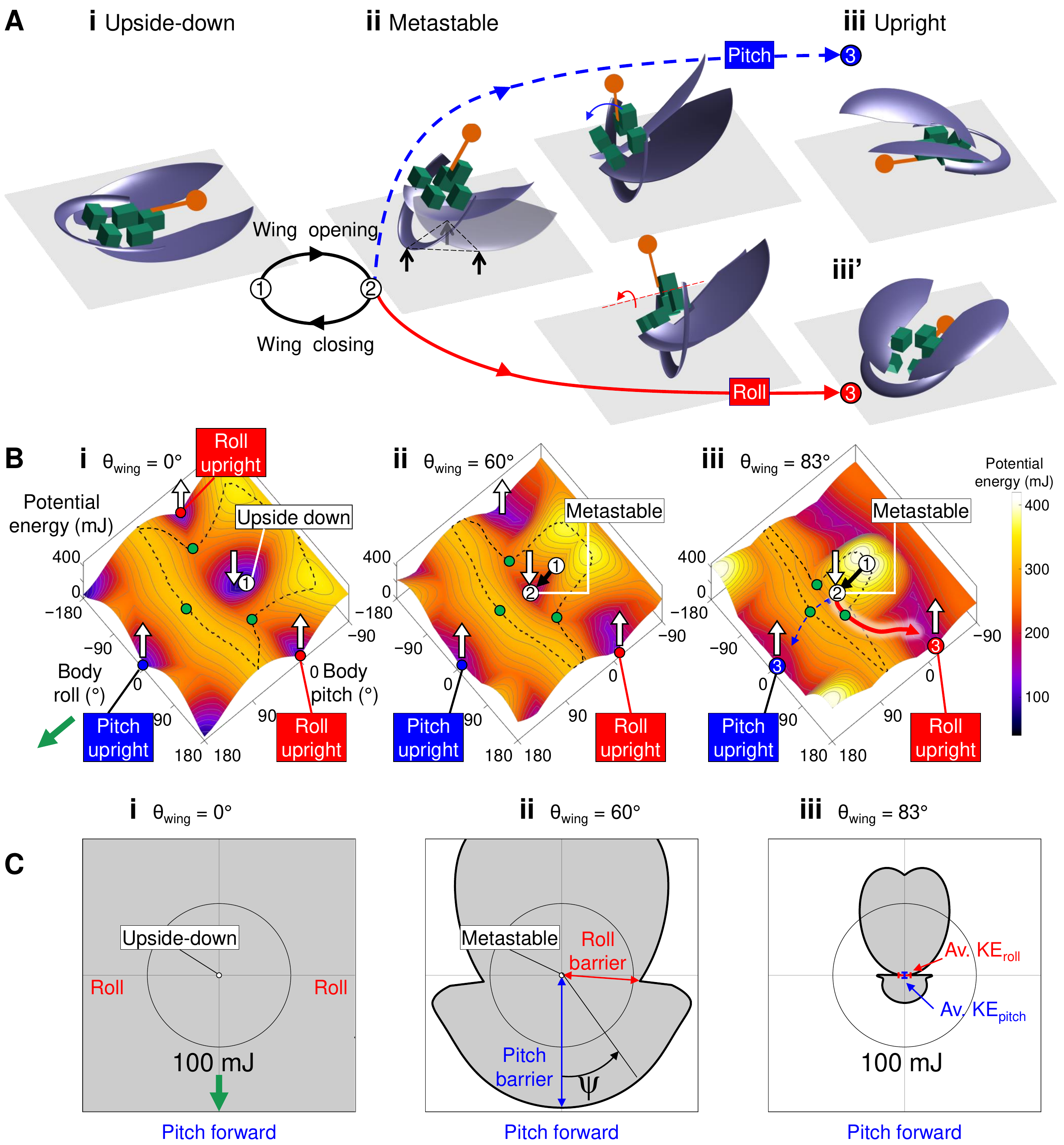}
\end{center}

\begin{figure}[t]
   \centering
   \includegraphics[width=1.0\linewidth]{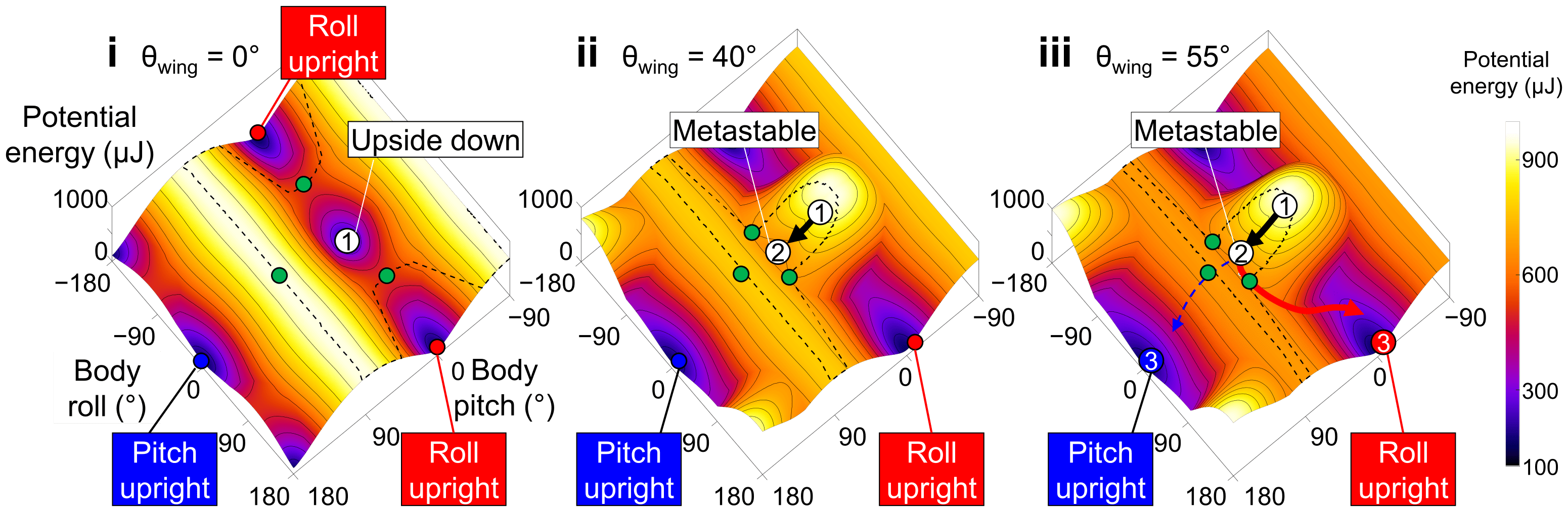}
   \caption[Animal’s potential energy landscape.] %
   {Animal’s potential energy landscape. Snapshots of potential energy landscape at different wing opening angles. Black curve is representative trajectories of failed attempts and dashed blue and red curves are for successful attempt by pitch mode and self-righting by roll mode, respectively. Thin black curves on landscape are constant potential energy contours. Dashed black curves show boundary of upside-down/metastable basins. Green dots show saddles between metastable basin and the three upright basins.}
   \label{fig:elife_f5_fs1}
\end{figure}

\clearpage
\subsection{Robot self-righting resembles animal’s}
The robot’s winged self-righting behavior resembled that of the discoid cockroach in multiple aspects (Figures \ref{fig:elife_f1}, \ref{fig:elife_f5}A, \ref{fig:elife_f5_fs1}). First, it often took the robot multiple attempts (Figure \ref{fig:elife_f4}D) to self-right probabilistically (Figure \ref{fig:elife_f4}C). In addition, as the wings opened, the robot’s body pitched up (Figure \ref{fig:elife_f5}Ai), and the head and two opened wings formed a triangular base of support in which the center of mass projection fell (metastable state, Figure \ref{fig:elife_f5}Aii). In failed attempts, after the wings opened fully, the robot was unable to escape this metastable state by either pitching over the head or rolling sideways and fell back to the ground upside-down as the wings closed (Figure \ref{fig:elife_f5}A). In successful attempts, the robot escaped the metastable state and always self-righted by rolling to either side (Figure \ref{fig:elife_f5}Aiii’-iv’, red). Moreover, the robot never lifted off the ground during self-righting. Finally, the robot’s motion trajectories in the space of body pitch, roll, and center of mass height were stereotyped for both failed and successful attempts (Figures \ref{fig:elife_f6}, \ref{fig:elife_f6_fs1}), although they are also stochastic with trial-to-trial variations in body pitch and roll.

\clearpage
\subsection{Robot and animal have similar evolving potential energy landscapes}
For both the animal and robot, the potential energy landscape over body pitch-roll space were similar in shape, and both changed in a similar fashion as the wings opened (Figures \ref{fig:elife_f5}, \ref{fig:elife_f5_fs1}). This is expected because the animal and robot were geometrically similar (Table 2). When the wings were fully closed, the potential energy landscape had a local minimum at near zero body pitch and roll (Figure \ref{fig:elife_f5}Bi). This is because either pitching or rolling of the body from being upside-down increases center of mass height and thus gravitational potential energy. Hereafter, we refer to this local minimum basin as the upside-down basin. The landscape also had three other local minima corresponding to the body being upright . One local minimum at (body pitch, roll) = (180$\degree$, 0$\degree$) could be reached from the upside-down basin by pitching forward (Figure \ref{fig:elife_f5}A, blue dot). Two local minima at (body pitch, roll) = (0$\degree$, $\pm$180$\degree$) could be reached by rolling left or right (Figure \ref{fig:elife_f5}Aiii’-iv’, red and blue curves are for roll and pitch modes respectively). Hereafter, we refer to these basins as pitch and roll upright basins, respectively . Transition from one basin to another required overcoming the potential energy barrier separating them (Figure \ref{fig:elife_f5}B, dashed black curve). As the wings opened, both the robot’s and animal’s potential energy landscape and its equilibria changed (Figure \ref{fig:elife_f5}B, \ref{fig:elife_f5_fs1}). The upside-down basin evolved  into a metastable basin around a local minimum with a positive pitch and zero roll (Figures \ref{fig:elife_f5}Bii, \ref{fig:elife_f5_fs1}Aii, white dot). This local minimum corresponded to the metastable state with the triangular base of support (Figures \ref{fig:elife_f5}Aii, \ref{fig:elife_f1}B). The more the wings opened, the higher the pitch of this local minimum was. To self-right via either the pitch (Figures \ref{fig:elife_f5}Aiii-iv, \ref{fig:elife_f1}Aiii-iv) or roll (Figures \ref{fig:elife_f5}Aiii’-iv’, \ref{fig:elife_f1}Aiii’-iv’) mode, the system state must escape from the metastable basin to reach either the pitch or a roll upright basin (e.g., Figure \ref{fig:elife_f5}Biii, blue and red curves).

\subsection{Self-righting transitions are destabilizing, \\barrier-crossing transitions on landscape}
Reconstruction of the robot’s 3-D motion on the potential energy landscape revealed that its self-righting transitions are probabilistic barrier-crossing transitions (Figure \ref{fig:elife_f6}). Except when the robot was upright, upside-down, or metastable, it was always statically unstable and its system state was strongly attracted to one of these three local minima basins. At the beginning of each attempt, the system state was in the upside-down basin. As the wings opened, it was attracted towards the metastable basin that emerged. In failed attempts, the system state was trapped in the metastable basin and unable to escape it (Figure \ref{fig:elife_f6}, black curves). In successful attempts, it crossed a potential energy barrier (Figure \ref{fig:elife_f5}B, dashed black curve) to escape the metastable basin and reach a roll upright basin (Figure \ref{fig:elife_f6}, white curves). These observations are in accord with the animal’s center of mass height measurements at the beginning, maximal pitch, and end of each attempt from the previous study \citep{li2019a} projected onto the animal’s potential energy landscape (Figure \ref{fig:elife_f2} C, D).

\clearpage
\begin{figure}[t]
   
   \caption[Robot state trajectories on potential energy landscape.] %
   {Robot state trajectories on potential energy landscape. (A) $\theta_{wing}$ = 60$\degree$. (B) $\theta_{wing}$ = 72$\degree$. (C) $\theta_{wing}$ = 83$\degree$. Columns i and ii show successful (white) and failed (black) self-righting attempts, respectively. n is the number of successful/failed attempts at each $\theta_{wing}$. Note that only the end point of the trajectory, which represented the current state, showed the actual potential energy of the system at the corresponding wing opening angle. The rest of the visualized trajectory showed how body pitch and roll evolved but, for visualization purpose, was simply projected on the landscape surface. Gray lines show energy contours. Green dots show saddles between metastable basin and the three upright basins.}
   \label{fig:elife_f6}
\end{figure}

\clearpage
\begin{center}
       \includegraphics[width=1.0\linewidth]{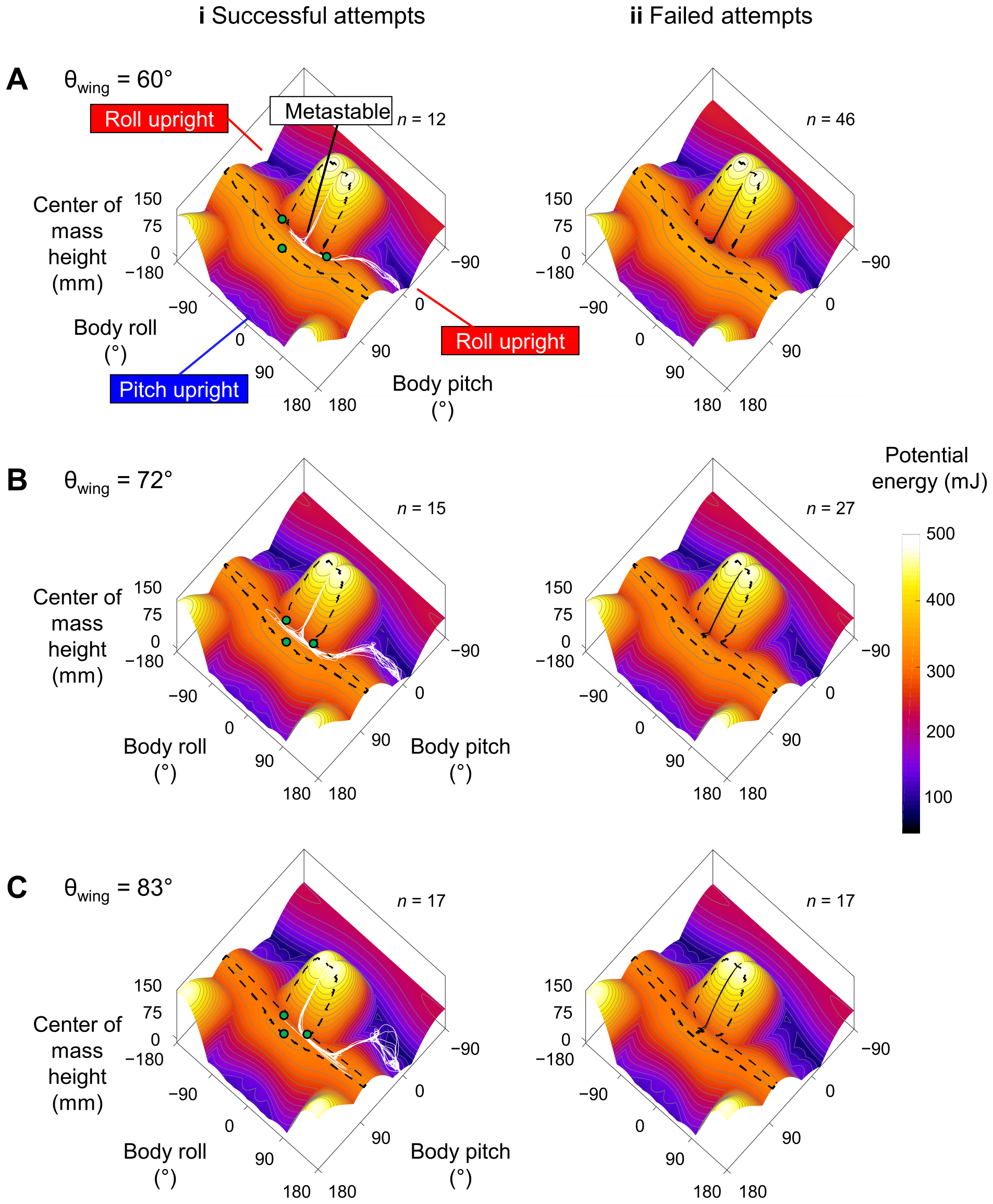}
\end{center}
\clearpage
\begin{figure}[t]
   \centering
   \includegraphics[width=0.7\linewidth]{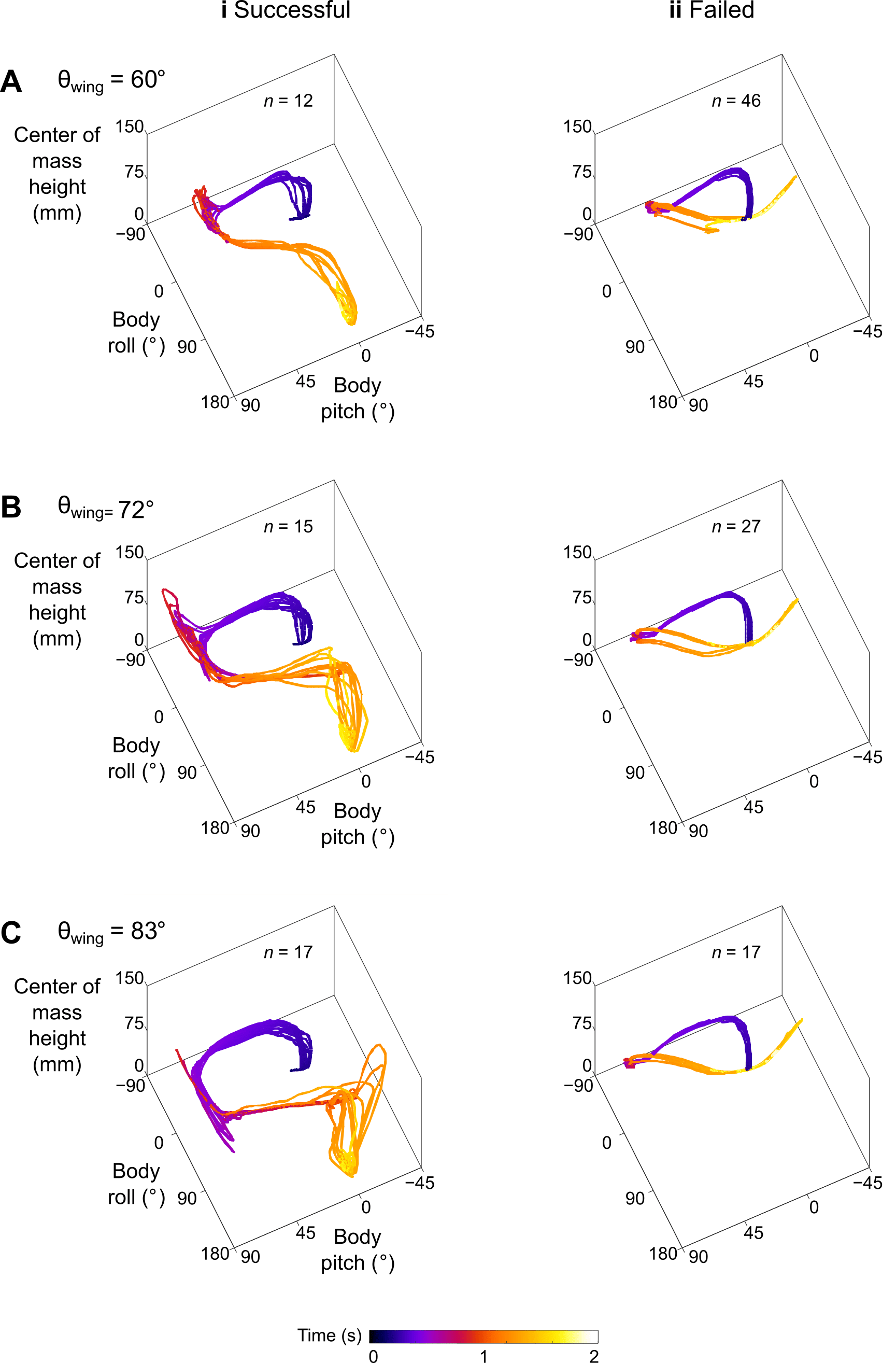}
   \caption[Robot’s stereotyped body motion during self-righting] %
   {Robot’s stereotyped body motion during self-righting. State trajectories in body pitch, body roll, and center of mass height space. (A) $\theta_{wing}$ = 60°. (B) $\theta_{wing}$ = 72°. (C) $\theta_{wing}$ = 83°. Columns i and ii show successful and failed self-righting attempts, respectively. \textit{n} is the number of successful or failed attempts at each $\theta_{wing}$. }
   \label{fig:elife_f6_fs1}
\end{figure}

\clearpage
\subsection{Self-righting via rolling overcomes smaller barrier than via pitching} 
For both the animal and robot, the potential energy landscape model allowed us to quantify the potential energy barrier for self-righting via the pitch and roll modes. The barrier to escape the metastable state to self-right varied with the direction along which the system moved in the body pitch-roll space (Figures \ref{fig:elife_f5}C, \ref{fig:elife_f7}C, \ref{fig:elife_f7_fs1}). We defined the pitch and roll barriers as the minimal barriers to escape from the metastable local minimum towards the pitch and roll upright basins (Figure \ref{fig:elife_f5}C, blue and red arrow). At all wing opening angles up to 90$\degree$, the roll barrier was always lower than the pitch barrier (Figures \ref{fig:elife_f5}C, \ref{fig:elife_f7}C, \ref{fig:elife_f7_fs1}C).

\clearpage
\subsection{Barrier reduction by wing opening facilitates self-righting via rolling}
\begin{figure}[t]
   \centering
   \includegraphics[width=1.0\linewidth]{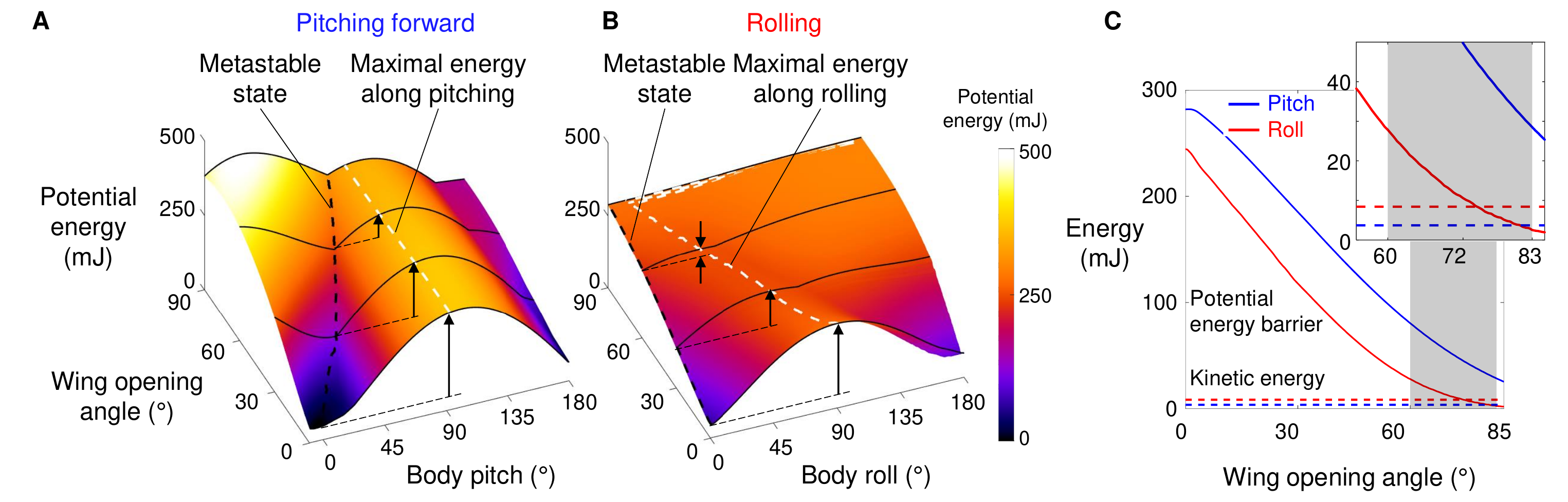}
   \caption[Robot’s potential energy barriers for self-righting via pitch and roll modes.] %
   {Robot’s potential energy barriers for self-righting via pitch and roll modes. (A) Potential energy during self-righting via pitch mode as a function body pitch and wing opening angle. (B) Potential energy during self-righting via roll mode as a function of body roll and wing opening angle. Dashed white curves in A and B show energy of metastable state. Dashed black curve in A and B shows maximal energy when pitching forward or rolling from metastable state, respectively. Vertical upward arrows define pitch (A) and roll (B) barriers at a few representative wing opening angles. (C) Pitch (blue) and roll (red) barrier as a function of wing opening angle. Blue and red dashed lines show average maximal pitch and roll kinetic energy, respectively. Gray band shows range of wing opening amplitudes tested. Inset shows the same data magnified to better show kinetic energy.}
   \label{fig:elife_f7}
\end{figure}

\begin{figure}[t]
   \centering
     \includegraphics[width=1.0\linewidth]{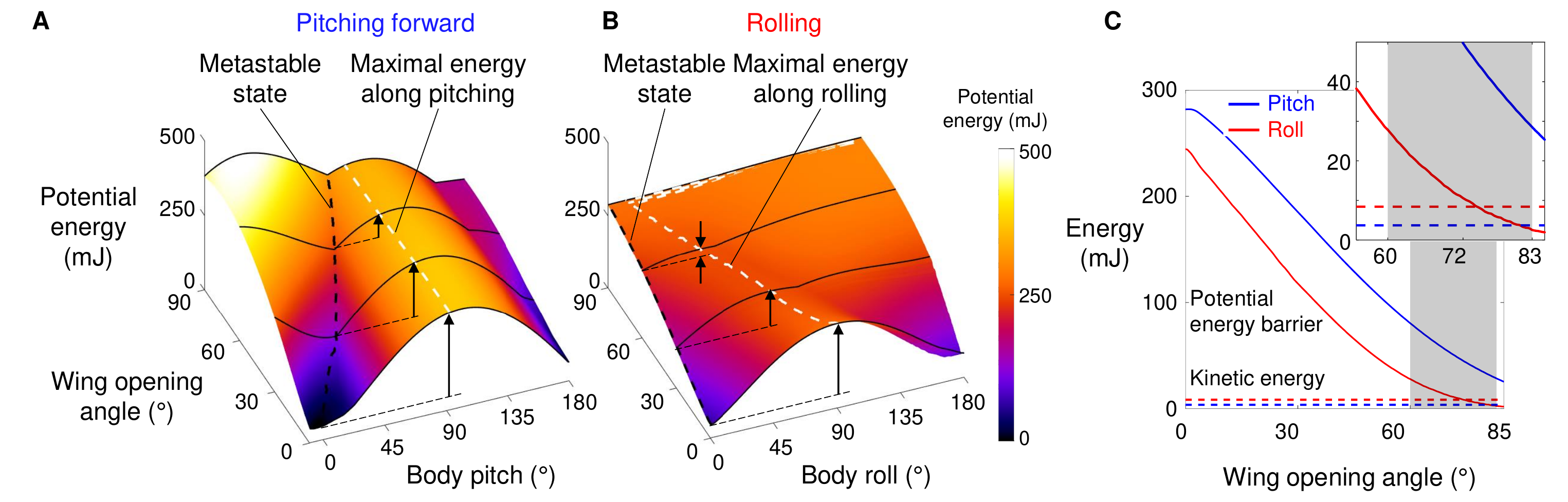}
   \caption[Animal’s potential energy barriers for self-righting via pitch and roll modes.] %
   {Animal’s potential energy barriers for self-righting via pitch and roll modes. (A) Potential energy of self-righting via pitch mode as a function body pitch and wing opening amplitude. (B) Potential energy of self-righting via roll mode as a function of body roll and wing opening amplitude. Dashed white curves in A and B show energy of metastable state. Dashed black curve in A and B shows maximal energy when pitching forward or rolling from metastable state, respectively. Vertical arrows define pitch (A) and roll (B) barriers at a few representative wing opening angles. (C) Pitch (blue) and roll (red) barrier as a function of wing opening angle. Dashed and solid horizontal lines show the intact (dashed) and modified (solid) animal’s average pitch kinetic energy (blue) and average roll kinetic energy (red), respectively. Inset shows the same data magnified to better show kinetic energy.}
   \label{fig:elife_f7_fs1}
\end{figure}

For both the animal and robot, as wing opening angle increased, both the pitch and roll barrier decreased monotonically (Figure \ref{fig:elife_f7}C). As the wings opened to the range of $\theta_{wing}$ tested (Figure \ref{fig:elife_f7}C, gray band), the pitch barrier was still much greater than the average pitch kinetic energy (Figure \ref{fig:elife_f7}C, \ref{fig:elife_f7_fs1}C, solid curve vs. dashed blue line). By contrast, the roll barrier was lowered to a similar level as the average roll kinetic energy (Figure \ref{fig:elife_f7}C, solid curve vs. dashed red line). This explained why the modified animal, with its higher average kinetic energy, self-righted at a higher probability than the intact animal (Figure \ref{fig:elife_f7_fs1} solid vs. dashed lines). These findings demonstrated that, even though wing opening did not generate sufficient kinetic energy to self-right by pitching (Figure \ref{fig:elife_f7}C), it reduced the roll barrier so that self-righting became possible using small, perturbing roll kinetic energy from leg flailing.

To further confirm this, we compared the robot’s kinetic energy with potential energy barrier along the pitch and roll directions respectively during each attempt (Figures \ref{fig:elife_f7_fs2}, \ref{fig:elife_f7_fs3}). The robot’s pitch kinetic energy was insufficient to overcome even the reduced pitch barrier in both failed and successful attempts (Figure \ref{fig:elife_f7_fs3}). By contrast, as wing opening and leg flailing amplitudes increased, the robot’s roll kinetic energy more substantially exceeded the roll barrier during successful attempts (\ref{fig:elife_f7_fs2}; \textit{P} < 0.001, nominal logistic regression), and the surplus enabled it to self-right via rolling.

\begin{figure}[t]
   \centering
   \caption[Comparison between robot’s kinetic energy and potential energy barrier along roll direction. ] %
   {Comparison between robot’s kinetic energy and potential energy barrier along roll direction. (A) Roll kinetic energy, (B) roll potential energy barrier, and (C) roll kinetic energy minus potential energy barrier along roll direction over time for a representative successful and failed attempt. Between two vertical dashed lines is when wings are held fully open. (D) Surplus along roll direction over time for all attempts from all trials. The attempts are grouped along vertical axis, based on increasing leg oscillation amplitude $\theta_{leg}$. For each $\theta_{leg}$, the attempts are further grouped along by different wing opening amplitudes $\theta_{wing}$ (increasing along upward direction). Columns (i) and (ii) are successful and failed attempts. Asterisk indicates significant difference in roll kinetic energy minus potential energy barrier between successful and failed attempts. (E) Average of maximal roll kinetic energy minus potential energy barrier, as a function of wing opening amplitude and leg oscillation amplitude. Red and blue show surplus and deficit of roll kinetic energy minus potential energy barrier, respectively. \textit{n} = 134 attempts. See Figure 7—source data 1 for results of statistical tests.}
   \label{fig:elife_f7_fs2}
\end{figure}

\clearpage
\begin{center}
     \includegraphics[width=1.0\linewidth]{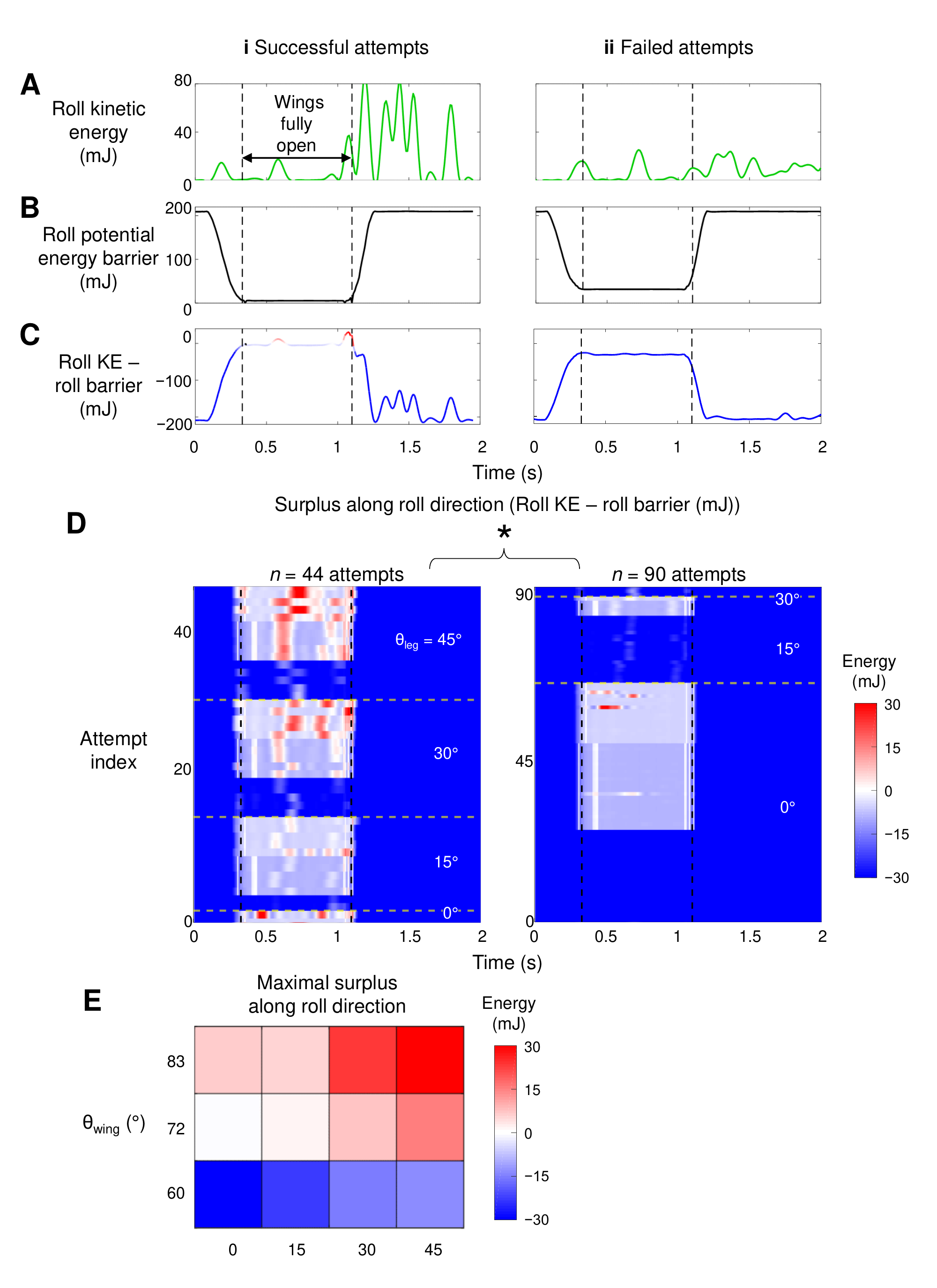}
\end{center}
\begin{figure}[t]
   \centering
   \caption[Comparison between robot’s pitch kinetic energy and pitch potential energy barrier.] %
   {Comparison between robot’s pitch kinetic energy and pitch potential energy barrier. (A) Kinetic energy, (B) potential energy barrier, and (C) kinetic energy minus potential energy barrier as a function of time for a representative successful (i) and failed (ii) attempt. Between two vertical dashed lines is when wings are held fully open. (D) Pitch kinetic energy minus potential energy barrier as a function of time of all attempts from all trials. Along vertical axis, attempts are grouped into increasing leg oscillation amplitude $\theta_{leg}$. For each $\theta_{leg}$, attempts are further grouped into increasing wing opening amplitudes $\theta_{wing}$. Columns (i) and (ii) are successful and failed attempts. Asterisk indicates significant difference in pitch kinetic energy minus potential energy barrier between successful and failed attempts. (E) Average of maximal pitch kinetic energy minus potential energy barrier when wings are fully open as a function of $\theta_{wing}$ and $\theta_{leg}$. Red and blue show surplus and deficit of pitch kinetic energy minus potential energy barrier, respectively. n = 134 attempts. See Figure 7—source data 1 for results of statistical tests.}
   
   \label{fig:elife_f7_fs3}
\end{figure}

\clearpage
\begin{center}
   \includegraphics[width=1.0\linewidth]{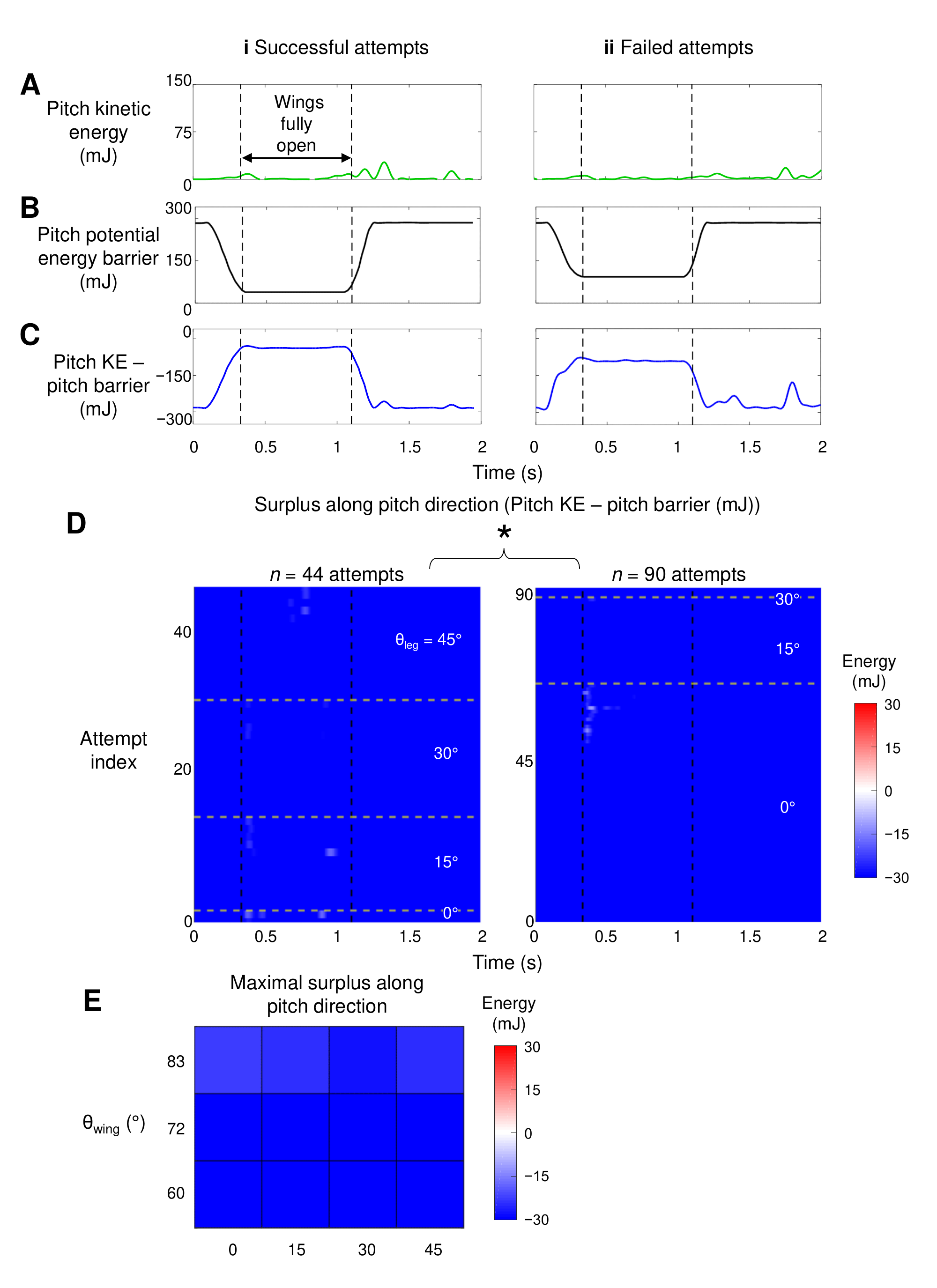} 
\end{center}

\clearpage
\section{Discussion}
We integrated animal experiments, robotic physical modeling, and potential energy landscape modeling to discover the physical principles of how the discoid cockroach use propelling and perturbing appendages (wings and legs, respectively) together to achieve strenuous ground self-righting. Ground self-righting transitions are stochastic, destabilizing barrier-crossing transitions on a potential energy landscape. Even though propelling appendages cannot generate sufficient kinetic energy to cross the high potential energy barrier of this strenuous locomotor task, they modify the landscape and lower the barriers in other directions sufficiently so that kinetic energy from perturbing appendages can help cross them probabilistically to self-right. Compared to only using propelling or perturbing appendages alone, using them together makes self-righting more probable and reduces the number of attempts required, increasing the chance of survival. 

Although the intact animal’s average kinetic energy from hind leg flailing was not sufficient to overcome the potential barrier at the range of wing opening observed, it still self-righted at a small but finite probability (Figure \ref{fig:elife_f3}B). This was likely because of the additional kinetic energy from flailing of fore and mid legs, small forces from legs scraping the ground, as well as abdominal flexion and twisting and passive wing deformation under load (\citep{li2019a}), both of which induce lateral asymmetry and tilts the potential energy landscape towards one side and lowers the roll barrier. This consideration further demonstrates the usefulness of co-opting a variety of appendages for propulsion and perturbation simultaneously to achieve strenuous ground self-righting. Such exaptation (\citep{gould1982a}) of multiple types of appendages that evolved primarily for other locomotor functions for self-righting is likely a general behavioral adaptation and should be adopted by terrestrial robots.

\subsection{Stereotyped motion emerges from physical interaction constraint}
Our landscape modeling demonstrated that the stereotyped body motion during strenuous leg-assisted, winged self-righting in both the animal and robot is strongly constrained by physical interaction of the body and appendages with the environment. The stereotyped repeated body pitching up and down during failed attempts and rolling during successful attempts directly result from the strong attraction of the system state to the landscape basins, which directly arise from physical interaction of body/appendages with the ground. This finding suggested that potential energy landscape modeling can be used to understand stereotyped ground self-righting strategies of other species (\citep{ashe1970a}; \citep{domokos2008a}; \citep{golubovi2013a}; \citep{li2019a}; O’Donnel, 2018)and even infer those of extinct species (analogous to (\citep{gatesy2009a})). Similarly, it will inform the design and control of self-righting robots (e.g., \cite{caporale2020a,kessens2012a})/

\begin{figure}[t]
   \centering
   \includegraphics[width=1.0\linewidth]{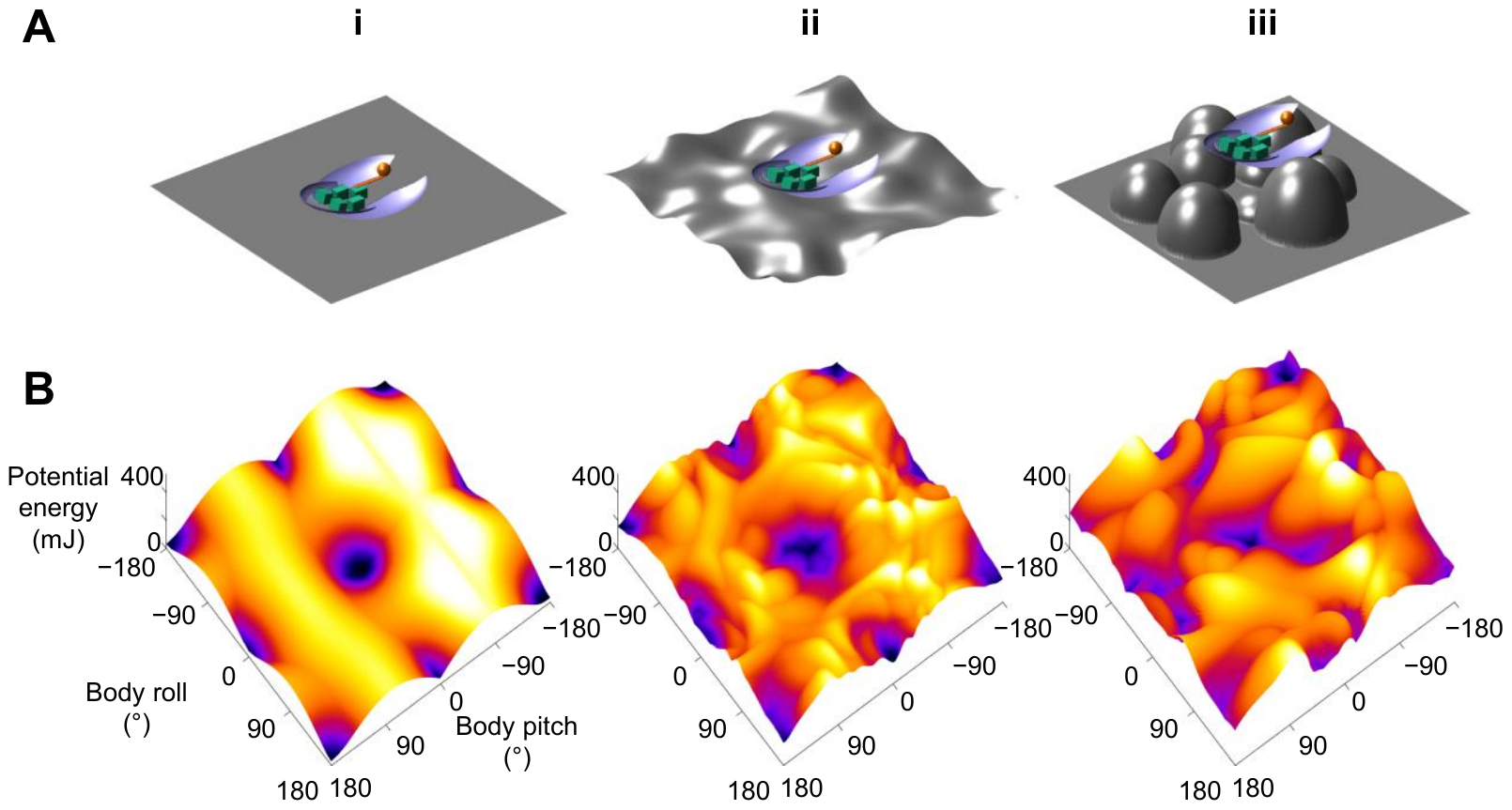}
   \caption[Dependence of potential energy landscape on ground geometry.] %
   {Dependence of potential energy landscape on ground geometry. (A) Grounds of different geometry. (i) Flat ground. (ii, iii) Uneven ground with small (ii) and large (iii) asperities compared to animal/robot size. (B) potential energy landscapes for self-righting on corresponding ground. In ii and iii, landscape is not invariant to robot body translation as in i. Landscape is show for robot at the geometric center of the terrain with wings closed. Robot shown for scale.}
   \label{fig:elife_f8}
\end{figure}

Although only demonstrated in a model system, the potential energy landscape approach can in principle be applied to more complex and different self-righting behaviors, as well as on ground of different properties (\citep{sasaki2016a}), to understand how propelling and perturbing effects work together. For example, as the ground becomes more rugged with larger asperities, the landscape becomes more rugged with more attractive basins (Figure \ref{fig:elife_f8}). In addition, for leg-assisted, winged self-righting, we can add degrees of freedom for fore and mid leg flailing, abdomen flexion and twisting, and even passive wing deformation due to load (\citep{li2019a}) to create fine-grained potential energy landscapes to understand how these motions may emerge from physical interaction constraints. We can also understand legged self-righting by modeling how the legs and deformable abdomen (\citep{li2019a}) affect the potential energy landscape when wings are not used. This broad applicability will be useful for comparative studies across species, strategies, and even environments, such as understanding why some cockroach species’ self-righting is more dynamic than others (\citep{li2019a}). However, this approach does not apply to highly dynamic self-righting strategies, such as those using jumping (\citep{bolmin2017a}; \citep{kovac2008a}) where kinetic energy far exceeds the potential energy barrier.

\subsection{Towards potential energy landscape theory of self-righting transitions}
The potential energy landscape model here does not describe self-righting dynamics. Recent dynamic modeling using multi-body dynamics simulations (\citep{xuan2020a}) and dynamical templates (\citep{xuan2020b}) in our lab revealed that wing-leg coordination affects self-righting by changing the mechanical energy budget (\citep{xuan2020b}) and that the randomness in the animal’s motion helps it self-right (\citep{xuan2020a}). However, these approaches have their limitations: multi-body dynamic simulations are effectively experiments on a computer; dynamical templates are increasingly challenging to develop as system degrees of freedom increases. Further development of a potential energy landscape theory that adds stochastic, non-conservative forces to predict how the system “diffuses” across landscape barriers (analogous to \citep{socci1996a}) may be a relatively simple yet intuitive way to model probabilistic barrier-crossing dynamics.

\cleardoublepage

\chapter{Tracking and reconstructing large obstacle interaction of small animals using a terrain treadmill }
\label{chap:jeb}

\let\thefootnote\relax\footnotetext{This chapter is a paper by Ratan Othayoth*, Blake Strebel*, Yuanfeng Han, Evains Francois, and Chen Li (*equal contributions) published in \textit{The Journal of Experimental Biology} (2022) \citep{Othayoth2022}}

\section{Summary statement}
A device keeps a small animal on top of a sphere while it traverses obstacles, creating a “terrain treadmill” to study locomotion over a long time and distance at high-resolution.

\section{Author Contributions}
Ratan Othayoth implemented 2-D tracking and 3-D reconstruction, analyzed data, created visualizations, and wrote the paper; Blake Strebel designed and constructed the treadmill, implemented the treadmill control system, and wrote an early draft; Yuanfeng Han designed the treadmill and assisted construction; Evains Francois collected animal data for testing treadmill performance; Chen Li. oversaw the study, designed the treadmill, created visualizations, and wrote the paper.

\section{Introduction}
In nature, terrestrial animals often move through spatially complex, three-dimensional terrain (\citep{dickinson2000a}. Small animals are particularly challenged to traverse many obstacles comparable to or even larger than themselves \citep{kaspari1999a}. By contrast, the majority of laboratory studies of terrestrial locomotion have been performed on flat surfaces \citep{alexander1983a,blickhan1993a,cavagna1976a,diederich2002a,ferris1998a,full1990a,koditschek2004a,li2012a,minetti2002a,moritz2003a,spagna2007a,spence2010a}, either rigid or with various surface properties (friction, slope, solid area fraction, stiffness, damping, ability to deform and flow, etc.).

Recent laboratory studies have begun to advance our understanding of animal locomotion in complex terrain with obstacles \citep{birn-jeffery2012a,blaesing2004a,collins2013a,daley2006a,duerr2018a,gart2018b,gart2018a,harley2009a,kohlsdorf2006a,li2015a,olberding2012a,parker2016a,sponberg2008a,theunissen2014a,tucker2012a}. Because of typical laboratory space constraints, the terrain arenas used in these studies are usually no larger than a few dozen body lengths in each dimension. Thus, they only allow experiments at relatively small spatiotemporal scales beyond ~10 body lengths and ~10 movement cycles. It remains a challenge to study animal locomotion in complex 3-D terrain with large obstacles at larger spatiotemporal scales.

Experiments at large spatiotemporal scales are usually realized by treadmills to keep the animal (including humans) stationary relative to the laboratory \citep{buchner1994a,darken1997a,full1987a,herreid1984a,jayakumar2019a,kram1998a,stolze1997a,watson2002a,weinstein1999a}. However, only small obstacles can be directly mounted on such treadmills \citep{voloshina2013a}; larger obstacles have to be dropped onto the treadmill during locomotion \citep{park2015a,snijders2010a}. Furthermore, such linear treadmills allow only untethered movement along one direction. Alternatively, spherical treadmills use lightweight spheres of low inertia suspended on air bearing (kugels) to allow small animals to rotate the spheres as they freely change their movement speed and direction, \citep{bailey2004a,okada2000a,ye1995a}. However, the animal is tethered . However,  the animal is tethered, and obstacles cannot be used.

\begin{figure}
    \centering
    \includegraphics[width=1.0\linewidth]{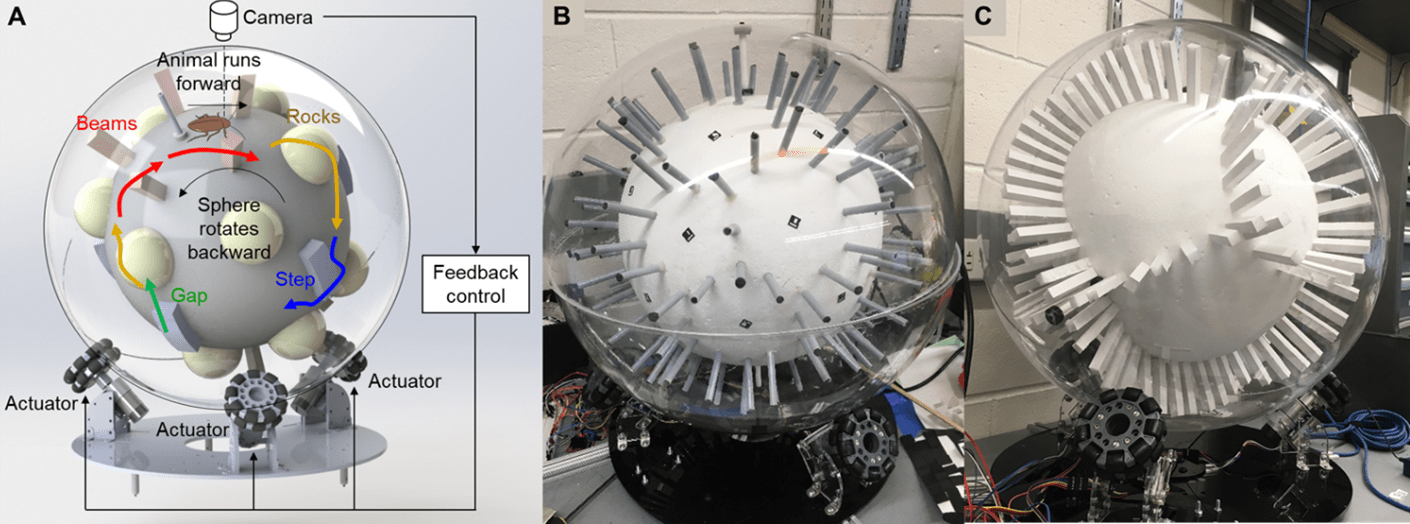}
    \caption[Terrain treadmill] %
    {Terrain treadmill. (A) Design of terrain treadmill. Colored elements show example modular terrain that can be used. (B, C) Terrain treadmill, with (B) sparsely and (C) densely spaced vertical pillars as example terrain modules. ArUCo markers attached on the inner sphere are also shown in (B).}
    \label{fig:jeb_1}
\end{figure}

Here, we used a terrain treadmill (Figure \ref{fig:jeb_1}A, B) that enabled large spatiotemporal scale, high-resolution observations of small animal locomotion in complex terrain with large obstacles. Our terrain treadmill design was inspired by a celestial globe model. The terrain treadmill consists of a transparent, smooth, hollow, outer sphere rigidly attached to a concentric, solid, inner sphere using a connecting rod (Figure \ref{fig:jeb_1}A). Terrain modules can be attached to the inner sphere (Figure \ref{fig:jeb_1}) to simulate obstacles that  small animals encounter in natural terrain \citep{othayoth2021b}. The outer sphere is placed on an actuator system consisting of three actuated omni-directional wheels (Figure \ref{fig:jeb_1}A). An overhead camera captures videos of the animal moving on top of the inner sphere, with an ArUCo \citep{garrido-jurado2014a} marker attached on its body. The animal’s position estimated from tracking the marker is used by a feedback controller to actuate the connected spheres with the opposite velocity to keep the animal on top (Figure \ref{fig:jeb_2}D, E) as it moves through the obstacle field (Figures \ref{fig:jeb_2}A-C, \ref{fig:jeb_3}A, B,). Finally, the reconstructed 3-D motion can be used to estimated different metrics such as body velocities and antennal planar orientation relative to the body heading (Figures \ref{fig:jeb_3}, \ref{fig:jeb_method},).

\clearpage

\section{Methods}

\subsection{Experiment and data collection}%
We used discoid cockroaches (\textit{Blaberus discoidalis}) to test the treadmill’s ability to elicit free locomotion and measure animal-terrain interaction over large spatiotemporal scales. We put the animal inside the outer sphere and then sealed it. To pick and place the animal onto the inner sphere, we attached a square magnet (16mm side length, 3.5g) on the animal’s dorsal side, with an ArUCo marker attached to it for tracking (\ref{fig:jeb_3}A, B). We used a larger magnet to pick up and move the animal to the top of the treadmill and dropped it onto the inner sphere.

We then started the control program to keep the animal on top. The images recorded by the camera were then sent to the ROS program, which first saved each frame in its native format (a bagfile) and then processed the image to track the marker position. Based on the tracked and then filtered marker position, which were used to calculate the velocity of the animal through forward kinematics, motor velocities required to keep the animal centered on were calculated and commanded to the motors. After each experiment, the bagfiles were retrieved and processed using custom MATLAB code to extract the saved images for post processing.

\begin{figure}
    \centering
    \includegraphics[width=1.0\linewidth]{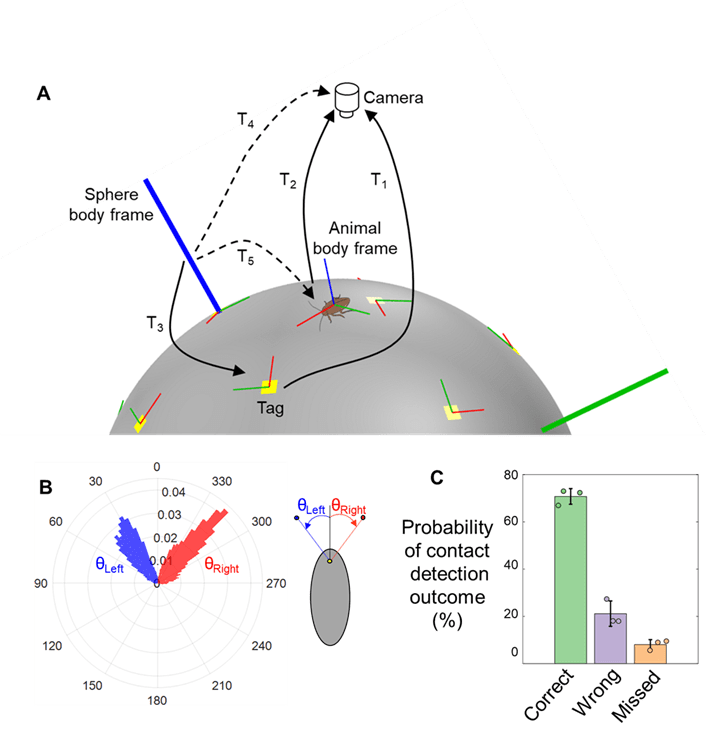}
    \caption[Measuring motion of animal exploring sparse pillar fields.] %
    {Measuring motion of animal exploring sparse pillar fields. (A) Coordinate frame transformation to measure animal motion relative to sphere. Solid black arrows are relative 3-D poses ($T_1$, $T_2$ and $T_3$) that are known or measured directly from acquired images. Dashed arrows are the two relative 3-D poses ($T_4$ and $T_5$) that are calculated from measurements to obtain animal motion relative to the sphere. Yellow squares with red and green lines show the markers attached to the sphere and their x and y axes, respectively. Thick green and blue lines show the y and z axes of the frame attached to the inner sphere. (B) Histogram of left ($\theta_{left}$, blue) and right ($\theta_{right}$, red) antenna planar orientation relative to body heading (see schematic on right for definition). (C) Accuracy of antenna-pillar contact detection outcomes. \textit{N} = 3 animals, \textit{n} = 3 trials. 
}
    \label{fig:jeb_method}
\end{figure}

\subsection{Measuring animal movement in obstacle field}
To measure the animal’s movement relative to the pillar obstacle field, we first measured the movement of the pillar obstacle field (i.e., treadmill rotation) relative to the camera. We attached 31 ArUCo markers to the inner sphere, with one each at the center of hexagonal and pentagonal regions of the soccer ball pattern projected on the sphere (Figure \ref{fig:jeb_method}A). We then separately created a map of all markers attached on the inner sphere (referred to as marker map) using ArUCo Marker-mapper application. Because each marker and its four corners were fixed relative to the coordinate frame attached to the inner sphere (i.e., $T_3$ is known, Figure \ref{fig:jeb_method}A), when one of the markers on sphere is tracked (i.e., $T_1$ can be measured, Figure \ref{fig:jeb_method}A), the relative pose between sphere body frame and the camera (Figure \ref{fig:jeb_method}A, $T_4$) can be computed. When more than one marker on the sphere is detected, relative pose of sphere and camera can be computed by solving the Perspective-n-points problem \citep{lepetit2009a}, which estimates camera pose from a known set of 3D points (marker corners) and the corresponding 2D coordinates in the image. The ‘solvePnP’ program in in image processing toolboxes in MATLAB or OpenCV may be used to for this purpose. Because the animal’s movement relative to the camera (Figure \ref{fig:jeb_method}A, $T_2$) is directly available from tracking via the calibrated camera, the animal’s pose relative to the sphere body frame and hence relative to the terrain obstacle field can be calculated (Figure \ref{fig:jeb_method}A, $T_5$). Because the ArUCo marker attached to the animal is not necessarily at its center of mass, a constant position and orientation offset must be manually determined and added. 

\subsection{Unwrapped 2-D trajectory}
	Considering that the sphere diameter is $\approx9\times$ that of animal body length, we approximated the immediate region surrounding the animal to be flat and estimated the animal’s equivalent 2-D planar trajectory. To obtain the 2-D trajectory, we integrated the body forward and lateral translational velocities and body yaw angular velocity (Figure \ref{fig:jeb_3}F-G) over time, with the initial position at origin and body forward axis along x axis. Because during portions of a trial the animal body marker was not tracked for a long duration, we did not consider those video frames. As a result, each trial was assumed to be composed of multiple segments, and each of their equivalent 2-D trajectories were assumed to have the same initial conditions as described above (Figure \ref{fig:jeb_3}E).

\clearpage
\section{Results}
\subsection{Free locomotion at large spatiotemporal scales}
\begin{figure}
    \centering
    \includegraphics[width=1.0\linewidth]{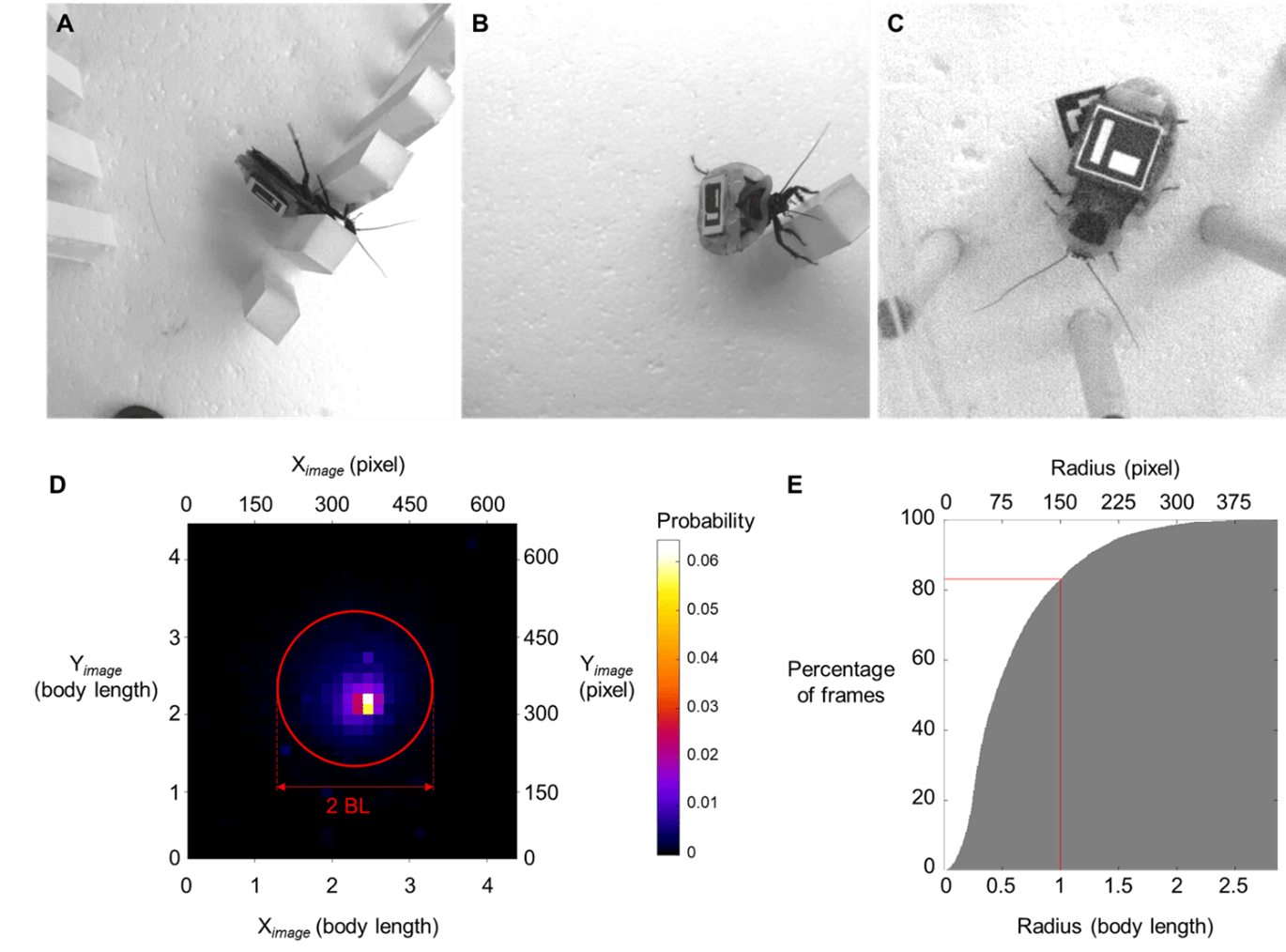}
    \caption[Animal behavior and performance of the treadmill.] %
    {Animal behavior and performance of the treadmill. (A-C) Representative snapshots of behaviors including (A) body rolling, (B) body pitching and pillar climbing, and (C) antennal sensing observed during free exploration of terrain.  (D) Probability of animal’s detected location in the image. Red circle of radius 2 animal body lengths is centered at the image center. (E) Cumulative histogram of animal’s radial position (in body lengths) from the center of the image. Vertical and horizontal red lines show a radius of red circle in (A) and the percentage of frames in which animal’s position was maintained within this circle. \textit{N} = 5 animals, \textit{n} = 12 trials.}
    \label{fig:jeb_2}
\end{figure}

We tested the terrain treadmill’s performance in eliciting sustained locomotion of discoid cockroaches (\textit{N} = 5 animals, \textit{n} = 12 trials, sparse obstacles) through both sparse (Figure \ref{fig:jeb_1}B) and cluttered (Figure \ref{fig:jeb_1}C) pillar obstacles (see ‘Experimental validation using pillar obstacle field’). Even with cluttered obstacles, where gaps between obstacles were smaller than animal body width, we were able to elicit continuous trials, in which the animal moved through pillars for 25 minutes ($\approx$ 2500 stride cycles) over 67 m ($\approx$ 1500 body lengths) (Video 2). For 83\% of the experiment duration, the terrain treadmill contained the animal within a circle of radius 4 cm (0.9 body length) centered about the image center (Figure \ref{fig:jeb_method}D, E) even at locomotion speeds of up to 10 body length/s (peak speed of 50 cm/s). We implemented a Kalman filter (\cite{harvey1990a}) to estimate the position of animal and reduce the noise and error in marker tracking . The Kalman filter continued to estimate the animal’s position even when the marker was obscured from body rolling (Figure \ref{fig:jeb_2}A) or the outer sphere’s seam. In addition, over the course of 12 trials, the animal freely explored and visited almost the entire obstacle field (Figure \ref{fig:jeb_3}C, D). Finally, the animal’s motion relative to the treadmill was used to estimate metrics such as body velocity components (Figure \ref{fig:jeb_method}F-H), antenna planar orientation relative to the body heading (Figure \ref{fig:jeb_method}B), and unwrapped 2-D trajectories (Figure \ref{fig:jeb_3}E).

\subsection{Animal-obstacle interaction}
\begin{figure}
    \centering
    \includegraphics[width=1.0\linewidth]{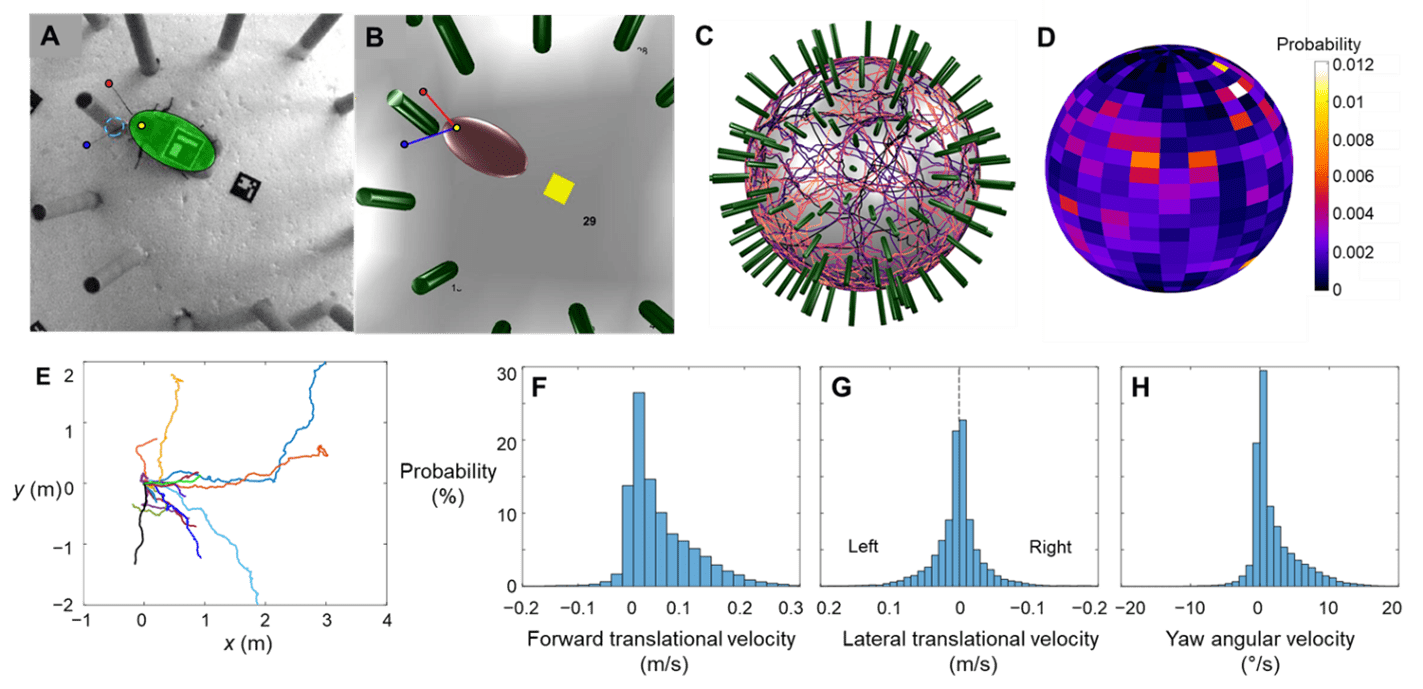}
    \caption[Representative metrics and 3-D reconstruction of animal exploring sparse pillar obstacle field. ] %
    {Representative metrics and 3-D reconstruction of animal exploring sparse pillar obstacle field.  (A, B) Representative snapshot and reconstruction of animal moving through sparse pillar obstacle field. Transparent green ellipsoid in (A) and brown ellipsoid in (B) show approximated animal body. Red and blue dots show antenna tips. Yellow dot shows the tracked point on animal’s head. Dashed cyan circle is the base of the two pillars with which the animal’s antenna is interacting. (C, D) Ensemble of trajectories (D) and probability density distribution of animal center of mass (E) during free exploration of sparse pillar obstacle field, \textit{N} = 5 animals, \textit{n} = 12 trials. (E) Unwrapped 2-D trajectories of animal, obtained by integrating forward and lateral translational velocities and yaw angular velocity over duration of trial. (F-H) Histogram of animal’s (F) forward and (G) lateral translational velocities and (H) yaw angular velocity.}
    \label{fig:jeb_3}
\end{figure}

We measured and reconstructed the animal-terrain interaction for 12 trials in which the animal freely explored the sparse obstacle field (Figure \ref{fig:jeb_3}). The ArUCo markers attached on the animal and the inner sphere, allowed measuring and reconstructing animal motion relative to obstacle field (see ‘Measuring animal movement in obstacle field’ in Methods). Because lighting was not optimized, the pillar shadow resulted in substantial variation of the background, and because the left and right antenna are visually similar and often moved rapidly, automated antenna tracking was accurate in only $\approx$ 40\% of frames after rejecting inaccurately tracked data (see ‘Automated animal tracking’ in Methods). However, this can be improved with refinement of our experimental setup in future (see Discussion).

We then detected which pillar the animal’s antennae contacted (Figure \ref{fig:jeb_3}A, B) by measuring the minimum distance from each antenna to all nearby pillars. To determine which pillar the antenna interacted with, we determined whether any pillars where within 3 cm from both antennae and which among them were closest to both antennae. We also manually identified the antenna pillar contact, which served as the ground truth. The antenna-pillar contact detected automatically was accurate in over 70\% of the contact instances (Figure \ref{fig:jeb_method}C). 

\subsection{Multiple behaviors and behavioral transitions}
In addition to walking or running while freely exploring the obstacle field, the animal displayed other behaviors during interaction with the terrain. For example, when moving in dense obstacle field, the animal often rolled its body in to the narrow gap between the pillars (Figure \ref{fig:jeb_2}A) to traverse and occasionally climbed up the pillars Figure \ref{fig:jeb_2}B). In sparse obstacle field, the animal often swept its antennae during free exploration (Figure \ref{fig:jeb_2}C). The animal also transitioned between these behaviors and occasionally stopped moving.

\clearpage
\section{Discussion}
We created a reconfigurable laboratory platform for large spatiotemporal scale measurements of small animal locomotion through complex terrain with large obstacles. Compared to existing locomotion arenas, our device increased the limits of experiment duration by $\sim100\times$ and traversable distance by ~100×. Such large spatiotemporal scales may be useful for studying spatial navigation and memory \citep{collett2013a,varga2017a} in terrain with large obstacles, and the larger spatial resolution may be useful for studying interaction of the animal (body, appendages, sensors) with the terrain in detail \citep{cowan2006a,duerr2018a,okada2006a}. There may also be opportunities to advance neuromechanics of large obstacle traversal by combining the terrain treadmill with miniature wireless data backpacks \citep{hammond2016a} for studying muscle activation \citep{sponberg2008a} and neural control \citep{mongeau2015a,watson2002a}. The treadmill design may be scaled down or up to suit animals (or robots) of different sizes. Our treadmill enables large spatiotemporal scale studies of how locomotor behavior emerges from neuromechanical interaction with terrain with large obstacles.

Our study is only a first step and the terrain treadmill can use several improvements in the future to realize its potential. First, we will add more cameras from different views to minimize occlusions and diffused lighting from different directions to minimize shadows, as well as increase camera frame rate to accommodate rapid antenna and body movement, to achieve more reliable tracking of the animal body and antenna through cluttered obstacles during which 3-D body rotations are frequent. Second, feedback control of the sphere can be improved to use not only position but also velocity of the animal to better maintain it on top. This will be particularly useful if the animal suddenly accelerates or decelerates when traversing obstacles. Furthermore, for longer duration experiments, animal could be perturbed when at rest to elicit movement by automatically moving the treadmill. Finally, we need to take into account how locomotion on the spherical treadmill may affect the animal’s sensory cues as compared to moving on stationary ground \citep{buchner1994a,stolze1997a}
\clearpage
\cleardoublepage

\chapter{Conclusion}
\label{chap:conclusion}

\section{General remarks}
In this dissertation, we integrated biological, robotic, and physics studies of beam obstacle traversal and ground self-righting to demonstrate our hypotheses of physical interaction-mediated locomotor transitions. We made the following discoveries.

\subsection{Reasoning about terradynamic favorability during physical interaction}
Physical interaction is easier via modes that are terradynamically favorable. Due to terradynamic favorability, the system state is more likely to be attracted to a more favorable mode (landscape basin) or transition to one when it exists. When traversing flimsy beams, it is physically easier to move forward by pushing down the beams (Figure \ref{fig:p1_1C}). By contrast, stiff beams are difficult to push against and deflect whereas transitioning rolling the body into the beam gaps enables easy traversal. While this was not explicitly demonstrated for ground self-righting, we can infer that both roll and pitch upright modes are equally favorable as they have the same system potential energy (Figure \ref{fig:elife_f5}). However, transition to pitch mode does not occur due to the large pitch barrier to be overcome compared to the pitch kinetic energy fluctuation (Figure \ref{fig:elife_f7}; see below). 

\subsection{Criteria for barrier-crossing locomotor transitions}
We demonstrated that locomotor transitions of animals and robots during beam obstacle traversal and ground self-righting are probabilistic barrier-crossing transitions on their potential energy landscape. Our observations from robotic physical modelling showed that transitions mediated by purely mechanical interactions occur stochastically from one mode (landscape basin) to a more favorable mode when system kinetic energy fluctuation is comparable to the barrier to be overcome. In beam traversal, the kinetic energy fluctuation was generated by oscillatory leg propulsion whereas in self-righting, vigorous leg flailing generated them.

Alternatively, when kinetic energy fluctuation is not enough to overcome potential energy barrier along a direction, potential energy landscape maybe modified by changing relevant parameters of locomotor-terrain system, to reduce the barrier along other directions and enable smaller kinetic energy fluctuation to probabilistically induce transitions. For example, in self-righting, wing opening reduces the barrier for the roll mode and enabled small kinetic energy fluctuation along roll direction (from leg flailing) to probabilistically induce body rolling. 

In general, transitions in animal beam traversal and self-righting followed the similar criteria as in passive mechanical robots. However, observations from the beam traversal studies provided evidence that they can often transition even when kinetic energy fluctuation is insufficient. This strongly suggests that transitions maybe facilitated by sensory feedback (see Future Directions). Because self-righting is an emergency response, it is likely that the self-righting occurs in a feedforward manner with minimal sensory feedback. 

\section{Stereotypy and Variations}
Locomotor modes in both beam traversal (pitch and roll) and self-righting (metastable and roll) are strongly attracted to energy basins on potential energy landscapes, leading to stereotypy in locomotor modes, transitions, and system state trajectories on landscape. Stereotypy emerges directly from animals' and robots' mechanical (and likely neural) systems directly interacting with the physical environment under constraints. In traversal of stiff beams, the animal and the robot are constrained to pitch-up into the gap due to the large restoring forces from beam deflection; because the robot’s other degrees of freedom are constrained except pitching and rolling, variations in movement will induce favorable rolling mode with no restoring forces. Similarly in self-righting, the metastable triangular base of support that emerges from wing opening constrains the robot from pitching or rolling, but sufficient kinetic energy can induce body rolling. Variations in movement that create kinetic energy fluctuation in both model systems lead to stochastic transitions resulting in beam traversal and self-righting and can be advantageous—in our case, for traversal and self-righting—when locomotor behavior is separated into distinct modes. 

\section{Implications for animal and robot locomotion}
\begin{itemize}
\item Methods developed to measure and analyze animal movement across large spatiotemporal scales in the existing treadmill will help begin to understand how transitions occur during slow, free exploration where sensory feedback is dominant (Chapter 4).
\item Variations such as kinetic energy fluctuation are usually considered as adverse to the robot performance. Our discovery of kinetic energy fluctuation-drive locomotor transitions suggest that embracing the variations from oscillatory self-propulsion may prove useful (to a certain degree) when moving in unpredictable natural environments with beam-like obstacles or self-righting (of course, too large a variation or fluctuation may begin to affect robot control). This also suggests that robots must destabilize themselves, at least momentarily, in order to transition to a different mode (Chapters 2 and 3).
\item Simultaneous use of multiple appendages can modulate kinetic energy fluctuation relative to potential energy barriers and facilitate transitions (Chapter 3).
\end{itemize}

\clearpage
\section{Future directions}
While the investigation in this dissertation have addressed a few of the knowledge gaps, more questions have been raised than those answered.  Here I briefly mention some of the related directions that I find intriguing and are possible extensions of the studies in this dissertation. 
\subsection{Neuromechanical interaction with during locomotor transitions} 
The studies in this dissertation (\cite{othayoth2021a}; \cite{othayoth2020a}) focused on locomotor transitions in the feedforward regime of locomotion (\cite{dickinson2000a}; \cite{nishikawa2007a}), by eliciting transitions during escape responses in which the role of sensory feedback is diminished due to the inherent neuromechanical delays (\cite{sponberg2008a}). However, an interesting observation from cockroach beam obstacle traversal was that the animal can make a barrier-crossing transition even if kinetic energy was not sufficient to overcome potential energy barriers (Chapter 2, Figure \ref{fig:p18_6}) (\cite{othayoth2020a}). Following this, a recent study led by Yaqing Wang (\cite{wang2021a}) demonstrated that during beam traversal, cockroaches actively adjust body and appendages, which facilitates locomotor transitions. An interesting question that is yet to be answered is how different neural and mechanical sensory streams (\cite{mongeau2021a}; \cite{roth2016a})—for example, antennal sensing (\cite{cowan2006a}; \cite{mongeau2014a}; \cite{mongeau2015a}), proprioception from legs (\cite{agrawal2020a}; \cite{tuthill2018a}) and strain sensors (\cite{tuthill2016a}) distributed across body—are integrated (\cite{dickinson2000a}; \cite{mongeau2021a}) to elicit or avoid locomotor transitions \citep{Xuan2021b}. 

\subsection{Decision making in natural environments}
In addition to the combined interactions of the animal’s internal states (\cite{calhoun2019a}) (such as hunger and fear) and external cues perceived by other modalities (\cite{kennedy2014a}) (such as such as incoming predators and escaping prey),  decision making, and more broadly behavior in natural environments, emerges also from neuromechanical interaction with terrain (\cite{dickinson2000a}). The locomotor modes that the animal may choose and the necessary or desired physical interaction possibly depends on its higher-level goals as well (\cite{cisek2010a})—for example, during escape responses, transitioning to modes by that requires overcoming higher energy barrier maybe terradynamically unfavorable, but beneficial in the sense that it reduces the risk of predation. By contrast, during free exploration of environment during foraging, animal may choose to make transitions to access ecological patches with more resources (\cite{fryxell2008a}). A challenge to studying decision making during locomotor transitions is that while physical interaction may be quantified concretely using first principles (e.g., biomechanical templates and anchors (\cite{full1999a}), potential energy landscapes (\cite{othayoth2021a}), modelling the effects of risk, rewards, and resources perceived by the animal is not yet clear. Recent data-driven and optimization-based approaches have begun to represent internal states (\cite{calhoun2019a}) and biomechanical risk (\cite{hackett2020a}) and holds promise towards developing comprehensive models that would inform how animals make decisions.

\subsection{Detecting and leveraging terrain affordances}
For animals moving in complex natural environments, physical interaction with terrain components is the rule rather than an exception. However, physical interaction in cluttered, obstacle-laden terrain have been traditionally deemed as unfavorable or hostile for robots. As a result, obstacle avoidance (\cite{khatib1986a}) has been the focus of various robot motion planning approaches (\cite{latombe2012a}). However, recent opinions in robotics (\cite{koditschek2021a}; \cite{roberts2019a}; \cite{roberts2020a}) inspired by James Gibson’s theory of affordances (\cite{gibson2014a}) posit that that obstacles in terrain, which are otherwise avoided, present an opportunity to apply and direct forces to elicit motion that is otherwise not possible on flat, featureless environments. For example, bump- or gap-like terrain elements (\cite{gart2018a}; \cite{gart2018b}) present surfaces against which the animal or robot can exert forces to climb or push-off(\cite{hunt2021a})—such contact interactions are often impossible on flat, featureless terrain. In other words, obstacles afford opportunities in locomotion by providing a physical element against which effort can be directed to generate desired movement, with the requirement that the predictive models of interaction physics must be available to predict the desired forces. Development tractable physics models (\cite{astley2020a}; \cite{holmes2006a}; \cite{li2013a}) combined with intelligent terrain detection and planning algorithms will enable robots to leverage the affordances provided by their environment to move better. 

\section{Closing remarks}
Working towards the projects explored in this dissertation have helped changed my outlook on doing science and engineering, and to a certain extent, life. These studies gave me an opportunity to acquaint myself with fascinating topics in fields both related and unrelated to locomotion. More importantly, it gave me an opportunity to meet and learn from amazing people of diverse backgrounds. 
In hindsight, some plans never materialized; some were “not great, but not terrible”; but I’d like to think that most of them worked out better than I anticipated. Nevertheless, it has been a reasonably constructive endeavor for me! I wrap up with a quote from the Twelfth Doctor:

“\textit{Things end. That's all. Everything ends, and it's always sad. But everything begins again too, and that's... always happy. Be happy.}”

\cleardoublepage

\bibliography{all_refs}



%
\end{document}